\documentclass[aps,twocolumn]{revtex4}
\input epsf
\usepackage{amsmath}
\usepackage{amsfonts}
\usepackage{amssymb}%
\usepackage{graphicx}
\usepackage{epsfig}

\newcommand\makefig[3]{\begin{figure}[htbp]  \begin{center}  \leavevmode
\epsfxsize=#1 \textwidth  \hskip 0in \epsfbox{#2.eps} \end{center}
\vskip -0.00in  \caption{#3}  \label{fig:#2}
\end{figure}}

\newcommand{\bel}[1]{\begin{equation}\label{#1}}
\newcommand{\ee}{\end{equation}}

\newcommand{\reffig}[1]{Fig.~\ref{fig:#1}}

\newcommand{\beq}{\begin{eqnarray}}
\newcommand{\eeq}{\end{eqnarray}}

\def\tr{{\rm\ Tr}}

\def\mn{{\mu\nu}}

\newcommand{\vev}[1]{\langle#1\rangle}

\newcommand{\ket}[1]{|#1\rangle}
\def\be{\begin{equation}}
\def\ee{\end{equation}}
\newcommand{\eref}[1]{(\ref{#1})}
\newcommand{\Eref}[1]{Eq.~(\ref{#1})}
\newcommand{\rem}[1]{}
\def\tr{{\rm tr}}

\def\OO{{\cal O}}
\def\ZZ{{\bf Z}}

\def\gsim{\agt}
\def\lsim{\alt}

\begin{document}

\title{Why Unparticle Models with Mass Gaps are Examples of Hidden Valleys}

\author{Matthew J. Strassler}
\affiliation{Department of Physics and Astronomy, Rutgers University, Piscataway, NJ 08901}

\begin{abstract}
Hidden valleys, hidden sectors with multi-particle
dynamics and a mass gap, can produce striking and unusual final states
at the LHC.  Unparticle models, hidden-sectors with
conformal dynamics and no (or a very small) mass gap, can result in
unusual kinematic features that indirectly reflect the conformal
dynamics.  When sufficiently large mass gaps are added to unparticle
models, they become hidden valley models.  Predictions using
unparticle propagators alone overlook the most striking signals, which
are typically of hidden-valley type.  Inclusive signatures often cannot be
predicted from unparticle dimensions, and exclusive signatures are
often visible and can be spectacular.  Among possible signatures are:
Higgs decays to pairs of particles that in turn decay to two quarks,
leptons or gauge bosons, possibly with displaced vertices; Higgs, top,
and neutralino decays to more than six particles; resonances below an
``unparticle'' continuum which produce multi-body final states; etc.
The Stephanov model is deconstructed, reconstructed, and shown to be a
hidden valley model.  Some effects of strong dynamics on hidden valley
observables, not predictable using unparticle methods, are discussed,
including resonances, reduced flavor symmetry
breaking, reduced supersymmetry breaking, and a strongly enhanced hidden
parton shower.
\end{abstract}

\maketitle

\section{Introduction to Hidden Valleys and Unparticles}

The ``hidden valley'' scenario \cite{HV1} envisions a hidden sector
coupling weakly (but not too weakly) to the standard model, with a
multi-particle production mechanism and a mass gap (or an equivalent;
see below.)  It encompasses a very wide class of models, since almost any
gauge theory from the last forty years, with a mass gap added, will
fit within this category.  A few examples of such models were given in
\cite{HV1}.  The phenomenological signatures, which can include novel
and sometimes spectacular signals at particle accelerators, were
explored in \cite{HV1,HV2,HV3,HVWis}, but much more study is needed.
The main point is that some of the particles which are produced in
abundance in the hidden sector can decay back to standard model
particles on detector time-scales, creating a directly observable
signal at the LHC.

The ``unparticle'' scenario \cite{Un1} (see also
\cite{SeibergNAD,RS2,HEIDI}) envisions a hidden sector coupling weakly
(but not too weakly) to the standard model, with conformal dynamics (itself a
multi-particle production mechanism) and, in the original papers, no
mass gap.  It encompasses a wide class of models; essentially any
conformal or near-conformal field theory known from the last forty years
will do.  In this case the effects described in \cite{HV1,HV2,HV3} are
absent, because the particles in the hidden sector distribute their
energy into massless modes (or, if there is a a tiny mass gap, into
states with lifetimes too long to observe.)  Then one must rely on
less spectacular but no less interesting inclusive processes, such as
kinematic distributions in events with missing energy, and effects on
production of small numbers of standard model particles.

Both scenarios are motivated in part by the fact
that hidden sectors coupling to the standard model at or near the TeV
scale are common in non-minimal beyond-the-standard-model theories,
for example in supersymmetry breaking models and in many string theory
constructions.  The Randall-Sundrum-2 scenario \cite{RS2} is dual (by
gauge-string duality) to such a model.  Other recent examples include
the original Twin Higgs \cite{TwinHiggs} 
and Folded Supersymmetry models \cite{FoldedSUSY}.  The other
motivation for considering the two scenarios is that they both give
unusual signals at the LHC.  In the hidden valley case, these
signals often create experimental challenges that have been considered
very little, if at all.  To avoid missing these signals altogether in
the complex environment of the LHC, it is vital that these signatures
be studied.

Soon after the term ``unparticles'' was introduced, mass gaps were
added to unparticle theories, for example in
\cite{unSteph,unFox,unQuiros}.  Somewhat surprisingly, the fact that
unparticle models with mass gaps are identical to hidden valley models
--- ones that have conformal dynamics above their mass gap --- seems
not to have been recognized.  

In particular, all known conformal field theories in four dimensions
are gauge theories, which all have the parton shower dynamics that
plays an important role in \cite{HV1}.  With a mass gap, many of these
models also have hadronization and/or cascade decays, as again is
common in hidden valleys \cite{HV1}.  The predictions of hidden valley
models are qualitatively the same, and many of the experimental
implications similar, independent of whether the multiparticle
dynamics above the mass gap is conformal or non-conformal,
and whether it is weakly-coupled or strongly-coupled.  The main differences
are ones of degree.  Thus an
unparticle model with a mass gap satisfies all the criteria for a
hidden valley model, and should have the same predictions, even though
the literature, up to now, has indicated otherwise.

In this paper I will show how to make the hidden valley phenomenology
of these models more obvious.  I will emphasize that focusing on the
inclusive ``unparticle'' phenomenology often overlooks the main
predictions for experiment.  The most dramatic signals typically come
from exclusive final states that cannot be studied using current
unparticle methods.  Even the most striking inclusive signals are
often not of unparticle type.  In any case, unparticle propagators are
not sufficient for prediction; much more detailed information about
the conformal sector and about the breaking of conformal invariance is
needed.

I will now summarize the main points of this paper.  These are a
combination of general cautionary remarks and predictions of LHC
signatures that may arise in this context.
\begin{itemize}
\item Within the unparticle language for describing conformal field
theory, there are two critical issues, often ignored, that strongly
impact phenomenology.  These are the imaginary part of the unparticle
propagator, and unparticle interactions, including both
self-interactions and interactions with other composite operators in
the conformal field theory.  With a mass gap, these effects become
even more central, typically dominating the phenomenology.
(Throughout this paper, I will generally refer to a mass ``gap'', but
in fact only a mass ``ledge'' --- where one or more particles get
stuck because they cannot decay within the hidden sector, forcing them
to decay via the standard model --- is necessary for many of the conclusions.)

\item Once a mass gap is introduced at the scale $M$, conformal
invariance and the dimensions of operators cannot predict much of the
physics in the vicinity of $M$.  The details of the hidden sector and
of the precise form of conformal symmetry breaking are critical.  Not
even the dominant cross-sections, much less distributions of kinematic
variables for the most common events, can be predicted from unparticle
propagators.  Unparticle methods are reliable only on tails of
distributions, where all the kinematics lies high above $M$.
\item The hidden sector often becomes much more visible 
once a mass gap is introduced and the sector becomes a
hidden valley.  Hidden valley phenomenology can be
spectacular.  However, the range of visible phenomena that can occur
in a hidden valley is enormous, as emphasized in \cite{HV1,HV2,HV3},
and cannot be predicted from the dimensions of the operators in the
conformal field theory above the mass gap.  Instead, many more details
of the hidden sector are required in order to make predictions for the
LHC, and in particular, to determine whether the hidden sector
phenomenology is invisible, is visible but challenging, or is visible and
spectacular.
\item As an example, the mixing of the Higgs boson with an
``unparticle'' can easily lead to spectacular Higgs decays, such as
Higgs decays to four leptons, Higgs decays to eight or more partons,
Higgs decays to two or more particles which decay with displaced
vertices, etc.  This is for the same reason as in \cite{CFW,HV2,CFW4}; 
see also \cite{manyhiggs, NMSSM, JHU}. 
Again, the details cannot be predicted from the dimensions
of operators; specific knowledge of the hidden sector and 
of its conformal symmetry breaking are required for any predictions.
\item It has been proposed that a mass gap in a conformal hidden
sector will lead to a tower of narrow states, with lifetimes
decreasing as a power law as one goes up the tower.  This is only true
in an extreme situation.  Instead, it is far more likely that all but
one or two of the lightest states will decay to one another with short
lifetimes not predictable from unparticle methods.  Only the lightest
one or two states in each tower will be extremely narrow and may have
long enough lifetimes to give displaced vertices.  This is exactly
analogous to what one sees in QCD, or in pure Yang-Mills theory
\cite{Morningstar}, and is also precisely what one expects in a
confining hidden valley model \cite{HV1}.
\end{itemize}

Meanwhile, strong dynamics in a hidden sector can have other,
non-unparticle, effects.  
\begin{itemize}
\item Strong couplings can lead to sharp resonances at a scale very
roughly of order $M$, but with a distribution of masses and widths
that cannot be predicted from any unparticle method.
\item
Large anomalous dimensions (of operators not necessarily coupled to
the standard model) can suppress global symmetry breaking and/or
supersymmetry breaking in the hidden sector, with potentially
observable consequences.  It is possible that supersymmetry might be
discovered first in the hidden sector before evidence for it in the
standard model sector is convincing.
\item
Operators of high spin and large anomalous dimension, which cannot be
given low-dimension couplings to the standard model, can strongly
affect the hidden parton shower, which often plays an important role
in hidden valleys; see for example \cite{HV1}.  The effect is that
high energy events, instead of producing two jets of
hidden-sector-particles, produce a large number of soft
hidden-sector-particles, in a distribution that probably is
quasi-spherical.  This has important experimental consequences.
\end{itemize}

We first review the hidden valley scenario and its predictions.  We
briefly review unparticles and discuss the effect of adding conformal
symmetry breaking.  In Sec.~\ref{sec:breakCFT} we explore some of the
physical phenomena which can arise when conformal symmetry is broken,
emphasizing its diversity.  We reconsider \cite{unQuiros} and come to
different conclusions.  In Sec.~\ref{sec:unSteph} we deconstruct the
Stephanov model of unparticles \cite{unSteph} and reconstruct it,
coming to very different conclusions from \cite{unSteph} about the
effect of conformal symmetry breaking.  Finally in
Sec.~\ref{sec:othereffects} we discuss some non-unparticle effects of
strong coupling in the hidden sector, and their possible consequences
for observable signatures at colliders.

\makefig{0.4}{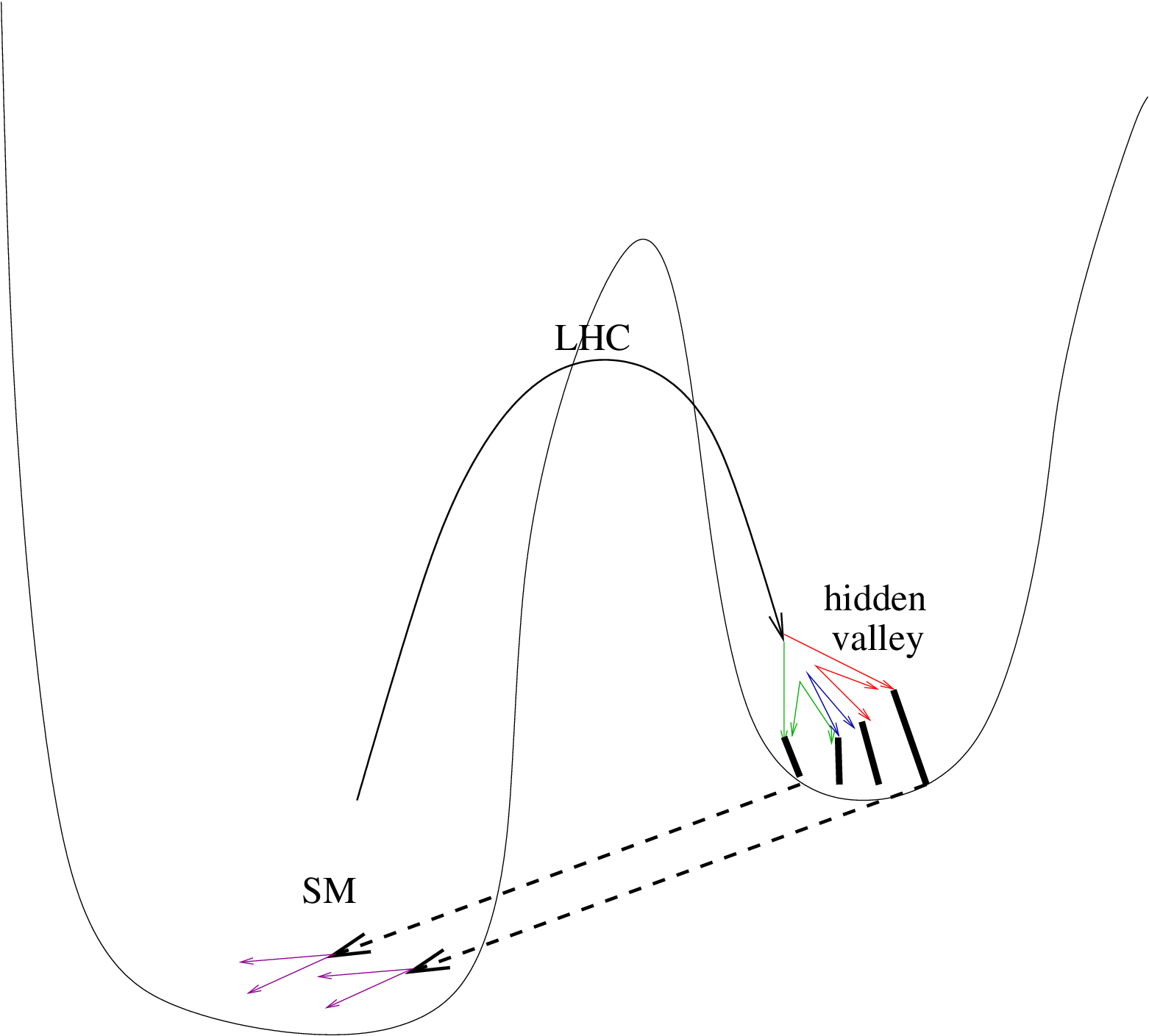}{In the hidden valley scenario, a hidden sector
couples at or near the TeV scale to the standard model sector.  In the
simplest hidden valleys, a barrier limiting production of
hidden-sector particles will be breached in the near future.  The
number of particles increases through a multi-particle production
process in the hidden sector.  A mass gap prevents decays within the
hidden sector, allowing hidden-sector particles to decay to visible particles,
often with long lifetimes due to the barrier.  Events with
high multiplicity and/or displaced vertices naturally result.}

\subsection{The Hidden Valley Scenario}

In the hidden valley scenario, a model must have three ingredients,
illustrated in \reffig{newvalley}:
\begin{itemize}
\item a coupling through a
``communicator'' (``mediator'', ``portal'') to a hidden sector,
\item a multi-particle production process in the hidden sector, and
\item a mass gap or ``ledge'' which prevents some particles in the
hidden sector from easily decaying to lighter particles in the hidden
sector
\end{itemize}
The specific realization in any model may take many forms.  

The
communicator could be any neutral particle, including 
\begin{itemize}
\item neutral gauge bosons, such as $Z$ or $Z'$
bosons \cite{HV1}, 
\item Higgs bosons \cite{HV1,HV2}, 
\item neutralinos
\cite{HV3}, 
\item right-handed neutrinos.
\end{itemize}
Communication could also be generated through 
a loop of particles charged under both standard model and
hidden sector gauge groups \cite{HV1}.  

The multi-particle production
mechanism might include \cite{HV1}
\begin{itemize} 
\item cascade decays of massive particles,
\item parton showering (but note this need not be QCD-like parton showering)
\item hadronization (but note this need not be QCD-like hadronization)
\end{itemize}
In QCD, let us note, all three mechanisms are operative, and all three
may be active in the hidden sector.  However, any one of the
mechanisms is sufficient for the predictions below.  Note also that in
the gauge-string correspondence (often called AdS/CFT
\cite{malda,GKP,WittenAdS} or AdS/QCD
\cite{WittenADSQCD,ReyTheisenYee,Sonnen,Csaki2,Nonestar,cascade},
with RS-type models \cite{RS1,RS2} arising in a certain limit), these
dynamical processes can be represented through equivalent processes in
five dimensions, as we will review in Sec. \ref{sec:unSteph}.

Finally, a mass gap (or its equivalent) could arise from several
sources:
\begin{itemize}
\item explicit masses from, for example, supersymmetry breaking or the
Higgs expectation value.
\item confinement (of a form similar to, or different from, QCD) \cite{HV1}
\item the Higgs mechanism (which is electric-magnetic dual to confinement)
\item compactification of an extra dimension (which is 
often dual to confinement or the Higgs mechanism,)
\end{itemize}
The standard model and its supersymetric extension exhibit examples of
the first three.  Meanwhile the fourth, if the extra dimension has a
warp factor, is known in some cases to be a crude dual description of
confinement in QCD, and
in some cases to be an {\it exact} dual description of confinement
effects (or of the Higgs mechanism) in some non-QCD-like gauge
theories.

The reason to take such a large scenario with so many classes of
models in it is that it makes a general class of novel predictions.
This is analogous to the way that one gathers many supersymmetric
models, extra dimensional models and little-Higgs models together
because of their basic mechanisms, which immediately imply missing
energy signals and heavy partners for some or all standard model
particles.  The predictions made in \cite{HV1}, which follow from the
general structure of hidden valley models, are
\begin{itemize}
\item new light neutral states, decaying to the standard model through
a variety of decay modes,
\item long lifetimes for the new states, including therefore the
probability of substantial missing energy and the possibility of
highly displaced vertices,
\item abundant high-multiplicity final states in high energy
processes, with large event-to-event fluctuations in multiplicity,
visible energy, event shape, and other quantities.
\end{itemize}
Other possible phenomena include
\begin{itemize}
\item non-standard multi-body decay modes for the Higgs boson, including
ones already discussed \cite{NMSSM,CFW,HV2,JHU} and beyond,
including the possibility of discovery channels involving highly
displaced vertices \cite{HV1,HV2} ;
\item new decay modes \cite{HV3} for the lightest R-parity-odd (or KK-
or T-parity-odd) particle in supersymmetric or extra-dimensional or
little Higgs models, and indeed in any model with a new global
symmetry; again these decay modes offer several possible sources of
highly displaced vertices and high-multiplicity final states.
\end{itemize}
Since these signals arise so easily and generally, and yet many of
them do not appear often or at all in the standard array of
most-studied models --- technicolor, supersymmetry and its cousins,
extra dimensions or little Higgs models --- they pose potentially new
challenges and opportunities for the Tevatron and LHC experiments.  A
little exploration of the experimental literature and internal notes
shows many of the associated issues have not yet been addressed.  More
than anything else, it is this point which makes the hidden valley
scenario important to consider.

\makefig{0.4}{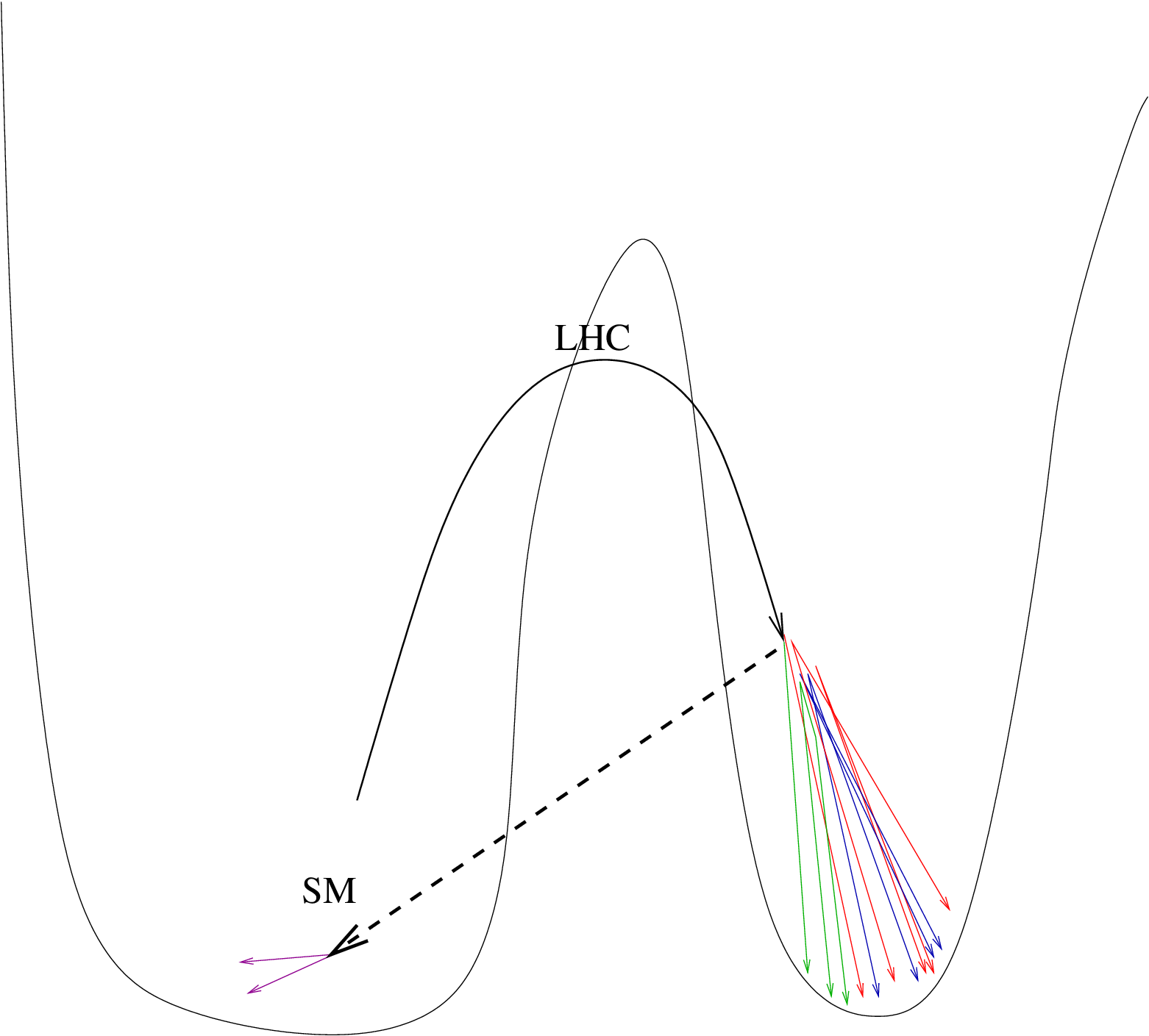}{The unparticle scenario is similar to the
hidden valley scenario, but specifically assumes the hidden sector is
conformal, which is not a necessary assumption for a hidden valley.
In standard unparticle models, the mass gap is very low, so that
standard model particles reflect the hidden physics only indirectly.
If the mass gap is higher, the unparticle model becomes a hidden
valley model, with the same signals.}

\subsection{The Unparticle Scenario and the Effect of an Added Mass Gap}

The unparticle scenario is equally general.  In its original form it requires only two ingredients:
\begin{itemize}
\item A coupling of one or more gauge-invariant local operator
$\OO_{sm}$ in the standard model to one or more gauge-invariant local
operators $\OO$ in the hidden sector.
\item A conformal field theory in the hidden sector with no mass gap, 
or a very low one.
\end{itemize}
The predictions of the scenario include missing energy signals with
unusual kinematic distributions for the visible particles that depend
on the dimension of the operator $\OO$ \cite{Un1}, and new production
mechanisms and interference effects for two-to-two scattering of
standard model particles \cite{Un2, allunparticle}, as well as in
multi-particle production \cite{multiun}.

The addition of a third ingredient --- conformal symmetry breaking
that generates a mass gap or its equivalent --- turns these theories
into hidden valley models, in particular ones with ultraviolet-conformal
dynamics.  The general predictions of the models are therefore, not
surprisingly, the same.  The detailed predictions of course depend on
exactly what the conformal sector contains, and especially, as we will see,
on how conformal symmetry is broken.  This is not easy to discern from
the literature, however.  

For example, in \cite{unFox} a strong breaking of conformal
invariance was shown to be inevitable in a large class of unparticle
models, and the physical effect of a mass gap was modeled.  In
\cite{unQuiros} the mixing of a Higgs boson with an unparticle sector
with a mass gap was explored.  In \cite{unSteph} a mass gap was added
to a toy unparticle model to illustrate some theoretical points, and
also the physical implications of the mass gap were briefly discussed.
In all these cases, as we will see, the full story was not told, and
the neglected visible signatures are of the type expected in
hidden valleys, as described in \cite{HV1,HV2,HV3,HVWis}.

\section{Hidden Valley Physics of Unparticles with a Mass Scale}
\label{sec:breakCFT}

In this section I will consider the addition of conformal symmetry
breaking at a scale $M$ to an unparticle sector, mostly concentrating
on the case where $M$ is of order
or greater than a few GeV.  Using a few toy models, I will illustrate
how the dominant phenomenology typically cannot be predicted using
unparticle techniques.  This is especially true at the Tevatron and
LHC: the rapidly falling parton distributions bias the phenomenology
toward the lowest accessible energies, where conformal symmetry
breaking is most manifest.  Even if the sector is largely invisible,
its production and interference effects are highly variable and cannot
be predicted from operator dimensions alone.  Instead the details of
the conformal sector and its conformal symmetry breaking allow for
enormous variety.  Moreover, the sector itself may become visible,
with all the phenomenology of \cite{HV1,HV2, HV3}.  If so, the
resulting exclusive events are often the leading observable at the Tevatron
and LHC, and often even more so at the ILC.

I will mostly focus on a popular scenario in the unparticle
literature, where a hidden sector is coupled to the standard model
though the Higgs boson \cite{unFox,unQuiros}.  Many conclusions drawn
are not specific to this case, although my presentation will highlight
certain details of this particular unparticle coupling.  I will
consider some other cases briefly in Sec.~\ref{subsec:others}.

\subsection{Three Toy Models of the Hidden Sector}
\label{subsec:toys}

In order to illustrate various physical points, I will introduce three
useful toy models of the hidden sector.  Model A, hidden scalar QED,
will have the advantage of extreme simplicity and great
phenomenological riches, at the cost of being fine-tuned and requiring
us to remain at weak coupling.  Model B, a scalar Banks-Zaks
theory, will be only slightly more complicated and will have even
richer physics; it is strictly conformal and fixed points might exist
at stronger coupling. However it too is fine-tuned.  Nevertheless, its
concepts can be extended to fermionic Banks-Zaks theories, which are
not fine-tuned.  Finally, model C, a supersymmetric Banks-Zaks theory,
is a theory very similar to model B, as far as its unparticle physics,
and suffers no fine-tuning.  It has physics far too rich to fully
explore in this paper, as it includes all of the phenomenological
possibilities of models A and B as a small subset.

Now I will describe the models in more detail, and how their
unparticles couple to the standard model.  Except in section
\ref{subsec:others}, where I will briefly consider other examples and
show that the conclusions are similar, I will focus on the case where
the unparticle couples to the Higgs boson.  This case is already
sufficient to uncover the exquisite complexity of hidden valley
phenomenology.

Model A is weakly-coupled scalar QED: a theory of a photon plus $N_f$
massless scalars $\phi_i$ of charge 1.
This theory has a small beta function if $\alpha N_f$ is small, as I
will assume, so it violates conformal invariance by a small amount.
But small violations of conformal invariance have small effects on
unparticle predictions, as the reader will easily verify in the
discussion below.  In this theory, the mass operators $\phi_i^\dagger
\phi_j$ develop small negative anomalous dimensions and can serve as
unparticles of dimension just below 2.

Model B is a scalar Banks-Zaks fixed point with an $SU(N)$ gauge
theory and $N_f\sim 22 N$ scalars in the fundamental representation.
Again the mass operators will serve as unparticles.  
Although fixed points at strong coupling
may exist, I will only consider the weakly-coupled cases.

I have chosen scalars rather than fermions here because I want
some weakly-coupled examples, and many authors claim unparticles only
make sense for $d_\OO<2$, to avoid divergences.  I disagree, but
do not want to be distracted by this controversy here.  The mass operator in a
scalar Banks-Zaks theory has $d_\OO= 2 - $order$(\alpha N/\pi)$,
an unparticle by any measure.  

The third toy model addresses the problem of naturalness which is
present in models A and B.  Model C is a supersymmetric $SU(N)$
Seiberg fixed point \cite{SeibergNAD}, {\it not} necessarily weakly
coupled, with $N_f$ flavors of quarks $\psi_i, \tilde \psi^j$ and
squarks $\phi_i,\tilde \phi^j$, in superfields $\Phi_i,\tilde\Phi^j$.
When $N_f\sim 3N$ the theory is a Banks-Zaks point and is weakly
coupled.  This theory is natural, and its squark-antisquark bilinear
$\phi_i\tilde \phi^j$ is an unparticle of dimension $3-3N/N_f$, which
approaches 2 from below as $N_f\to 3N$.

As mentioned earlier, the first two theories are highly fine-tuned;
they involve scalars $\phi$ and gauge bosons, and both $\phi^\dagger
\phi$ and $(\phi^\dagger\phi)^2$ are relevant operators.  However, one
can check that all results obtained using these models apply also to
the third case, which is a natural theory without fine-tuning.  Many
also apply to fermionic Banks-Zaks models, albeit for a fermion
bilinear unparticle, whose dimension is close to 3 if the coupling is
weak, though it may be much smaller at strong coupling.

\subsection{Important General Observations}
\label{subsec:generalobs}

Before I put these models into action, I would like to make a few
observations which are very important for many applications to
phenomenology, and are very typical in conformal or near-conformal
sectors.
\begin{itemize}
\item All physically reasonable conformal field theories have non-zero
three- and higher-point correlation functions.
\item All conformal field theories have composite operators of higher
spin and dimension; these will not couple to standard model fields in
the Lagrangian but can still play an important role in the physics.
\item Many conformal field theories, including these toy models for $N_f>1$,
have flavor symmetries under which 
the unparticles transform as one or more multiplets.
\end{itemize}

On the first point, note the following.  In models A and B the three
point function for $\vev{\OO\OO\OO}$ is obviously non-zero; in
perturbation theory this is a non-vanishing loop diagram.  {\it
Importantly, this three-point function does not go to zero as the
hidden-sector gauge coupling goes to zero}.  Even as the hidden
gauge theory becomes free, the unparticles do {\it not} become free.
Composite operators have interesting $n$-point functions even in free
theories, and these ``unparticle interactions'' must not be neglected,
especially once conformal symmetry is broken.  We will see this
shortly.  The only way around this is to take a conformal gauge theory
with $N$ colors and take $N$ strictly to infinity.  At any finite $N$,
the $1/N$ corrections change the physics drastically, and we will see
later that even $N\sim 10000$ is not large enough that one can ignore
these interactions for phenomenological predictions.

On the second point, there are operators in the hidden sector which
must be present.  Obviously these include the stress tensor and any
conserved flavor currents.  Less obviously, any gauge theory has high-spin
high-dimension operators; for example, the so-called ``DGLAP''
operators are always present.  These cannot serve as unparticles, since
the couplings to the standard model would be highly irrelevant, but they
can affect the phenomenology, as we will see in
Secs.~\ref{sec:unSteph} and \ref{sec:othereffects}.

On the third point, models A, B and C all have multiple unparticles,
as do many reasonable conformal field theories, transforming under a
flavor symmetry.  The Higgs may couple to one linear combination of
these operators.  This is the case in all of our examples with
$N_f>1$.  If there are no interactions other than the gauge
interactions (and their supersymmetric partners, where appropriate),
then the first two theories would have $U(N_f)$ symmetry, and model C
has $SU(N_f)\times SU(N_f)\times U(1)$ symmetry.  In models A and B,
then, the operators $\OO_i^j$ break up into two subclasses: the
operator $\OO\equiv \sum_i \OO_i^i$, which is a singlet under
$U(N_f)$, and the remaining operators, which form an adjoint of
$U(N_f)$.  Since the $U(N_f)$ currents commute with the dilatation
operator, the members of the adjoint all have a dimension $d_{A}$, and
the singlet in general has a different dimension $d_{0}$.  (For
example, in ${\cal N}=4$ Yang-Mills, theory, there are six scalars in
an $SO(6)$ global symmetry; the adjoint bilinear has dimension 2 for
any coupling while the singlet has a positive anomalous dimension; see
\cite{lightscalars} for an application to the hierarchy problem.)  In
model C, however, the operators $\OO_i^j=\phi_i\tilde \phi^j$ are in
the bi-fundamental representation of $SU(N_f)\times SU(N_f)$, and all
share the same dimension, unless additional interactions are added.

The interaction with the Higgs may break these flavor symmetries,
splitting the degeneracies within the multiplets and allowing
processes that would be otherwise forbidden.  Such effects are very
important for the phenomenology, and very model-dependent, as we will
see shortly.  In general, if multiple standard model operators couple
to multiple unparticles, they will couple to different linear combinations.
One must keep track of the breaking of any global symmetries, as it
will affect the observed phenomenology.

Thus an unparticle coupled to the standard model cannot be
treated in isolation.  It may transform non-trivially under exact or
approximate flavor symmetries in the hidden sector.  It is an
interacting object, both with itself and with other composite
operators that may not be coupled to the standard model; in fact a
free unparticle of dimension above 1 is not consistent.  It will
interact with the energy-momentum tensor and with higher-dimension and
higher-spin operators in the theory.  These
interactions become extremely important when conformal symmetry is
broken.

\subsection{The Predictions of Conformal Invariance}
\label{subsec:predCFT}

Now let us couple the Higgs boson to the operator $\OO\equiv \sum_i
\OO_i^{\ i}$ (the flavor trace of the mass operator) in any of the toy
models.  In models A and B we simply add $f H^\dagger H \OO$ to the
Lagrangian; see for example \cite{unFox} for a discussion.  Recall
that the mass operator has a small negative anomalous dimension (at
least if the gauge coupling is the largest coupling in the theory,
which we will assume).  Of course this interaction will badly
destabilize the fixed point in our first two toy models, since
relevant operators $\OO$ and $|\OO|^2$ will be induced.  But these
models are indeed toys, so we accept some severe fine-tuning, as in
the standard model, in return for simple-minded illustrations that
can easily then be generalized to realistic cases.

In model C there are two choices.  We could introduce a
term supersymmetrically, by writing $\Delta W=yH_uH_dS + \zeta S\OO$,
where $S$ is a singlet and $H_u,H_d$ are the two Higgs doublets; this
induces the term $|y H_uH_d+\zeta\OO|^2$ in the Lagrangian.
Alternatively we could introduce $\Delta W= yH_uH_d\OO$, though by
unitarity this is an irrelevant operator, so its coefficient might be
suppressed.  In either case we will also assume that supersymmetry
breaking adds additional terms to the scalar potential.  Rather than
treat these carefully, we will speak in more general terms about the
low-energy dynamics, which will be sufficient to illustrate the
complexity and diversity of possible phenomenology.  Also, note that
supersymmetry breaking itself can break conformal invariance in the
hidden sector.  In this case, a small coupling to the Higgs boson does
not necessarily imply a low conformal breaking scale.

Now when the Higgs gets an expectation value, a``tadpole'' term
$fv^2\OO$ will appear in the Lagrangian.  This is simply a mass term for
the scalar fields $\phi_i$.  Obviously the conformal symmetry of the
hidden sector is broken.  Also, through the term $(fv) h\OO$, the
physical Higgs and the operator $\OO$ will mix, which allows the
process $gg\to h^*\to \OO$, \reffig{gg2O}.  This process was studied in
\cite{unQuiros}.

\makefig{0.4}{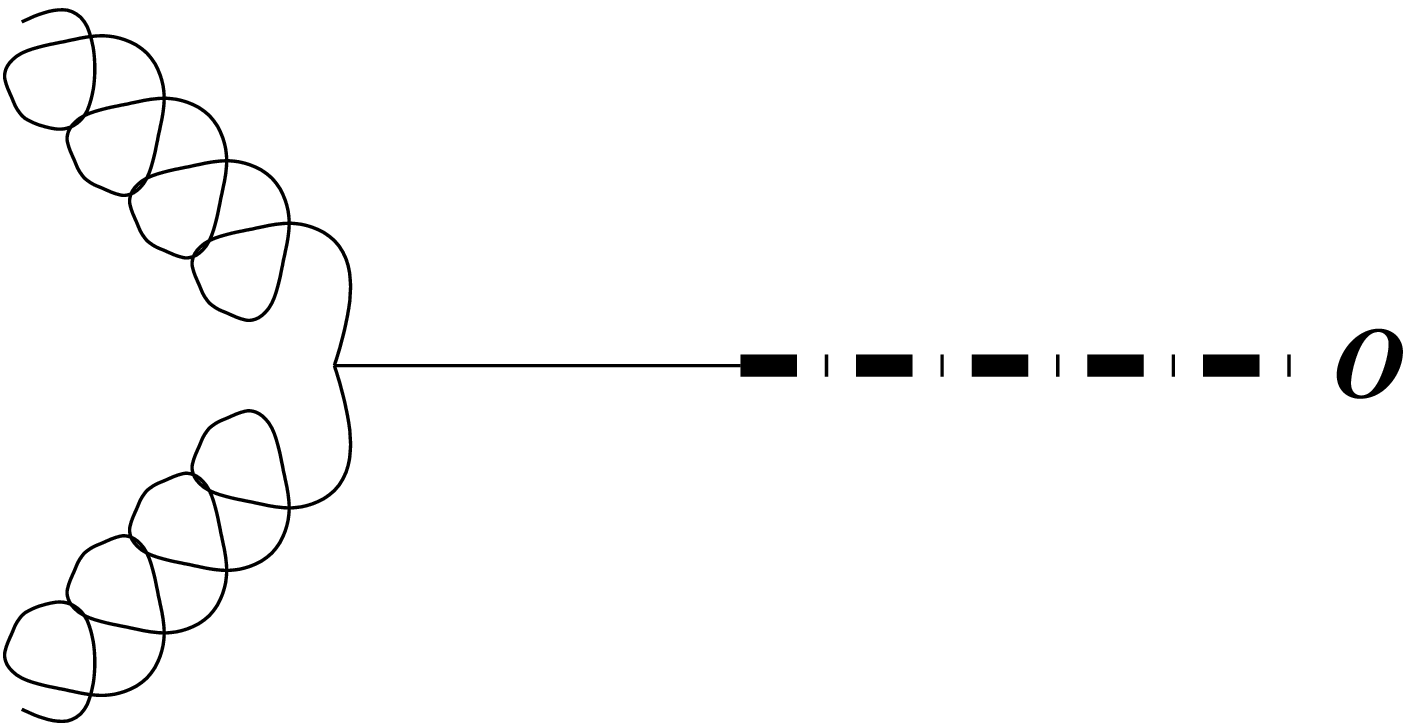}{An unparticle mixing with the Higgs boson is produced in 
$gg$ collisions.}

Note there is nothing mysterious here in our weakly-coupled toy
models.  The production of a single unparticle corresponds in more
familiar language to the process $gg\to h^*\to \phi^\dagger_i \phi^i$,
pair production for the $N_f$ scalars in the hidden sector via an
off-shell Higgs boson $h^*$.  If we consider the $\OO$ propagator
ending at a vertex, then the $\phi^\dagger$ and $\phi$ must come
together to form a loop, as in \reffig{Oproduce}.  
Alternatively, if we consider the imaginary
part of the $\OO$ propagator, this contains the $gg\to h^*\to
\phi^\dagger_i \phi^i$ process.

\makefig{0.4}{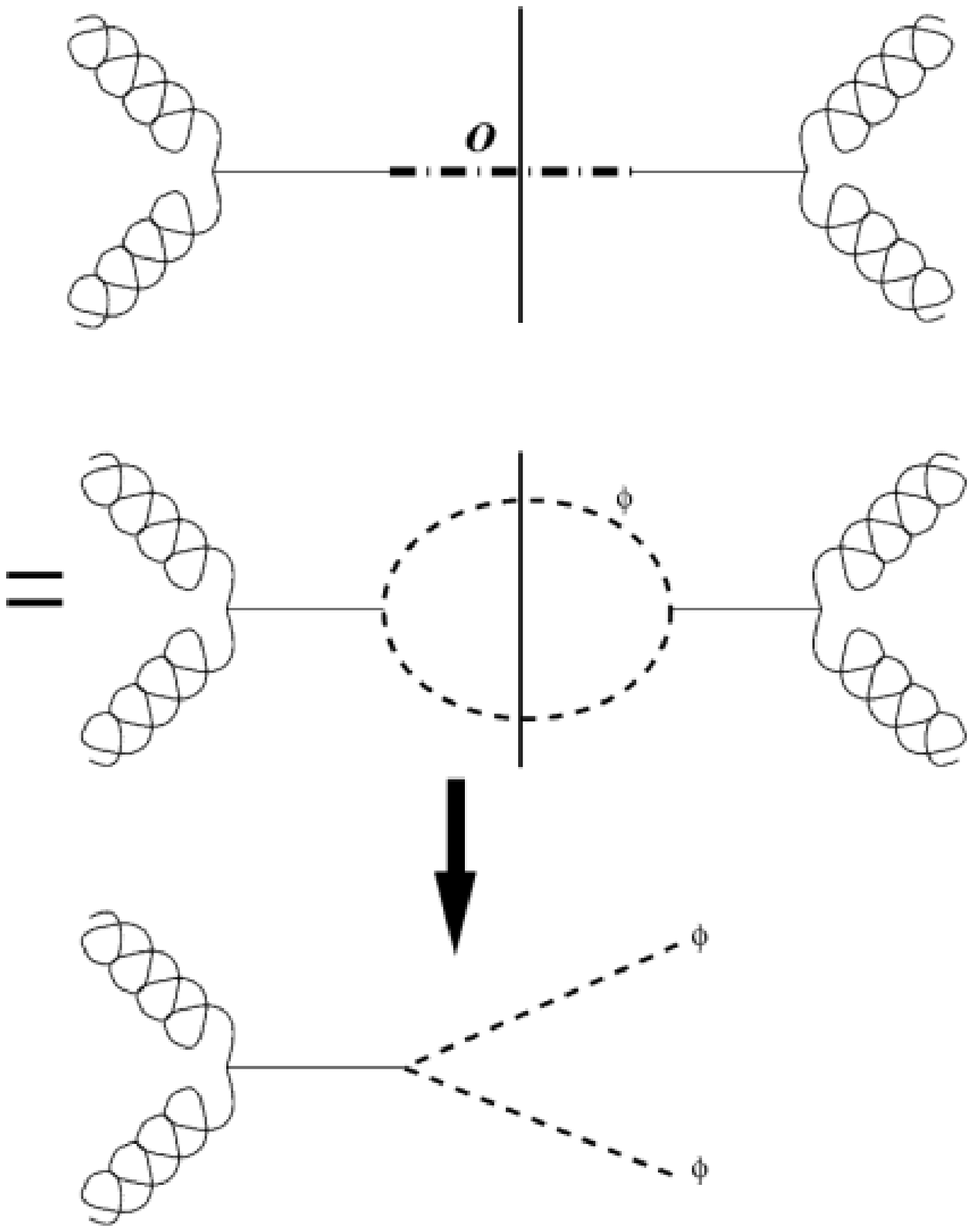}{In our toy models at weak coupling, the
unparticle propagator is a $\phi$ loop, corrected by hidden
gauge boson exchange, and 
the imaginary part of the unparticle propagator
contains the process $gg\to \phi\phi^\dagger$.}

The changed dimension of $\OO$ is also clear from this viewpoint.  The
departing $\phi$ and $\phi^\dagger$ are attracted to each other by the
hidden gauge interactions, modifying the cross-section and cause it to
fall faster than the $1/s$ behavior that would be expected if the
scalars were free.

\makefig{0.4}{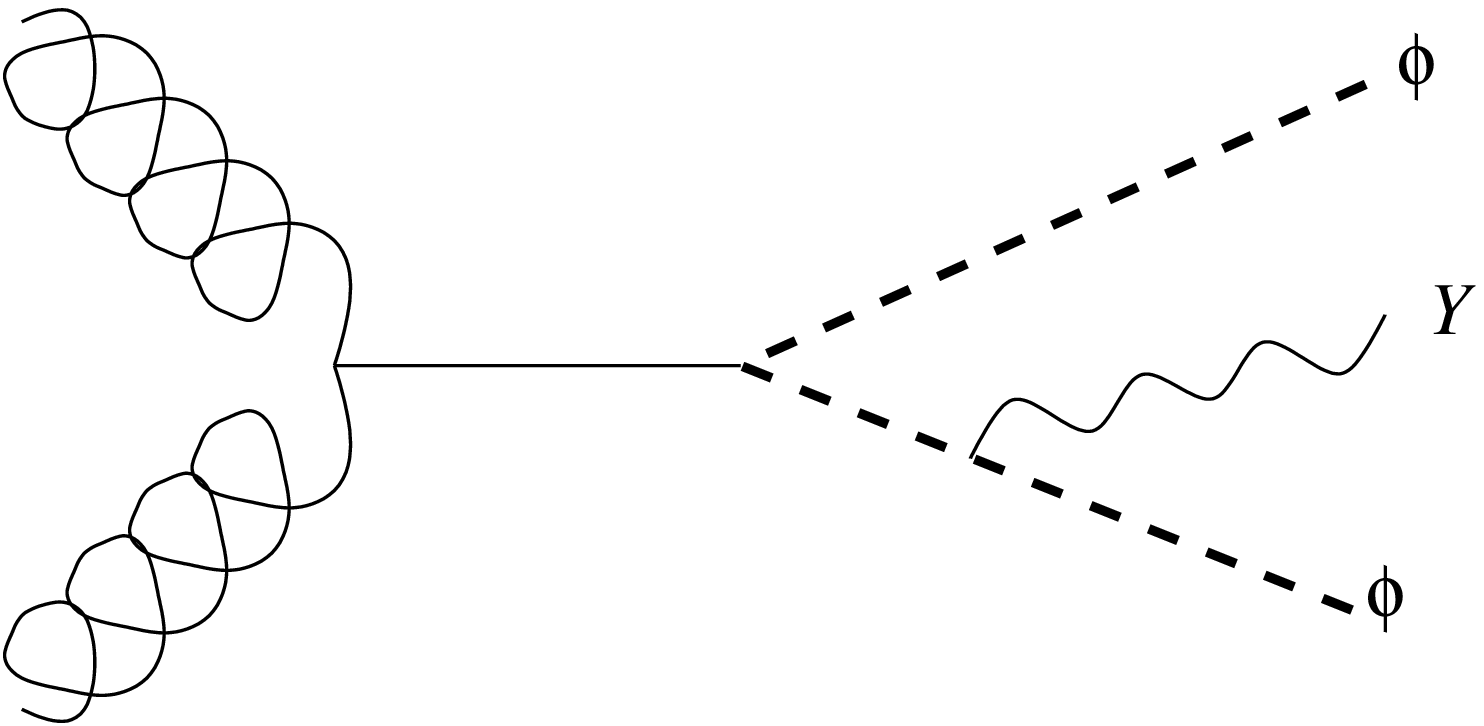}{The hidden sector is interacting:
a hidden gauge boson can be radiated off the $\phi$ particles.  This is 
{\it also} present in the imaginary part of the unparticle propagator.}

In models A, B and C at weak coupling, not only do virtual gauge
bosons at the production vertex change the scaling law, through
ultraviolet effects, but also they can be emitted in the production
process, along with $\phi_i$ and $\phi^\dagger_i$, as in
\reffig{gg2phiphiA}.  How should we represent this in terms of
unparticles?  It is another contribution to the imaginary part of the
$\OO$ propagator, which in fact will get contributions from an
infinite number of processes involving the scalars and gauge bosons.
If we only consider exclusive questions, we can treat the $\OO$
propagator as a black box and ignore exactly what is going on inside
the imaginary part.  But we will see in the next section that the
phenomenology depends crucially on looking inside this box.

\rem{\tiny
In model A, the photon can be expressed through a
gauge invariant operator $\OO'_{\mu\nu}=F^{(h)}_{\mu\nu}$, the field
strength of the hidden photon, 
but the diagram drawn in
\reffig{ggphiphiA} cannot be written as an correlation function of
gauge-invariant local operators, that is, as an unparticle scattering
amplitude, since $\phi_i$ and $\phi_i^\dagger$ are not at the same
point.  And 
in models B and C we cannot even express single gluons as
local gauge invariant operators.  Instead, we are forced to think of
\normalsize
}

The two-point function of $\OO$ must be modified at
energies where conformal symmetry breaking is important, and its
precise form cannot be determined without detailed understanding of
the hidden sector.  In \cite{unFox} a form for this two-point function
was proposed, valid for any $d<2$.  The functional form has a sharp
cutoff at some value of $q^2$; it would be visible in experiments, as
noted in \cite{Rizzo}.  But we can immediately see that this is not
the form which applies for any of our toy models.

\subsection{The Unbroken Phase: A First Look}
\label{subsec:unbroken1}

There is an important question at the first stage, which is whether
the breaking of conformal symmetry, which gives the $\phi_i$ a mass,
might also give them expectation values.  Let us first assume that we
are in an ``unbroken phase'' where the $\phi_i$ are massive but the
hidden gauge symmetry is not broken.

In models A and B, at weakly coupling, we can immediately see what
happens.  Let $m$ be the physical mass of the $\phi$ field, which at
weak coupling differs only slightly from $\sqrt f v$.  The calculation
of the cross-section is almost that of a free theory, which means that
there is a phase-space suppression of the cross-section at $q^2$ just
above $m^2$.  Here the continuum production smoothly ends.  But this
is not all: there are $\phi_i^\dagger \phi_i$ bound states, which give
a set of resonances below the cut.  These are not infinitely narrow,
because they can both radiate and annihilate to hidden gauge bosons,
so only a finite number of resonances are resolvable.  They are weak
and closely spaced if the hidden gauge coupling is small, but strong
and spread out for more strongly coupled theories.  Thus the
cross-section for ``unparticle'' production for $d_\OO$ near 2 might
resemble \reffig{CS1}.  Observation of these resonances allows for an
alternative measure of strong coupling, one complementary to the 
measurement of the power law obeyed by the falling cross-section.

\makefig{0.48}{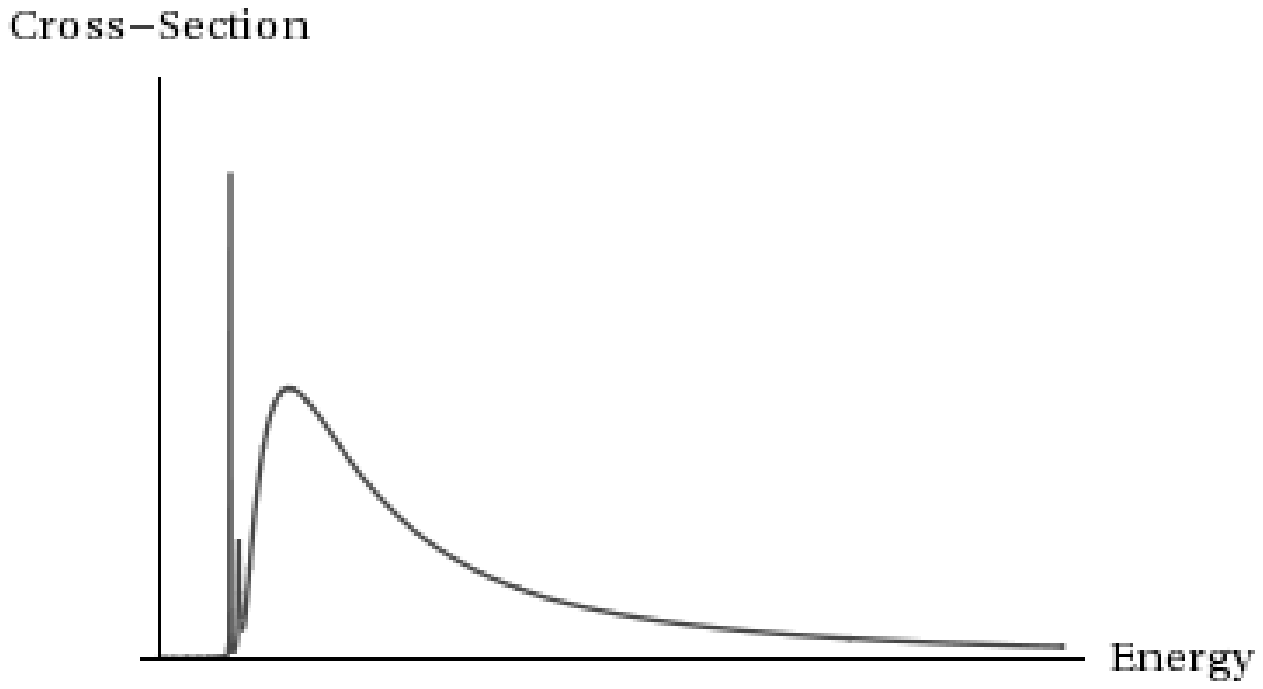}{One possible shape for the cross-section
$\sigma(s)$ to produce hidden sector particles, versus partonic
energy; parton distribution functions are not included here.  Note the
smooth turnoff, due to decreased phase space, as $s$ decreases, and
the resonances from hidden-sector $\phi^\dagger \phi$ bound states.}

But if the Higgs couples with different coefficients to the various
$\phi_i$ fields --- if it has small couplings to unparticles
$\OO_i^{j}$ other than $\OO \equiv \sum_i \OO_i^{\ i}$ --- then the
breaking of conformal invariance may be yet more complicated.  Rather
than one threshold with one set of bound states, there may be several.
This could lead to a cross-section given by the bold curve in
\reffig{CS2}.  The curve in \reffig{CS1} is shown as a thin curve on
the same plot; note that these differ strongly near the peak
cross-section, yet both match the unparticle prediction perfectly at
high energy.  These cross-sections are simple partonic cross-sections;
the cross-section at LHC, \reffig{CS3} (a log-linear plot!), where the
parton distributions are folded in, strongly de-weights the
high-energy region, where the unparticle prediction is valid.

\makefig{0.48}{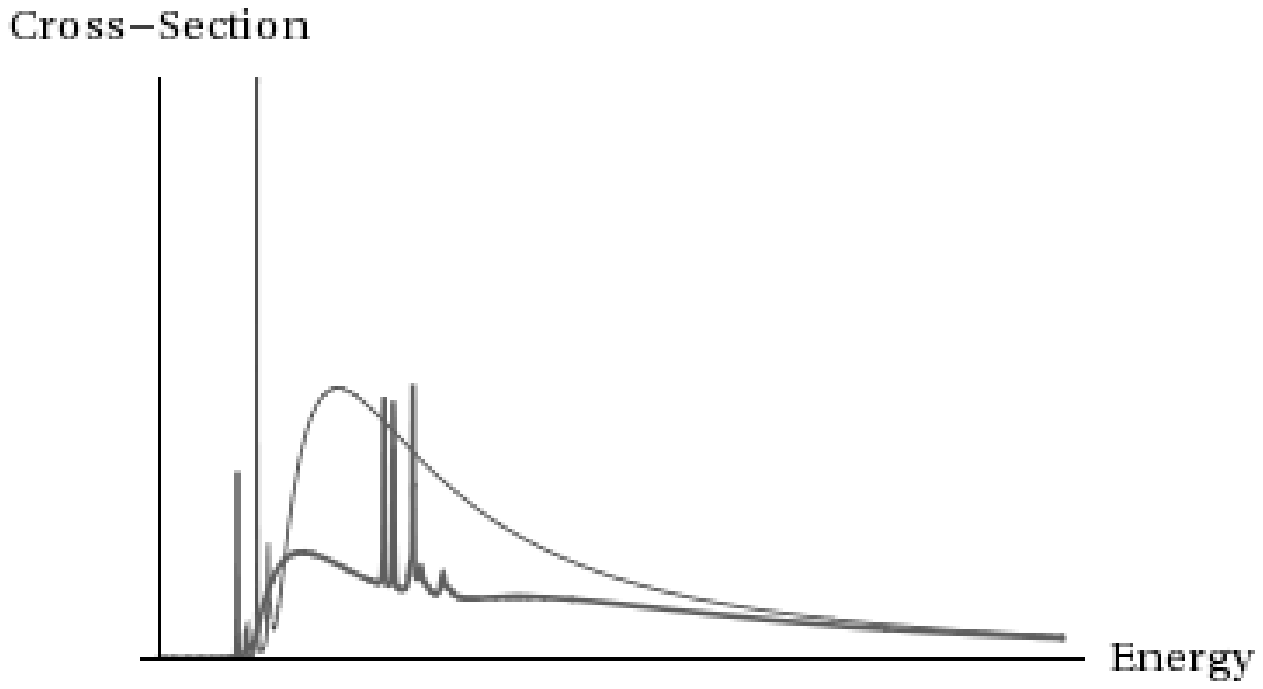}{Another possible shape for the cross-section $\sigma(s)$
versus the partonic energy, with the curve of \reffig{CS1} shown for comparison; parton distribution functions are not
included here.  Both curves are appropriate for $N_f=4$; the thick curve shows the result of having four different masses, one somewhat lighter than the other three, while the thin curve has equal masses.  Note the curves agree exactly
at high energy but differ greatly near the point of maximum cross-section.}

Moreover, in this case {\it the unparticles may not be invisible.}  If
any of the diagonal flavor symmetries are broken, either by the Higgs
couplings to the unparticles --- for example, if there is an
$H^\dagger H \OO_{12}$ coupling, even a very small one --- or by
flavor-violating $\OO$ or $\OO^2$ terms in the Lagrangian, then
nothing forbids the decay $\phi_2\to h\phi_1$.  Here the Higgs may be
on- or off-shell, depending on whether $m_2-m_1$ is larger or smaller
than $m_h$.  It is possible that this visible decay may be overwhelmed
by an invisible flavor-changing coupling of the $Y$ boson, namely
$\phi_2\to Y\phi_1$, but this is obviously model-dependent.  Certainly
we should not assume that unparticle production is invisible, and we
must reconsider the experimental implications and whether this helps,
or hinders, a measurement of the unparticle cross-section.

That the decay $\phi_2\to h\phi_1$ is possible is clear as day in the
languge of the scalars and gauge bosons in models A, B and C, at least
when the coupling constant is weak and the unparticle has dimension
just below 2.  Yet how unclear it becomes when one tries to write it
in terms of the gauge invariant operators $\OO_{i}^j$!  The otherwise
obvious process becomes quite obscure.

\makefig{0.48}{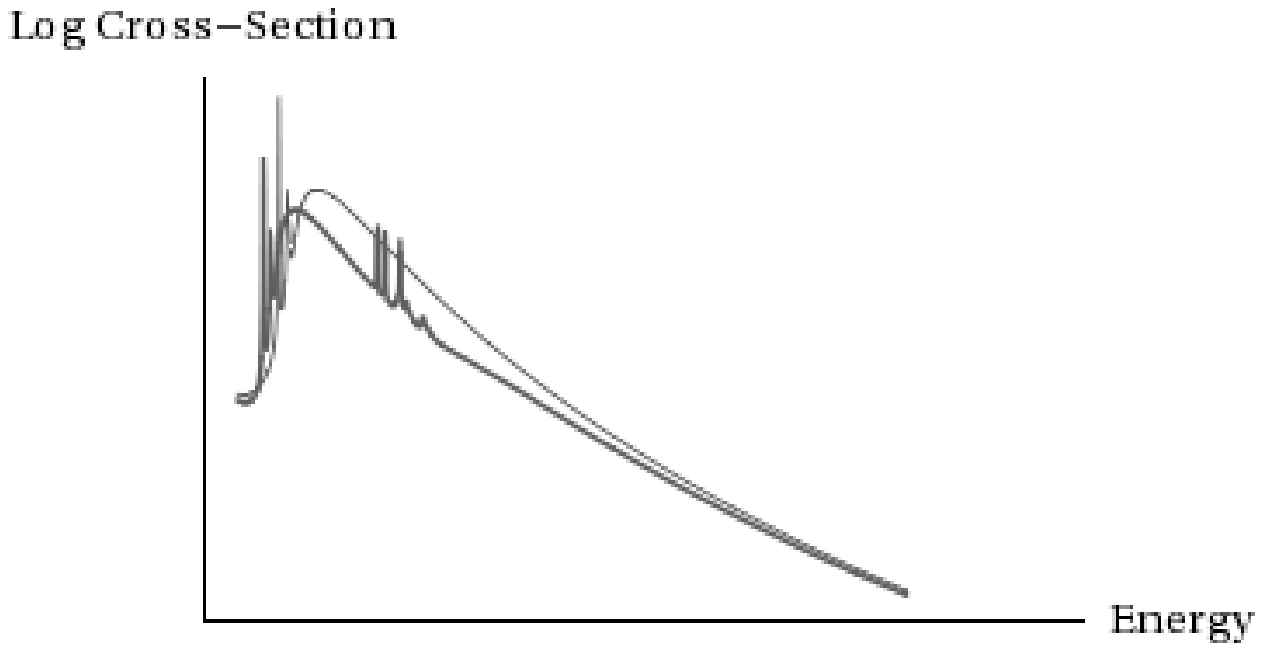}{A log-linear plot of $\log[\sigma(\sqrt{\hat s})]$
versus $\sqrt{\hat s}$ at the LHC, for the same two models shown in
\reffig{CS2}.  The gluon distribution functions greatly enhance the
region of greatest difference between the two models.}

We can try to write it down anyway, as an exercise, and also to see that
the process does not vanish when the coupling becomes stronger and
perturbative intuition fails.  In unparticle language, there is a
$\vev{\OO_{22}\OO_{21}\OO_{12}\OO_{11}}$ four-point function, even in
the limit that the hidden gauge coupling goes to zero.  This induces a
process $gg\to h^*\to \OO_{22}\to hh\OO_{11}$.  A cut through this
process includes the process $gg\to h h\phi_1\phi_1^\dagger$,
\reffig{ggOhhO}.  This would be very small in a conformal field theory
coupled to the Higgs, but when conformal symmetry is broken, the
analytic structure of the $\OO_{i}^j$ propagators is drastically
altered, making the process now unsuppressed.  The existence of
flavor-symmetry-breaking spurions, and the change in the analytic
structure of the $\OO_{i}^j$ propagators (whose cuts now end at finite
timelike values of $q^2$, not at $q^2=0$), are enough to guarantee
that processes with Higgs emission are enhanced compared to the case
with unbroken conformal symmetry.

\makefig{0.4}{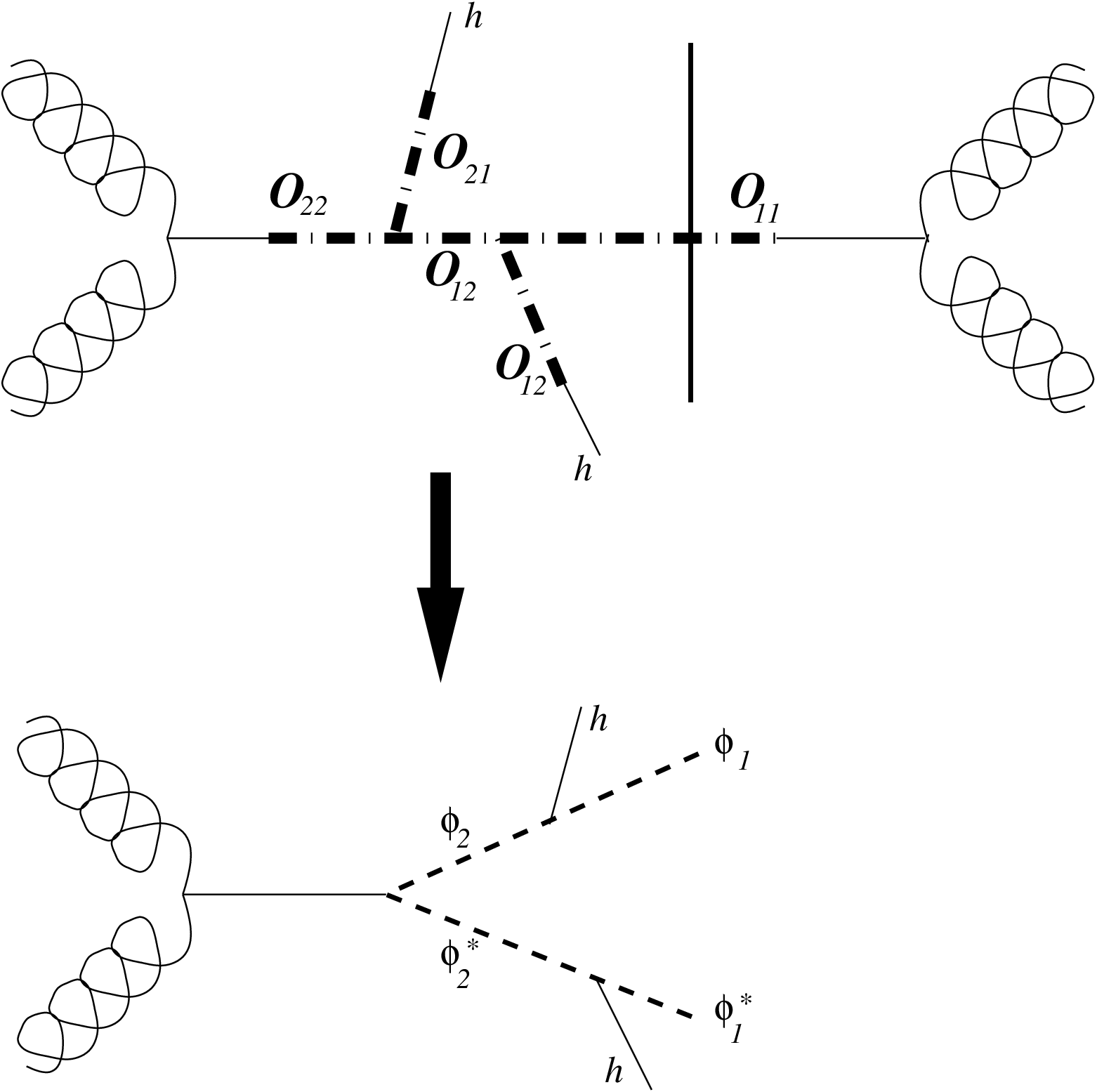}{In the imaginary part of the four-point
function hides the decays of $\phi_2\to h\phi_1$ and $\phi_2^\dagger
\to h\phi_1^\dagger$.}

\subsubsection{Possible Signatures}

Despite the fact that there are visible effects, this model may pose a
considerable challenge for the LHC.  At best one obtains the following
interesting but difficult signatures:
\begin{itemize}
\item If $m_h<2m_2$ and $m_2-m_1>m_h$ then in production of $\OO_{22}$
($gg\to \phi_2\phi_2^\dagger$) one might observe a final state with
two Higgs bosons plus MET from the invisible $\phi_1$ and $\phi_1^\dagger$.
The best channel might well be
$b\bar b\tau^+\tau^-$ plus MET.
\item If $m_h<2m_2$ and $m_2-m_1<m_h$, then the decay of $\phi_2$ must
go via off-shell Higgs bosons; the rates are suppressed.  It is
possible the potentially invisible channel $\phi_2\to\phi_1 Y$ will
dominate, but this may also be suppressed. The likely signal would
involve soft nonresonant $b$ quark
pairs or $\tau$ pairs plus MET.
\item
If $m_h> 2m_2$, then the decay $h\to \phi_2\phi_2$ is allowed,
followed by the decay $\phi_2 \to h^* \phi_1$ with an unknown
branching fraction.  From such decays, the visible energy may be quite
small, so triggering on $gg\to h$ may be impossible. In vector boson
fusion  events one might observe forward jets plus
soft jets or leptons and MET from the two $\phi_2$ decays.  Similar
signals would arise in $Wh$ and $Zh$ production.
\end{itemize}
All of these modes are challenging or perhaps even impossible at the
LHC.  Indeed these are signals best found at an ILC, or perhaps the
Tevatron.  But in any case, {\it the hidden sector is not generally
invisible.}  In particular, a search aimed at unparticles
assumed to be invisible, such as for events with a $W$ and large MET
with no central jets, might throw away the signal.

This is not the complete list of possibilities, and at larger $N_f$
the possible signatures multiply, as cascade decays ensue, giving
high-multiplicity final states.  These signatures are a minimal type
of hidden valley-like phenomenology, in which a $\phi_i$ that is trapped on
a ``ledge'' cannot decay rapidly within its own sector, and instead
decays through a visible-sector Higgs boson.  This kind of decay
should be common to many models with multiple sterile scalars mixing
with the Higgs boson \cite{manyhiggs}; however these signatures do not
seem to have been explored.

Let us compare this conclusion with that of \cite{unQuiros}, in which
a Higgs boson coupling to the hidden sector was considered.  In
figures 3 and 6 of \cite{unQuiros}, cross-sections for unparticle
production take the form of a pole below a continuum, if the Higgs is
lighter than $2m$, and of a broad resonance inside a continuum, if the
Higgs is heavier.  In the second case we recognize this as the typical
behavior of a particle mixing with a continuum -- a pole mixing with a
cut.  (See also \cite{manyhiggs}.)  In
\cite{unQuiros} it was stated that the signal is invisible, unless the
unparticle itself can decay by mixing back through the Higgs.

Why were the conclusions of \cite{unQuiros} so different from those of
this section?  A particular model for the unparticle was employed, a
specific regulated form (Eq. 2.8 of \cite{unQuiros}) of the
unparticle/Higgs coupling was introduced to assure stability of the
vacuum, and unparticle interactions were neglected.  These choices are
simple, but they are not characteristic of typical hidden-sector gauge
theories.  Most importantly, their lack of realism precisely assumes
away all the signals discussed above.  (Moreover, the assumptions made
cause the unparticle to develop an expectation value.  This has other
important effects that we will see in a moment, in
Sec.~\ref{subsec:broken}.)

In more realistic models, there are likely to be interesting
resonances, as in \reffig{CS1}, below the continuum, and possibly more
structure, as in \reffig{CS2}, if global symmetries are broken.  And
if there is any flavor mixing, as could be induced by a mismatch
between the Higgs couplings and hidden sector couplings, then we
expect decays within the hidden sector via Higgs boson emission, as in
\reffig{ggOhhO} and in the bullet points above. The rates and
branching fractions depend upon the rest of the hidden sector, not
specified in \cite{unQuiros}.

In sum, the hidden sector is easily made visible once conformal
symmetry breaking occurs.  But we are not done, by any means, at least
not in general models.  In model A, we may have identified all the
phenomenology of the unbroken phase, since the theory may simply
consist of massive scalars and a massless hidden photon.  The massless
scalars have a $\ZZ_2$ symmetry which forbids the lightest state
$\phi_1$ from decaying to the standard model, and the same is true of
the hidden photon $Y$, which also therefore remains invisible.  But in
models B and C we must address what happens at lower energy to the
so-far massless hidden gluons.  We will do this in
Sec.~\ref{subsec:unbroken2}.  

Before we do this, let us consider a
completely different possibility.  

\subsection{The Broken Phase}
\label{subsec:broken}

After conformal symmetry breaking, the theory may also end up in a
``broken phase'', where the gauge symmetry is broken.  The presence of
$|\OO|^2$ terms, either ab initio or induced when $H$ gets an
expectation value, may cause one or more of the $\phi_i$ (and thus
$\OO_{i}^i$) to develop an expectation value $w_i$.  This gives a mass
$m_Y\sim g_H w_i$ to some or all of the hidden gauge bosons $Y$.  Let
us assume, purely for simplicity, that they are all massive.

The nonzero $w_i$, aside from their effect through $f H^\dagger
H\vev\OO$ on the Higgs boson mass, also cause mixing of $H$ and $\phi$
itself.  Said another way, the operator $\OO$, after symmetry
breaking, is now just an ordinary particle at leading order
\be
\OO_i^j = w_i^\dagger w_j + w_i^\dagger\ \delta\phi^j + w^j\ \delta \phi^\dagger _i + \delta\phi^\dagger_i\ \delta \phi_j
\ee
and thus the mixing between $h$ and $\OO$ is ordinary particle mixing
between $h$ and $\delta\phi$.  Therefore the unparticle propagator
develops poles at $m$ and at $m_h$, with non-zero imaginary parts, and
with residues which depend on $f$, $v$ and $w_i$.  If the $\phi_i$
have different masses, or the $w_i$ are not all equal, there will be a
separate pole for each $i$.  

This has the effect that $\OO$ (or $\delta \phi_i$, to be less
obscure) has an amplitude to decay by $\delta\phi_i\to h^*$ to any
kinematically allowed final state of the Higgs boson, such as $b\bar
b$, $\tau^+\tau^-$, $WW$, $\gamma \gamma$, etc.  This is precisely
what happens in models 
\cite{manyhiggs,NMSSM} where there are standard-model singlets that
can mix with the Higgs boson.  This is also what can happen in hidden
valley models \cite{HV1,HV2}, which easily produce sterile scalars
that can mix with the Higgs.

But 
there is another
important effect to consider.  Just as the Higgs boson can decay to
$WW$ and $ZZ$ because it gives them their masses, the $\delta\phi_i$
can decay, if $w_i$ is nonzero and the $Y $ is sufficiently light, to
$Y Y $.  And since the Higgs can mix with the $\delta\phi_i$, it too
can decay by $h\to YY$.

More precisely, the Higgs and the $\phi_i$ form a mixed system of
scalar mass eigenstates $\hat\phi_a$, each of which can decay to
standard model fermions, standard model gauge bosons, and hidden gauge
bosons, if kinematically allowed.  Moreover, they can decay to each
other, because of the $H^\dagger H \phi_i \phi_i^\dagger$ coupling,
which after symmetry breaking induces many three-particle couplings,
such as $h\to \phi_i \phi_i$ and $\phi_j\to h h$, or more generally
$\hat \phi_a\to \hat \phi_b \hat \phi_c$.  (There are also in general
psuedoscalar states as well, but to keep the discussion under control
we ignore them here; a more serious study might reveal additional
signals.)  Whether these three-point couplings dominate over the
decays to standard model particles or to $Y Y $ depends strongly on
the couplings and the masses and the mixing angles of all the states.

Moreover, the $Y $, now massive, may not be invisible.  It will mix
with the $Z$ boson, with a model-dependent mixing angle, and decay
with a model-dependent lifetime to standard model fermions: quark
pairs or lepton pairs.  Its mass is often small and its mixing must
be small, for consistency with direct and indirect LEP bounds, so it may well decay with a displaced vertex.
For instance, if
the $Y$ field is heavier than $2m_b$, then its lifetime is roughly of
order that of the $Z$ (about $10^{-24}$ seconds), divided by the square of
a mixing angle, times $(m_Z/m_Y)^5$ \cite{HV1}.  For lighter $m_Y$
some decay channels are kinematically forbidden and the lifetime
becomes longer.  For a 20 GeV $m_Y$ to decay inside the
detector requires a mixing angle larger than $10^{-6}$, which is
certainly permitted by experimental constraints.

In \cite{unQuiros}, where the unparticle developed an expectation
value, these issues were not considered.  Admittedly it is difficult
to express or even recognize these phenomena in purely unparticle
language.  It is tricky, at best, to write a shift from one vacuum to
another, and the Higgs mechanism, in terms of gauge-invariant local
operators.
For example, unparticle interactions clearly cannot be neglected when
computing the two-point function of an unparticle in a shifted vacuum,
since three- and higher-point functions in the unshifted vacuum will
contribute to the two-point function in the new vacuum.  As all known
nontrivial conformal field theories in four dimensions are gauge
theories, an unparticle's interactions with other unparticles that
contain the hidden gauge fields $Y$, such as $\phi^\dagger_i D_\mu
\phi_j$ in our toy models, must be considered.  And in the toy models
above, three-point functions, which contribute to decays, get new
contributions from the unparticle interactions and from
Higgs-Higgs-unparticle couplings.

\subsubsection{Possible Signatures}

With all of the above effects accounted for, the potential for striking
phenomenology emerges.  Instead of an
invisible sector with a cross-section given by one or another of
Figs.~\ref{fig:CS1}--\ref{fig:CS3}, or in the figures of
\cite{unQuiros}, we may have a flurry, or perhaps even a blizzard, of
visible final states.  Several of these are illustrated in
Figs.~\ref{fig:gg2h2bbbb}--\ref{fig:gg5Y}.  Decays of any of the
scalars $\hat\phi_a$ (the mass eigenstates which are mixtures of the
Higgs and the $\phi_i$) may generate final states that range from the
well-known to the exceptional.  Examples (with resonant pairs shown in
brackets) include
\begin{itemize}
\item
$gg\to\hat \phi_a\to \hat\phi_b \hat\phi_b \to (b \bar b) (b\bar b) $ (\reffig{gg2h2bbbb}.)
\item 
$gg\to\hat\phi_a\to Y Y \to (\ell^+\ell^-)(\ell^+\ell^-)$ (\reffig{gg2h2YY}.)
\item 
$gg\to\hat\phi_a\to Y Y \to (q\bar q)(\ell^+\ell^-)$
\item
$gg\to\hat \phi_a\to \hat\phi_b \hat\phi_b \to (YY)(YY)\to[(q\bar
q)(\ell^+\ell^-)][(\nu\bar \nu)(\ell^+\ell^-)]$ (\reffig{gg4Y}.)
\item
$gg\to\hat \phi_a\to \hat\phi_b \hat\phi_c \to (YY)(b\bar b)\to[(q\bar
q)(\ell^+\ell^-)](b\bar b)$
\item
$gg\to\hat \phi_a\to \hat\phi_b \hat\phi_c \to (YY)( \hat\phi_d  \hat\phi_d )\to[(q\bar
q)(\ell^+\ell^-)[(b\bar b)(b\bar b)]$ (\reffig{gg2YYphiphi}.)
\end{itemize}
Clearly this is not the entire list.  These decays are in addition to
classic decay modes, such as $\hat \phi_a \to W^+W^-$ or $\to b\bar
b$, if kinematically allowed.  Note the decays are mainly on shell;
the fermions in the final states form resonances pairwise, so plots of
the invariant mass of the dileptons will show peaks, making the
four-lepton channel completely spectacular.  (As this paper was
completed, I learned that model A, investigated already in
\cite{SchabWells, BowenWells}, is being further studied in \cite{GW},
where some of these decay modes of the Higgs boson are also noted.)
Displaced vertices from $Y$ decays (or perhaps even from $\hat\phi$
decays) may also be present, adding additional spice to the story and
reducing backgrounds.

\makefig{0.4}{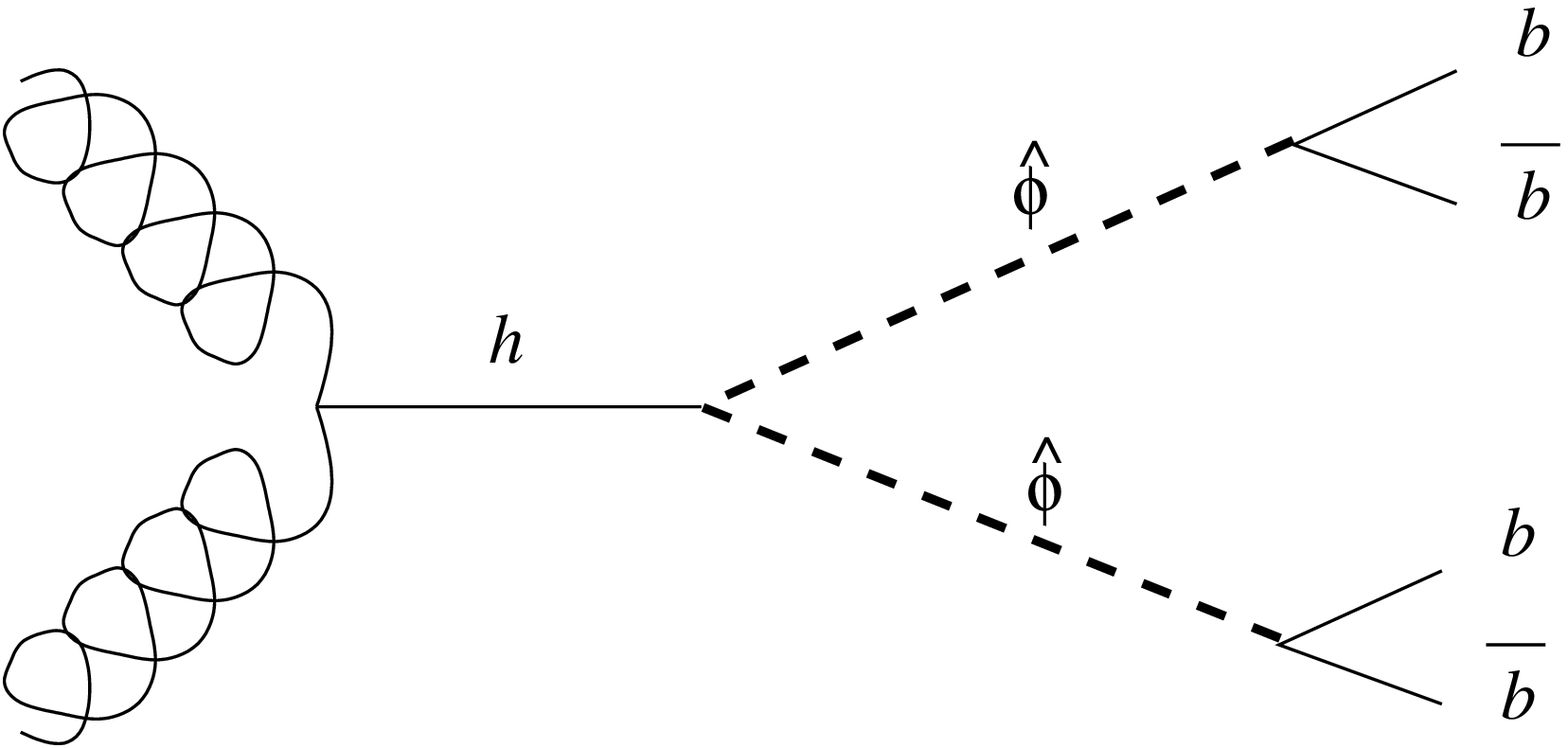}{The Higgs (or any sufficiently heavy $\hat
\phi^a$) may decay to two $\phi$ particles, which each decay to heavy
flavor; see \cite{NMSSM} for similar examples.}

\makefig{0.4}{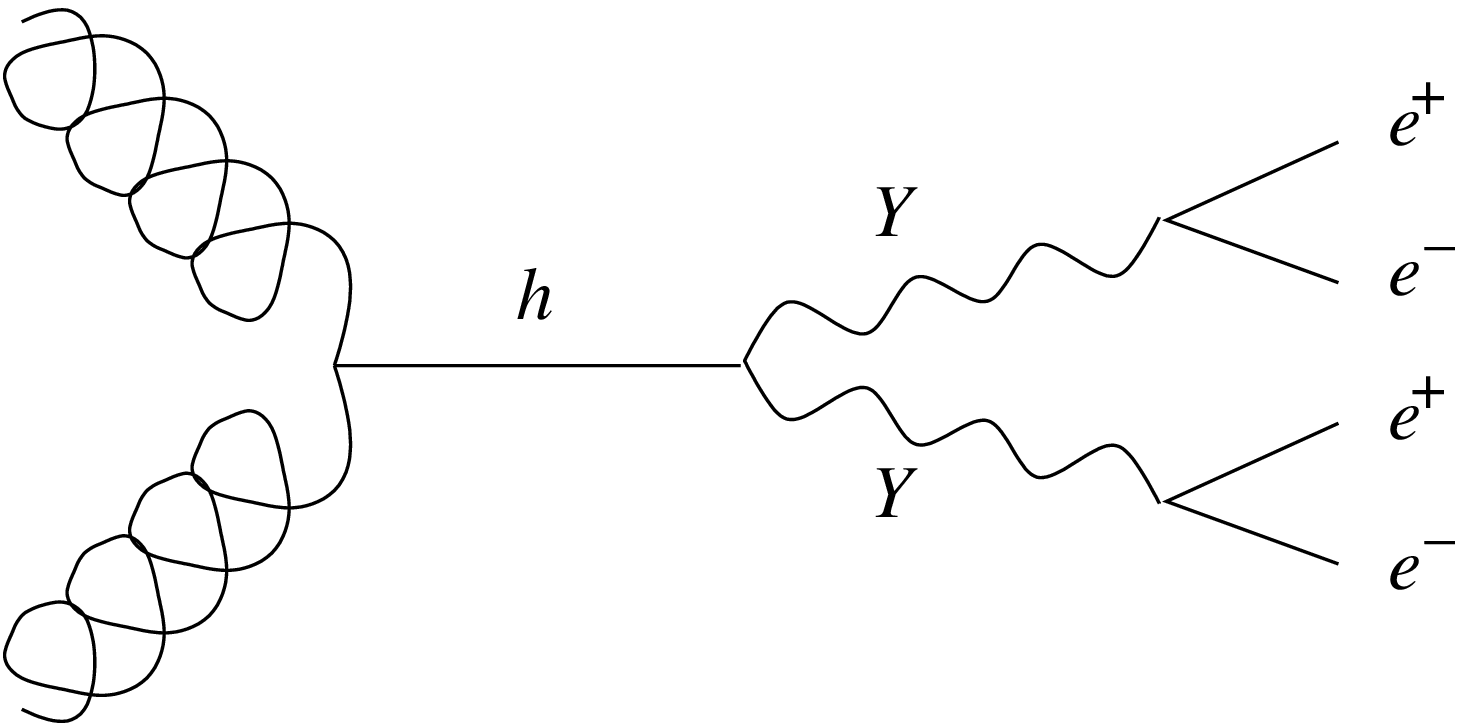}{The Higgs (or any sufficiently heavy $\hat \phi^a$)
may decay to two $Y$ bosons,
which decay to two standard model fermions each, often resulting in
two-lepton-two-jet or four-lepton final states.}

\makefig{0.4}{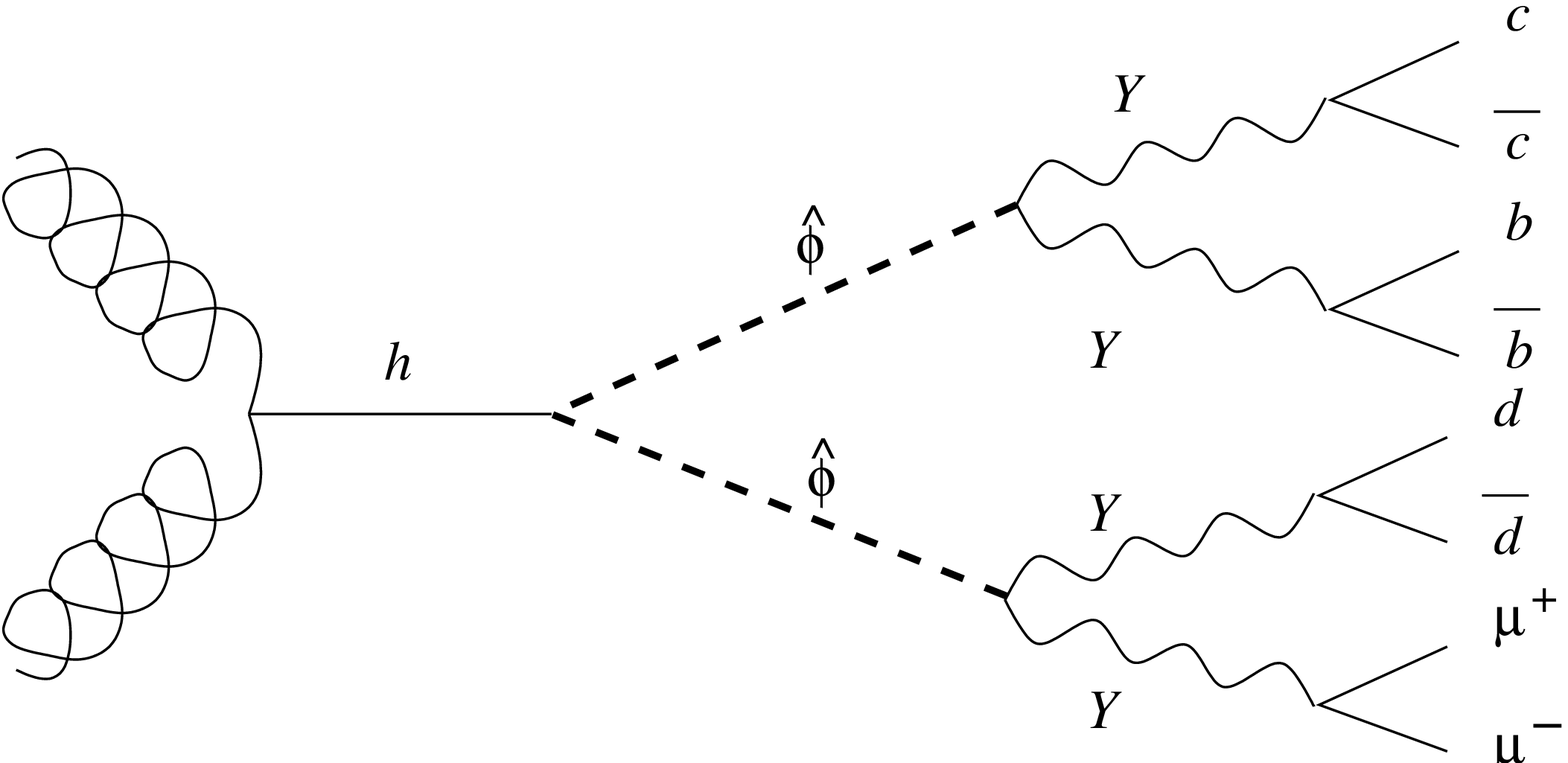}{The Higgs (or any sufficiently heavy $\hat \phi^a$)
may decay to two $\phi$ particles, which decay to two $Y$ bosons,
which decay to two standard model fermions each.  
}

The physics of multi-particle decays of the Higgs boson, and of mixing
of the Higgs with multiple scalars, has a long history.  In the NMSSM
it was noted that the decay $h\to aa$, $a\to b\bar b$ or
$\tau^+\tau^-$, where $a$ is a light pseudoscalar, often arises
\cite{NMSSM}.  Mixing of the Higgs boson with invisible sterile
scalars, spreading the Higgs signals among many resonances, has also
been considered \cite{manyhiggs}.  But the fact that multiple cascade
decays can so generically lead to multiple resonances and high
multiplicity in the final state has only been recently emphasized
\cite{CFW,HV1} and the possible relevance of displaced vertices for
discovering the Higgs boson has apparently also been overlooked until
recently \cite{HV1,HV2,JHU} (see also comments in \cite{CFW}.)

\makefig{0.4}{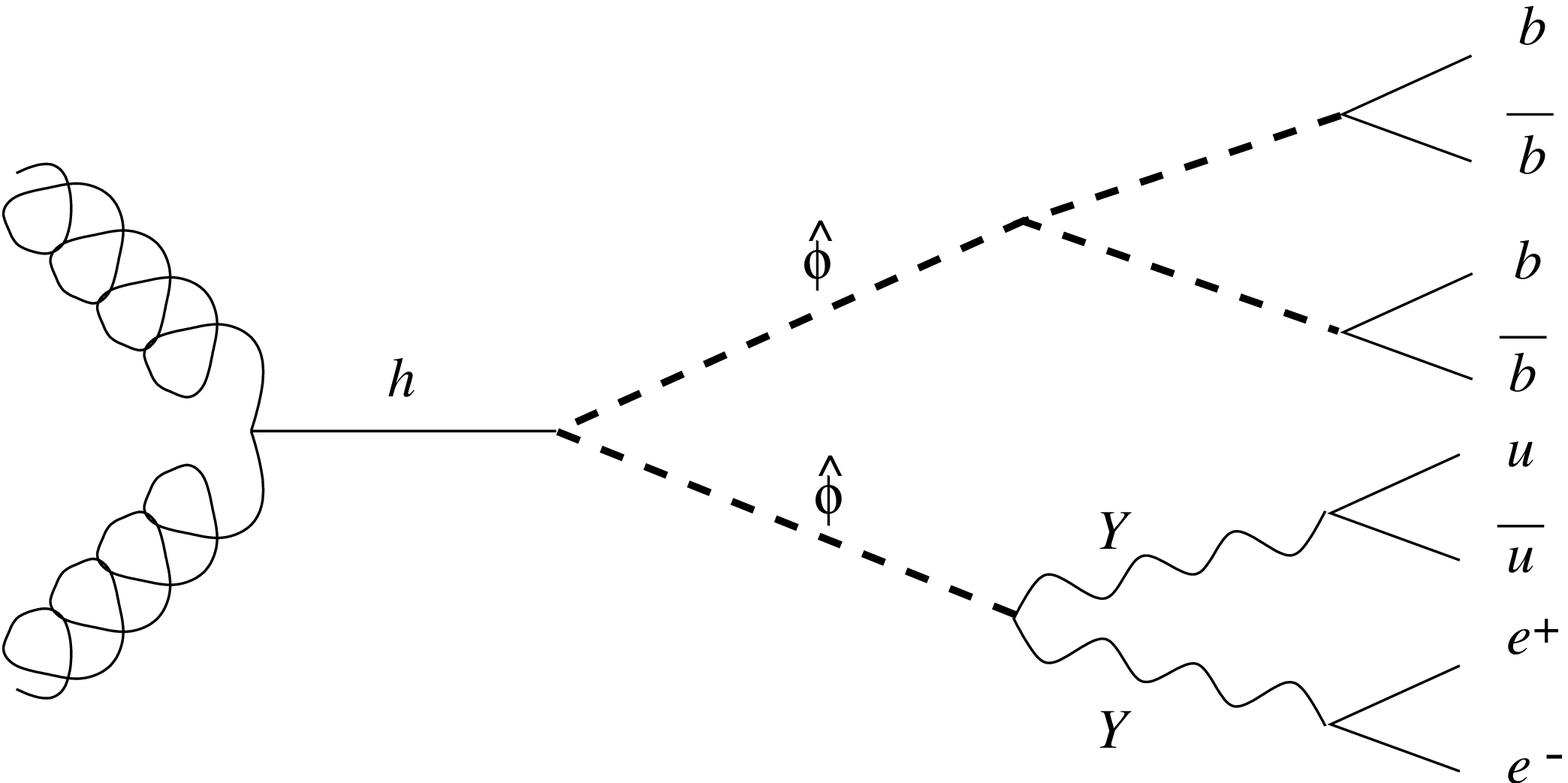}{The Higgs (or any sufficiently heavy
$\hat \phi^a$) may decay in a cascade to eight-fermion final
states; see \cite{CFW} for a similar example, and also \cite{HV2}.}

Even this is not all.  If the $Y$ is rather light, but the hidden
gauge coupling is not small, then it is easy for one or more $Y$
particles to be radiated in the production process, as in
\reffig{gg2phiphiA}.  These can then decay to additional standard
model particle pairs, as in \reffig{gg5Y}, again perhaps with a
displaced vertex.  And still more complexity may arise in models B and
C if the nonabelian $Y$ fields, whose masses are related to the $w_i$,
can decay to one another, increasing the multiplicity of final-state
particles still further.

We should not forget the unparticle production cross-section near
threshold for the lightest $\hat\phi$.  We still have a continuum
above $2m$ and $\phi$-onium resonances just below $2m$.  But the
number of resonances depends on $m_Y$; as the Compton wavelength of
$Y$ decreases, so does the number of resonances, eventually to zero.
As the resonances decay, or annihilate, they may produce mainly $Y$
bosons, whose decays would make for striking events.  Or if the $Y$ is
too heavy, decay may occur through an off-shell Higgs boson.  However,
the rate for this resonance production may be very small.


\makefig{0.4}{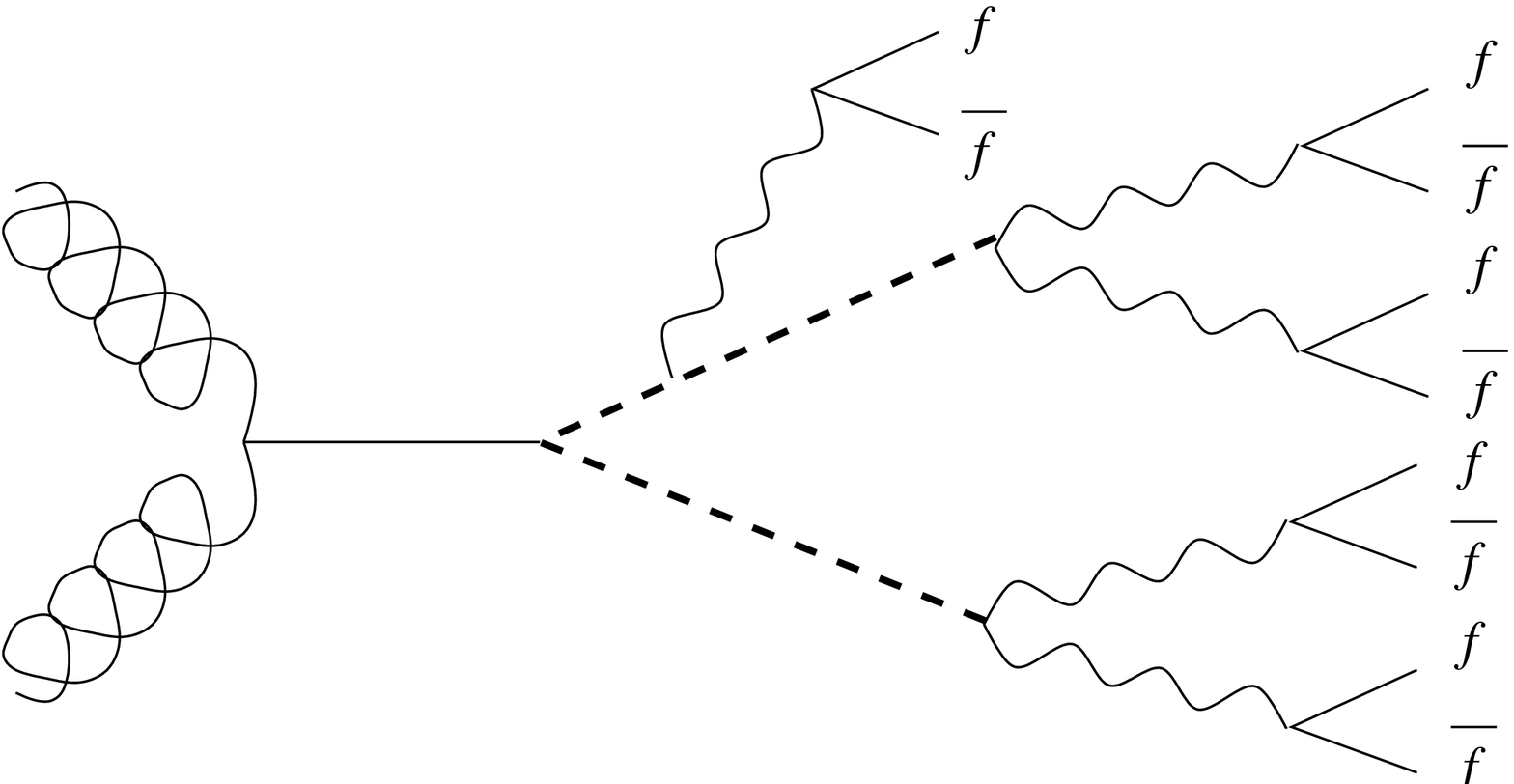}{As in \reffig{gg4Y}, with additional radiation of
a $Y$ boson; the probability for this process is proportional to the
hidden sector gauge coupling, possibly further enhanced by soft
logarithms.  Here $f$ represents any kinematically-accessible standard
model fermion, $u,d,s,c, b,t, e, \mu, \tau, \nu_i$.}

By now the fact that this is a hidden valley model should be fully
clear.  We have new light neutral resonances, with long lifetimes,
appearing in potentially large numbers through decay cascades and
through radiation \cite{HV1}.  There are new Higgs decays
\cite{HV1,HV2}.  If this is a supersymmetric world, as in model C, we
will also have the physics of \cite{HV3}, whose details depend on the
relative masses of the standard model and hidden sector LSPs.  Amid
all of this, the measurement of the conformally invariant high-energy
tail on the production process may seem less urgent, though it remains
a very important probe of the hidden sector.  Whether it is easy or difficult to measure this tail clearly depends on the visible signal, which may either
brightly illuminate or badly obscure the kinematic variable whose behavior
is controlled by the ``unparticle'' dimension.

\subsection{The Unbroken Phase: A Second Look}
\label{subsec:unbroken2}

Now we should revisit the unbroken phase, to see what happens to the
massless hidden gluons in models B and C.  Of course the unbroken
phase is a confining phase at low energy.  If the conformal hidden
gauge coupling is weak, then the confinement scale $\Lambda$ is far
below $M$, and any hidden-sector hadrons may be invisible.  But if the
gauge coupling at the fixed point is fairly strong, as is usually the
case when there are large anomalous dimensions for any operators, then
confinement may kick in within an order of magnitude or two of the
scale $M$; see for example \cite{springloaded,Nonestar}.  In this
case, we will see the remarkable phenomenology of a confining hidden
valley model, a few examples of which were given in \cite{HV1}.

There are a number of different possibilities, depending on the
precise nature of the matter in the hidden sector and the masses of
the matter.  We will cover just a couple of them; this is not the full
range of possibilities!

\subsubsection{Light Matter in the Fundamental Representation}

Suppose first that $N_f>1$, and that the Higgs couplings to the hidden
sector break its $U(N_f)$ symmetry, so that $k$ of the $\phi$ fields
end up heavy, while $N_f-k$ are light.  (As stated, this is a bit
fine-tuned.  A more natural scenario is that other couplings in the
hidden sector break the flavor symmetry, so that the Higgs couples
most strongly at low energy only to those scalars whose mass operator
has the lowest dimension.)  A simple version of this case, with
fermions instead of scalars and $N_f=2$, $k=1$, was discussed in \cite{HV1}.

We will continue to call the heavy fields $\phi_i$, but will rename the
light ones $\chi_r$, $r=1,\dots,N_f-k$.  
The heavy-heavy mesons $\sim
\phi^\dagger_i\phi^j$ will be rarely produced, except for $\phi$-onium states,
which we will ignore until the next section.  But the heavy-light
$\phi\chi$ mesons, analogous to $B$ and $D$ mesons, will always be
produced in open $\phi$ production, through the process
$gg\to h\to\phi^\dagger \phi$.  (Here the Higgs boson may be on- or
off-shell.)  The light-light $\chi\chi$ mesons, analogous to $\rho$, $K$ 
and $\pi$ mesons, will also be produced in open $\phi$ production.
Thus, just like open-charm production in the standard model,
$\phi^\dagger \phi$ pair production typically leads to two stable (and
invisible) heavy-light mesons, along with some number of light mesons.
The higher the center-of-mass energy, the more light mesons produced.
At high enough energy this process is driven by a parton-shower of
hidden sector $Y$ bosons.

The visibility of the signal then depends on whether there are
$\phi_i\to h\phi_j$ decays, as discussed in
Sec.~\ref{subsec:unbroken1}, and on whether the light mesons can decay
to the standard model.  By assumption the Higgs has smaller couplings
to the $\chi_r$ than to the $\phi_i$, but they need not be zero, so
the Higgs can mediate decays of some light meson states.  Other small
couplings, possibly higher dimension operators that couple
hidden-sector operators involving $\chi$ to standard model operators,
may also mediate light meson decays.  One example was given in
\cite{HV1}; many other examples, with varying phenomenology, may be
invented.  If one allows oneself the freedom to introduce all possible
operators, as in \cite{allunparticle}, then essentially any decay
modes for the light mesons allowed by symmetry, with a vast range of
lifetimes, are possible.

Thus, the signatures in models B and C may include the following (see
\reffig{gghhvhads}):
\begin{itemize}
\item For  each of the signatures outlined in
Sec.~\ref{subsec:unbroken1}, and the invisible production
processes in that section, we should add one or more light mesons.
\item As illustrated in \cite{HV1}, these mesons are most likely to be
scalars and pseudoscalars decaying to heavy quarks and leptons, or to
gluon/photon pairs, and possibly spin-one mesons decaying to leptons
and quarks more democratically.
\item Displaced decays are possible for light pseudoscalars and very
light vectors, or for any light state if the $\chi_r$ couple much more
weakly than $\phi_i$ to the standard model.
\end{itemize}

Other hidden hadrons with more complicated decays are certainly
possible \cite{HV1}.  For instance, in model C, we also have
supersymmetric partners of these states, which if supersymmetry is not
badly broken may allow for fermionic hadrons with additional
three-body decays.  Decays to four particles are also not uncommon.
Indeed there is much more to say about the supersymmetric case, and
the interplay of supersymmetry breaking with hidden-valley
phenomenology.

\makefig{0.4}{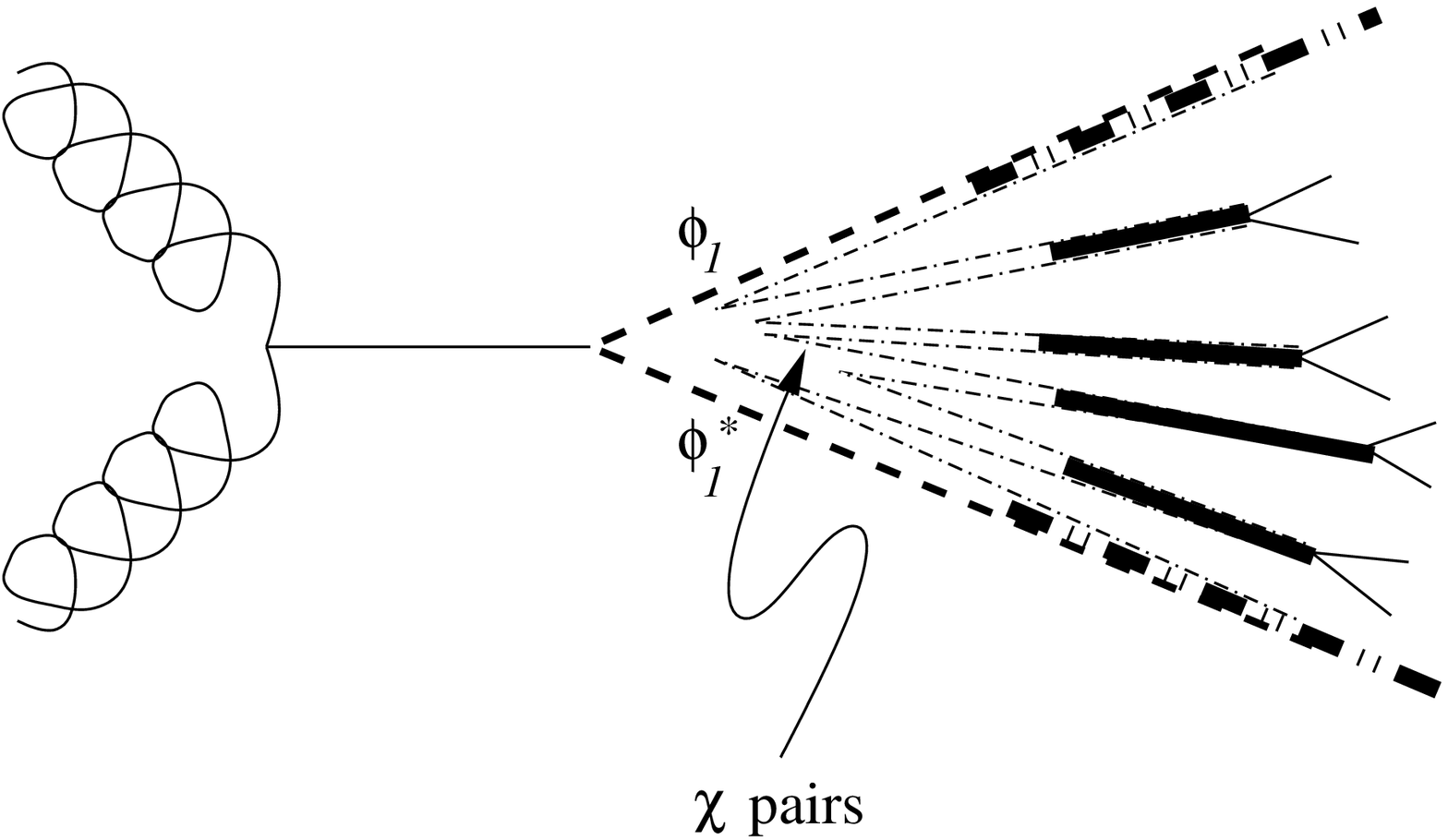}{In a confining theory with light $\chi$
fields and heavy $\phi$ fields, the production of $\phi$ particles
leads through fragmentation to two heavy-light mesons (which if stable
are invisible) and a number of light-light mesons, some of which may
decay to the standard model, possibly with displaced vertices.  The
standard model final states are variable and are shown here
schematically as sets of unmarked thin solid lines.}

Note that from the unparticle point of view, the production of $n$
light mesons and two heavy mesons must be described using an
$n+3$-point function involving gauge invariant operators built from
$\chi$ and $\phi_i$.  Again, this correlation function would be highly
suppressed in the conformal limit compared to other processes, but
here it benefits from the cuts and resonant enhancements that arise
from the production of many light on-shell states with relatively
narrow widths, along with the large phase space for decays of this
class.  Without including unparticle interactions and treating them
with great care, one might overlook these complex yet dominant
processes.

\subsubsection{Heavy Matter Only}

Now suppose instead that all the matter in the hidden sector becomes
massive relative to $\Lambda$ after conformal symmetry breaking.  Then
the low-energy limit is a pure hidden Yang-Mills theory, with a hidden
confining flux tube.  This very generic situation was also considered
in \cite{HV1}.  The low-lying hidden hadrons are hidden glueball
states, of mass $\sim \Lambda$.

When a pair of $\phi_i$ is produced, they cannot escape one another;
they are bound by a string that cannot easily break.  This has a
dramatic effect on the unparticle propagator: in such a theory, the
propagator is not a cut above $2m$ but is instead a sum of bound state
resonances continuing up to very high energy, with a cut setting in
only at $q^2=(4m)^2$, where two such resonances can be pair produced.
Note these states are not stable: all can decay by emission of hidden
glueballs except for the lightest, which can decay via annihilation to
two or more hidden glueballs. The widths of the resonances grow as one goes up
the tower, just due to the phase space for the decays, so eventually
they blur together to form a continuum which must satisfy the
contraints of conformal invariance.  Where this happens depends on the
details of the theory, but above this scale the
calculations done with conformal unparticle propagators will apply.

As an aside, note that in the case
where the fields $\phi$ carry electric charge or color, they are often
called ``quirks'' \cite{KLN, HV1}.  In this case their bound states may
also radiate standard model gauge fields.  This was very briefly
considered in \cite{HV1} for confinement scales well above 1 GeV, but
the physics is very subtle, and the question of the
expected LHC signals is still under study \cite{HW}.  The case of quirks with
confinement scales well below 1 GeV has many special features and has
been considered by \cite{KLN,HW}.

Clearly the observability of the physics depends in this case on
whether the hidden glueballs can decay visibly.
According to lattice simulations \cite{Morningstar} there will be
numerous glueballs, of various spins, parity and charge conjugation,
which cannot decay to other glueballs. How each one decays --- its
decay mode, lifetime, and branching fractions --- depends in detail on
whether there are nonzero couplings between hidden gauge invariant
operators $\tr F_{\mu\nu} F^{\mu\nu}$, $\tr F_{\mu\nu} \tilde
F^{\mu\nu}$, $\tr F_{\mu\nu} F^{\nu\rho}$, etc., and operators in the
standard model.  Because of the high dimension of these operators it
is easy to arrange that all these decays would occur outside LHC
detectors.  But if one or more of the lifetimes is sufficiently short,
the result will be a classic signature of a hidden valley: long-lived
light neutral resonances produced in abundance, with likely missing
energy and large event-to-event fluctuations. 

In particular, hidden $\phi$-onium production will produce some number
of low-$p_T$ glueballs, with various $J^{PC}$ quantum numbers, emitted
as a $\phi$-onium state relaxes to the ground state, followed by a
blast of glueballs produced in the annihilation process; see
\reffig{gghhonium}.  The signatures are then
\begin{itemize}
\item A number of high-$p_T$ glueballs with $p_T\sim m$,
and a number of low-$p_T$ glueballs with $p_T\ll m$.
\item Decays of the various glueballs, whose masses span a range of
about a factor of three, to different final states, such as $gg$, $b\bar
b$, $\gamma\gamma$, $ggg$, etc.
\item Possible displaced vertices from one or more glueball decays.
\item The $\phi$-onium annihilation may occasionally
occur through $h^*$, producing any final state of the Higgs, such as $ZZ$, in
place of the high-$p_T$ glueballs.
\end{itemize}
Unfortunately there is no known method for obtaining reliable predictions
for the $p_T$ spectrum of the glueballs and their standard model daughters.
Much additional work on this example is needed.

\makefig{0.4}{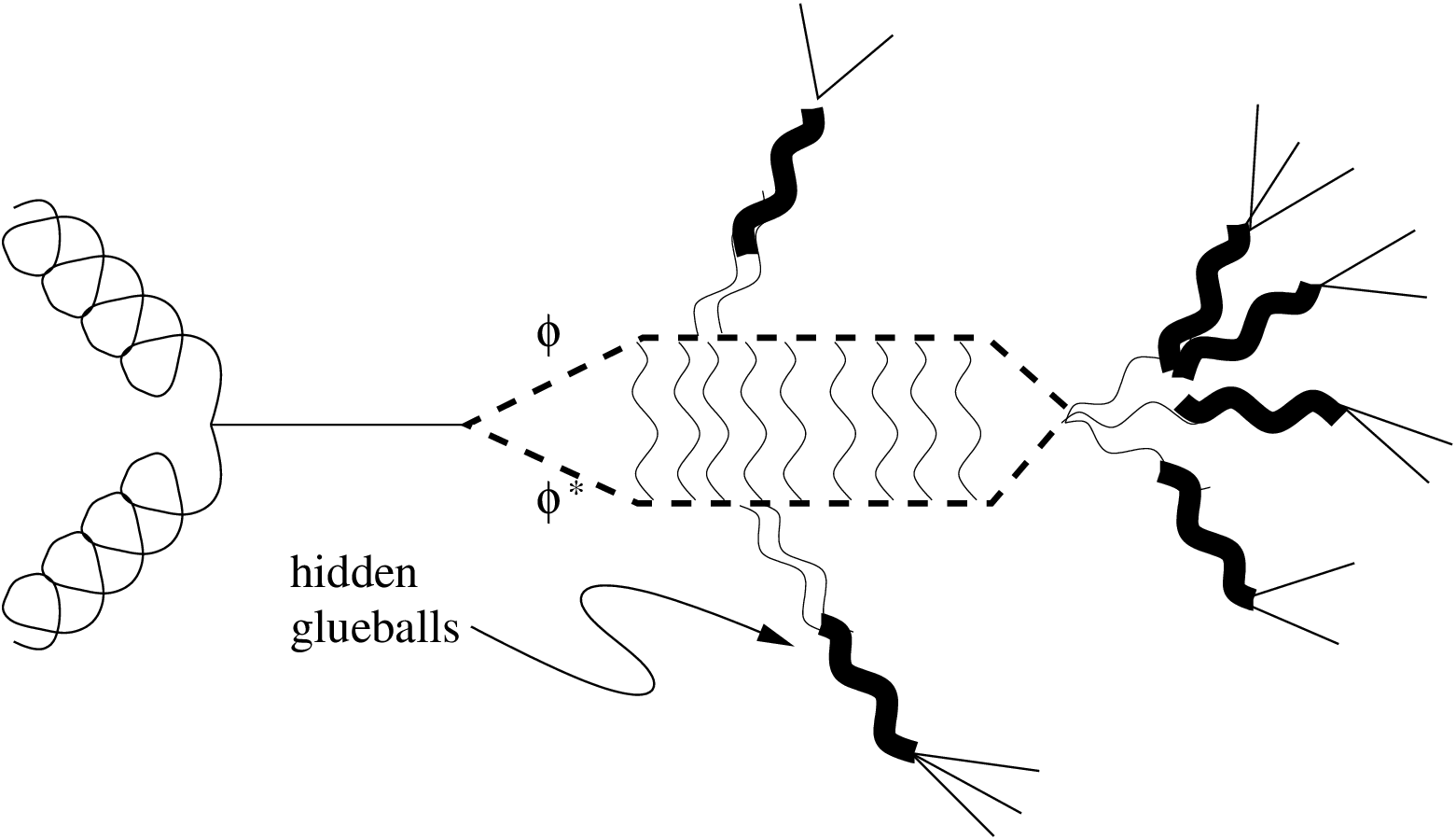}{In a confining theory with only heavy $\phi$
fields, the production of $\phi$ particles produces bound $\phi$-onium
states which decay toward their ground state via hidden glueball
emission, finally annihilating to multiple hard hidden glueballs.  The
hidden glueballs may decay to the standard model, possibly with
displaced vertices.  The standard model final states are variable and
are shown here schematically as sets of unmarked thin solid lines.}

\subsubsection{Possibilities for Future Study}

There are many other possibilities.  The hidden gauge group may not be
$SU(N)$; either the $\phi$ fields or the light $\chi$ fields, if any,
may not be in the fundamental representation; the phase of the theory
may be partially broken and partially confined.  Each of these
possibilities will change the details drastically, but in the majority
of cases the basic features of the hidden valley scenario, and its
experimental implications, will be retained.  While the conformal
invariance of the theory and the dimensions of the operators do
constrain some inclusive observables, the most dramatic effects on the
phenomenology are beyond the easy reach of unparticle methods.

\subsection{Other Models, Other Couplings}
\label{subsec:others}

We have seen that a wide variety of signals can arise even in simple
toy models.  There is an enormous diversity of phenomenological
possibilities, as is typical in the hidden-valley scenario.  But we
have only discussed a very small set of unparticle models, those with
a scalar unparticle with dimension somewhat below 2, with a coupling
to the Higgs boson.  Are these cases special?  Also, we have often
used weak-coupling intuition for guidance.  Is this misleading?

\subsubsection{Stronger coupling and $d_\OO\ll 2$}

In model C, as we decrease $N_f/N_c$, it is known that the hidden
gauge coupling becomes stronger and $d_\OO$ decreases from 2 toward 1.  
Should we expect something
completely different from our discussion above as $d_\OO$ approaches 1?

The reader is invited to consider the case $d_\OO=1+\epsilon$.  There
the operator $\OO$ can be treated a scalar that couples weakly to a
conformal sector, for example through a linear coupling to an operator
of dimension $3-\epsilon$, or quadratically to an operator of
dimension slightly less than 2.  The arguments can be repeated using
toy models.  Excellent toy models are Banks-Zaks supersymmetric fixed
points with additional gauge-singlet superfields, as in Seiberg
``magnetic'' fixed points \cite{SeibergNAD}, coupling in the
superpotential to squark-antisquark bilinears.  Not surprisingly,
since this is just model C with additional scalars, and since we
already considered model C with additional scalars (the Higgs boson itself!)
in our examples above, the physics in this case has the same features,
somewhat rearranged.  The resonances (both strong and weak), mixing,
cascade decays and complicated multibody final states that we have
seen here may arise there as well, though with rates that decrease to
zero as $\epsilon\to 0$.  The exercise is left for the reader.

\subsubsection{If there is no coupling to the Higgs boson}

If the $|H|^2 \OO$ coupling is absent, then some of the effects
discussed in the earlier sections may be absent as well.  In
particular, we may not see unusual Higgs boson decays, or the effect
of Higgs mixing with the unparticle sector.  However, we may still see
unusual decays of other heavy particles, including very rare decays of
$Z$, $W$ or $t$, or very common decays of new particles, such as
supersymmetric partners, little Higgs partners, Kaluza-Klein partners,
$Z'$ bosons, right-handed neutrinos, or new scalars other than the
Higgs boson.  What matters more than the Higgs coupling is whether
there is a mass gap which is sufficiently large.

How might a large mass gap (larger than a few hundred MeV, at least)
naturally arise without the Higgs boson coupling?  This is very easy
to imagine, since we know there must exist some mechanism that
generates the electroweak scale in the standard model sector.  For
instance, supersymmetry breaking in our own sector, if generated via
gauge mediation or supergravity mediation, will naturally generate
supersymmetry breaking in other sectors as well.  While this breaking
may be somewhat suppressed, it may still lead to a mass gap (or ledge)
at or somewhat below the 100 GeV scale.  Technicolor models may easily
break symmetry groups larger than that of the standard model,
including those of hidden sectors.  A further possibility is that the
very mechanism that leads to the local couplings at low energy between
the two sectors is precisely the same as that which generates the mass
gap; for example, if there are massive particles charged under both
groups, the masses at 1 to 10 TeV may destablize a quasi-fixed-point,
causing the hidden gauge coupling to run strong and confine at a scale
at 1 to 100 GeV.  The point is that it is very easy to imagine models
with a large mass gap without invoking the coupling to the Higgs boson.

The physics that ensues can, as before, leave the theory in any number
of phases: broken, confined, or partly both, with any number of
possible light modes that can decay via couplings to the standard
model.  The question of whether the hidden sector is invisible is very model-dependent.

\subsubsection{Effects in top quark decays}

If there are couplings to the top quark, then even without couplings
to the Higgs boson one may find remarkable signals.  For example, in
the $t\to c\OO$ transition considered in \cite{Un1}, one may have,
instead of missing energy, a decay to many particles, such as in
\reffig{t2c4Y}, where the top quark decays to nine particles,
including four hidden-sector resonances decaying to two particles
each.  This particular process can occur if the hidden sector is in a
broken phase, where the $\phi\to YY$ decay can occur.  
Note this is a fully reconstructable top quark decay.  

The total differential rate for this process may still
given by the unparticle prediction \cite{Un1}, because the energy
released in the decay $t\to c\OO$ may be large, compared
to the masses of the $\phi$ particles.  But unfortunately the
kinematic variable which one needs in order to measure the unparticle
dimension involves identifying which of the jets is the charm quark.
This may in some cases be the most energetic jet, but clearly it will
be very challenging to make the inclusive measurement and determine
the unparticle dimension directly.  A similar problem would arise in
other visible signals, such as $t\to c b\bar b b\bar b$, or $t\to c
\gamma\gamma g g$, with or without missing energy.  However, this
problem may be absent if the hidden sector particles (such as the
$Y$ in \reffig{t2c4Y}) decay with a displaced vertex.  

\makefig{0.4}{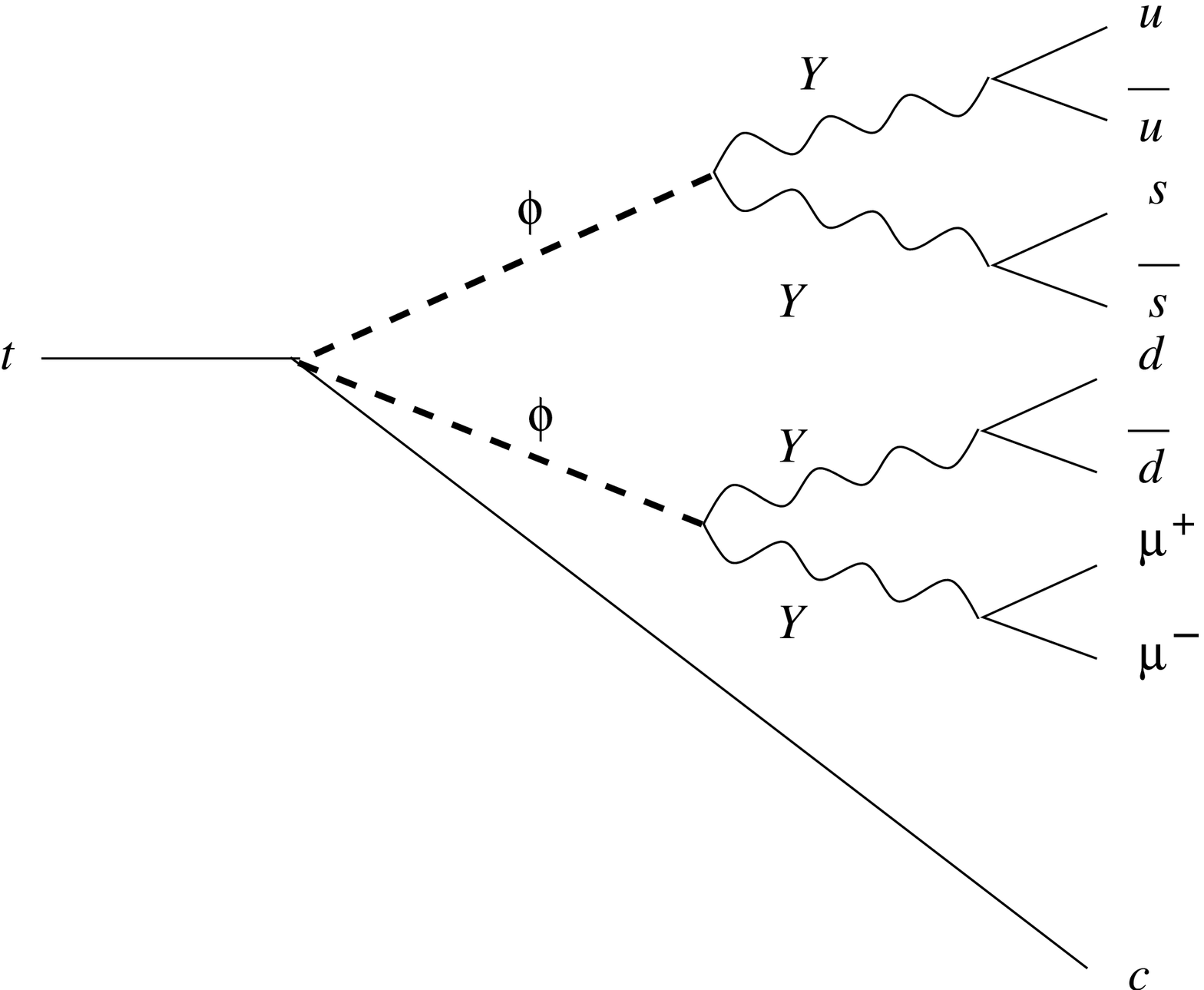}{The decay of a top quark to a charm quark plus an
unparticle may result in a decay to more than three visible particles
and little or no missing energy.  Many other final states are
possible, depending on the details of the hidden sector.  The
unparticle measurement of \cite{Un1} requires identifying the charm
quark.}

\subsubsection{Supersymmetric decays, and analogous cases}

With supersymmetry, unparticle couplings to supersymmetric particles
can lead to challenging decays of the lightest standard model
superpartner, as discussed in \cite{HV3}.  For instance, a neutralino
-- perhaps, in our toy model C, through its Higgsino coupling to the
unsparticle -- may decay to $\phi$ and its superpartner, which in turn
decays to $ \phi$ and $\tilde Y$, the invisible hidden gaugino
(\reffig{LSP24Y}.)  This type of decay can significantly reduce the
missing energy signal which is typically used to find supersymmetry,
and can replace it with soft jets and leptons \cite{HV3}.  More research
to understand how to find such a signal is needed. 

Another interesting possibility \cite{HV3} is that the lightest
standard model superpartner is not neutral.  For example, if this
lightest standard model particle is a stau, then the decay $\tilde
\tau\to \tau \tilde \OO$ to a unsparticle is similar to the $t$ decay
of \cite{Un1}, in that the $\tilde\tau\to \tau$ kinematics may reflect the
dimension of $\tilde \OO$.  However, as with the top quark decay
mentioned above, the visibility of the kinematical power law will
depend on the details of the final state emerging from the
unsparticle, as it is converted into visible particles.  Note also
that the $\tilde\tau$ may easily be long-lived (this is less likely in the
neutralino case) and may decay with a displaced vertex \cite{HV3}.

Essentially the same physical phenomena can arise in any model with
new particles that carry a new conserved global symmetry, such as
KK-parity in extra-dimensional models, or T-parity in little-Higgs
models.  As long as there are particles in our sector and in the
hidden sector carring the new charge, the possibility of interesting
cross-sector decays exists.  The scenario of a new heavy particle in
the standard model (the lightest particle carrying the new charge) and
an unparticle sector with conformal invariance and a mass gap below
100 GeV (which will have a lighter particle carrying the new charge)
virtually {\it guarantees} the decays studied in \cite{HV3} will
occur.  The only question is whether the decays are visible, and this
depends on the details of the hidden sector and the size of the mass
gap.

\makefig{0.4}{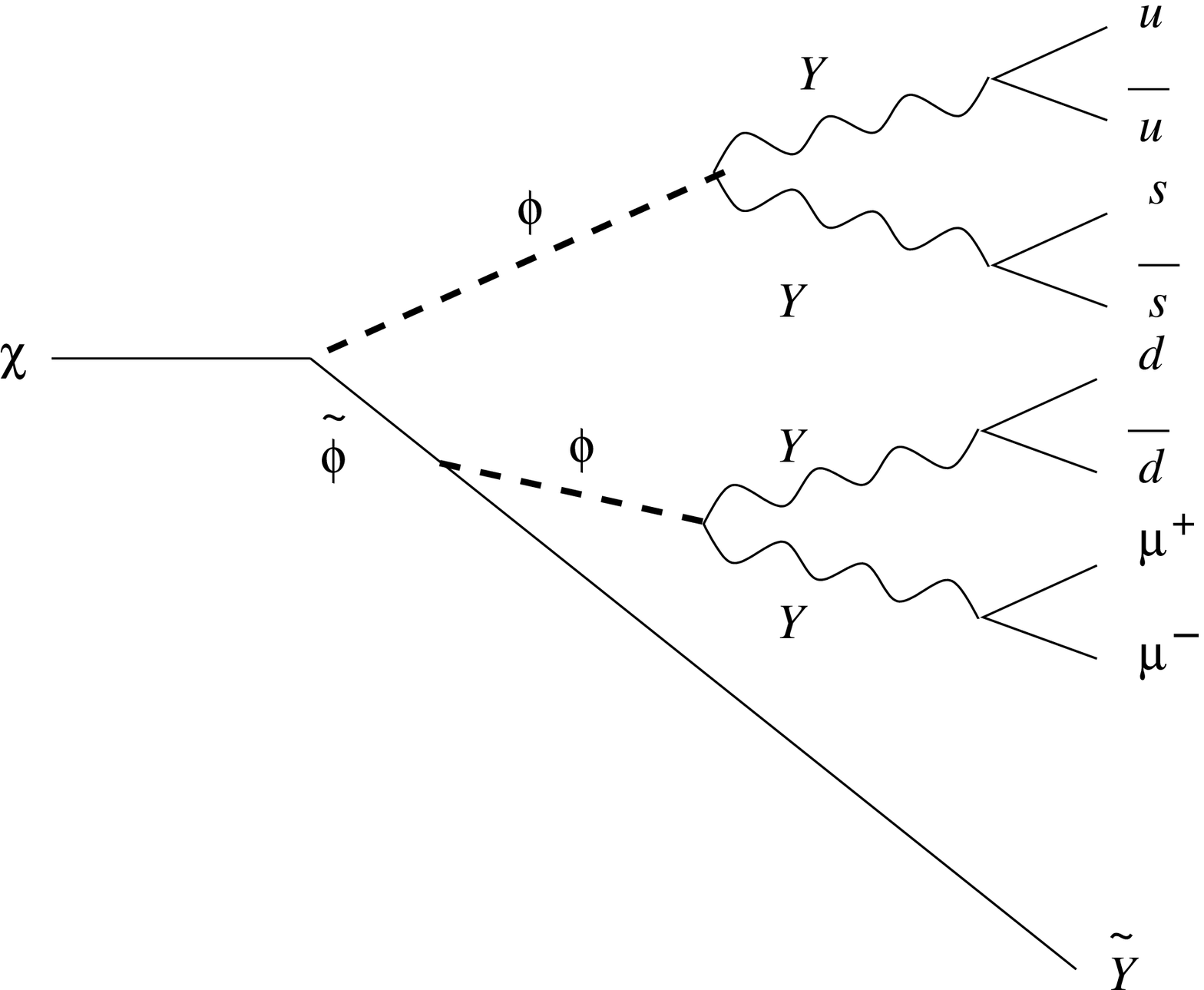}{The decay of a neutralino through an unsparticle
$\chi\to\tilde \OO$ can result in many visible particles plus one
stable invisible particle, which here is the hidden gaugino $\tilde
Y$.  The decay $\tilde\tau\to \tau \tilde \OO$ can have a similar
final state.}

Obviously there is much more to do in this arena.  Exploration of 
the possibilities is important as a first step to ensuring
that none of these challenging signals will
escape detection.

\section{The Stephanov Model and Hidden Valleys}
\label{sec:unSteph}

An interesting approach to unparticles was provided by Stephanov
\cite{unSteph}, where an unparticle was modeled by appealing to the
five-dimensional language inspired by the AdS/CFT correspondence, also
known as ``gauge/string'', ``gauge/gravity'', or ``boundary/bulk''
duality \cite{malda,GKP,WittenAdS}.  (See also the work of Randall and
Sundrum \cite{RS1,RS2}.)  The five-dimensional language, as we will
see, is indeed instructive, but only will guide us if we use it fully.
We will need the full gauge-string correspondence, not the
watered-down gauge/gravity version, in order to capture all the
physics and see the hidden valley in full (s)unshine.

\subsection{The Stephanov Viewpoint and Hidden Valleys}

In order to regulate and then interpret the unparticle propagator,
Stephanov broke the conformal invariance with an infrared cutoff.
(Note this is not deconstruction, but compactification
\footnote{Deconstruction \cite{ACG} refers to the discretizing of a
space, as in \cite{SonSteph}, and representing it as a gauge theory in
one lower dimension.  The number of Kaluza-Klein modes in a
deconstructed theory is finite.  The introduction of an infrared
cutoff while retaining the continuity of the five-dimensional space is
a form of compactification; the number of Kaluza-Klein modes is
countably infinite.  In this context this cutoff is known as the
``hard wall'' model, a model often used in the description of a
confining gauge theory, which is why ``Randall-Sundrum 1'' \cite{RS1}
can be viewed as dual to technicolor.}.)  This cutoff is often used in
the gauge/string literature as a model of how to describe confinement
in the gauge theory in terms of a five-dimensional (5d) theory on a
warped space-time.  In short, Stephanov's method of regulating the
unparticle propagator is simply this: a model of a two-point function
in a hidden sector which is confining in the infrared and conformal at
larger energies.  Many examples of such theories are known; one
explicit example with a dual string description is given in
\cite{Nonestar}.

Models of the same type were already considered in
Sec.~\ref{subsec:unbroken2}. But the predictions presented in that
section are very different from those of Stephanov.  This makes it far
from obvious that we are dealing with a hidden valley.  Let us review
these predictions and see where the difficulties with them lie.

First, Stephanov predicted a narrow tower of states.  Second, he
suggested these states would decay to standard model particles through
the coupling to the unparticle.  Third, he suggested these states
could have very long lifetimes and could be detected from displaced
vertices.  Finally, he suggested the lifetime $\tau_n$ of the $n^{th}$
state would be related to the dimension of the unparticle by
$\tau_n\sim m_n^{p-2d_\OO}$, where $p$ is a (known) positive number.  

This tower of long-lived states, with a directly measurable
lifetime-to-mass relation set by the dimension of the unparticle, is
an impressive prediction of unparticles not shared by typical hidden
valley models.  But there is a good reason for this.

The point is that Stephanov's formulas apply for a hidden gauge theory
in the limit that the number of colors $N$, and the 't Hooft coupling
$\lambda \equiv \alpha N$, where $\alpha= g^2/4\pi$ is the hidden
gauge coupling, are extremely large --- how large will be explored
below. (Meanwhile $N_f$, the number of flavors of matter fields, must
remain finite and small, so these are not classic Banks-Zaks-type or
Seiberg-type fixed point theories.)  In this limit, a confining gauge
theory has a spectrum that is merely an infinite tower of stable
non-interacting hadrons of spin $\leq 2$.  Many familiar aspects of
gauge theory would be absent, including all high-spin hadrons, BFKL
dynamics, parton showers, and the like.  This differs so dramatically
from QCD, and even from gauge theories with dozens of colors, that it
was not recognized as a hidden valley.  But the interpretation is
this: if one were to take a hidden valley model into this extreme
regime, it would eventually resemble the narrow-tower model.  Less
extraordinary models will have, not surprisingly, more ordinary
predictions similar to those of \cite{HV1}.

\subsection{Stephanov's approach}

In a companion paper \cite{myunSteph}, I will more carefully and
pedagogically add in the $1/N$ and $1/\sqrt\lambda$ corrections to the
``narrow-tower'' model.  I will also consider some variants of this
model and see how easily the predictions may be altered.  (For
instance, in many realistic models, the tower may have a finite or
infinite number of states extending only over a finite range of
energies, with a continuum above. The continuum may also have
additional embedded resonances. Alternatively, the density of narrow
states may suddenly change by a finite factor at some scale.  The
spacing of the states need not be uniform; there may be
degeneracies of states that grow with $n$; etc.) I will also consider
further how hard one must push the theory to make the original
narrow-tower model of \cite{unSteph} appropriate to the physics. Here,
I will keep things short and state some of the more subtle claims
without proof.

First some notation.  The radial coordinate in $AdS_5$ will be called
$r$, running from a boundary at $r=\infty$ to a horizon at $r=0$; the
five-dimensional metric is
\be
ds^2 = {r^2\over R^2} (-dt^2+dx^2+dy^2+dz^2)+{dr^2\over R^2} \ .
\ee 
 For RS experts, my coordinate is chosen such that if I cut off the
space at $r=r_{{\rm UV}}$ for large $r$ and at $r_{{\rm IR}}$ at small
$r$, then in RS1 the Planck brane is at $r_{{\rm UV}}$ and the TeV
brane at $r_{{\rm IR}}$.  In RS2 the interpretation would be slightly
different; but in any case, the space represents in this case a theory
which is conformal between the two energy scales $\mu_{{\rm UV}}\sim
r_{{\rm UV}}/R^2$ and $\mu_{{\rm IR}}\sim r_{{\rm IR}}/R^2$.  With no
UV or IR cutoff, and with the addition of a well-behaved five
dimensional compact space $X$ to make a total of ten dimensions, a
superstring theory on this space would, according to gauge/string
duality, precisely represent a conformal gauge theory.

Stephanov considered an AdS space with $r_{{\rm UV}}=\infty$ and
$r_{{\rm IR}}$ finite.  In a fully consistent setting, this would
correspond to a theory which is conformal at high energy but at low
energy has some sort of conformal symmetry breaking.  Cutting off the
space sharply, without any nuances, is called the ``hard-wall'' model.
It has been used extensively for study of confining gauge theories at
large 't Hooft coupling in gauge-string duality \cite{hardscat, DIS,
haduniv, rhouniv, Brodsky, BPST}.  It has its limitations, but is
often useful.  In such a model, the scale $\mu_{{\rm IR}}\sim r_{{\rm
IR}}/R^2$ is of order $\Lambda$, the confinement scale.  Other models
\cite{softwall} give similar structure, though the details differ.

It is a natural conjecture (independent of $\lambda$) that in the
$N\to\infty$ limit of a confining gauge theory the two-point function
of a reasonable operator can be exactly written as
\bel{twopointAdS}
 \vev{\OO(q)\OO(-q)} = \sum_n {|F_n|^2\over q^2-m_n^2 + i\epsilon }
\ee
where the masses $m_n$ are those of the confined hadrons $|n\rangle$
created by acting with the operator $\OO$ on the vacuum,
\be
\OO|0\rangle = \sum_n F_n |n\rangle\ .
\ee
Within the hard-wall model and its cousins, and using the low-energy
five-dimensional gravity theory, these equations can be shown to be
true without subtleties.  Thus this equation (and its analogues for
higher spin) is correct for $\lambda\to\infty$, $N\to\infty$,
$\lambda/N$ fixed and not too large, at least for primary operators
with spin and dimension of order 1.

One can see that within the hard-wall model, and many of its variants,
the only significant change as $d_\OO$ changes is the
$n$-scaling of the $F_n$, as reviewed in \cite{unSteph}.  The $m_n$
change; their $n$-scaling typically does not.  
However, this precise feature is a property of a particular model, and
as noted in \cite{unSteph} the constraints are rather weak.  

Conformal invariance only requires that the two-point function
approaches a particular power law.  In the limit that $r_{{\rm IR}}\to
0$, as reviewed in \cite{unSteph}, the spacing between the modes goes
to zero, and the two-point function must regain its conformal form.
But this requirement imposes only a single relation between the
large-$n$ behavior of the $m_n$ and that of the $F_n$.  As emphasized
in \cite{unSteph}, the requirements of conformal invariance on
\Eref{twopointAdS} do not permit the mass spectrum of the tower to be
predicted from the operator dimensions alone.  Thus to measure
$d_\OO$, one must measure something else.

\subsection{Finite $N$ Effects on the Spectrum}

It \cite{unSteph} it was proposed that one should measure the lifetimes
of the states.  Let us review the calculation.  Although there are
technical problems with the particular case chosen, the basic logic is
correct; with a coupling in the Lagrangian $\sim \hat c \OO_{SM}\OO$, where
$\OO_{SM}$ is a standard model operator, the decay of a state $\ket n$
in the tower to a standard model state such as $\mu^+\mu^-$ is
proportional to
\be
\big| \hat c \ \vev{\mu^+\mu^-|\OO_{SM}|0}\ \vev{0|\OO|n} \ \big|^2
\ee
For example, in the case \cite{unSteph} considered,
one obtains (converting to a notation in which
$m_n\sim\Lambda n^{\sigma}$ at large $n$) a decay rate of the form
\be
\Gamma_{\ket{n}\to\mu^+\mu^-} \sim \alpha^{(n)}_{{\rm eff}} m_n
\ee
where the effective coupling is
\be
\alpha_{{\rm eff}}^{(n)} = {c^2 A_\OO \over 16 \pi^2} 
\left({m_n\over M_Z}\right)^{2(d_{\OO}-1)} 
\ee
Here $A_\OO$ is a normalization constant for the unparticle, of order
1, and we have converted $\hat c$ into a dimensionless constant $c$
times the appropriate power of $M_Z$, following \cite{unSteph}.  This
rate can be enormously suppressed if either (1) $c$ is very small, or
(2) $\Lambda, m_n \ll M_Z$ and $d_\OO$ is significantly above 1.

We see, therefore, that if the lifetimes of the states can be measured,
then so can $d_\OO$.  But how can they be measured?

\subsubsection{A necessary condition}

Can these states decay with visibly displaced vertices?  This
requires lifetimes in the picosecond range or longer.  For decays to electrons
(muons), the electron (muon) mass is of order 1 (100) MeV, so let us
take $m_n\sim 1\ (100)$ MeV as well, to lengthen the lifetime as much as
possible.  Then we must have 
\be
\alpha_{{\rm eff}}^{(n)}\lsim 10^{-9}\  \ ({\rm decay\ to\ electrons})
\ee\be
\alpha_{{\rm eff}}^{(n)}\lsim 10^{-11}\  \ ({\rm decay\ to\ muons})
\ee
These tiny numbers are already an issue since production
rates, compared to ordinary electromagnetic processes, are very small. 
However, there is a more serious issue.

\subsubsection{Why the previous condition is not sufficient}

The above condition for displaced vertices, while necessary, is
not sufficient!  {\it It is only appropriate if the state $\ket{n}$ has no
other decay modes.}  And this is not true, except typically for the
ground state and perhaps the first excited state in every tower.

The narrow-tower model implicitly assumes that the 5d scalar field
that represents the scalar unparticle is non-interacting.  This is
equivalent to assuming that the unparticle has no 3- or higher-point
functions.  This is true in the $N\to\infty$ limit (and also in the
free $d_\OO\to d_{min}$ limit, where $\OO$ becomes an ordinary free
particle.)  But the infinite $N$ limit is very misleading (as is
the $d_\OO\to d_{min}$ free particle limit.)  At any finite $N$ a
conformal gauge theory will have $k$-point functions for $k>2$.
Equivalently, the 5d scalar field {\it will have self-interactions}
(if $d_\OO>d_{min}$).  As a result, the unparticle now has 
higher-point functions.

Also, one cannot treat one unparticle tower in isolation.  There are
always other fields in the bulk.  At the very least, the 5d graviton
{\it must} be present, because it represents the energy-momentum
tensor of the hidden sector, which is part of any conformal theory.
Each such field, in a confining gauge theory, will have its own tower.
And of course the graviton interacts with itself and with all other
fields in the bulk.  In fact, in any conformal gauge theory one
expects many fields in the bulk, with many quantum numbers.  Any
conserved currents in the theory, for instance, will be represented as
5d massless gauge fields, and they too will interact with themselves,
with the graviton, and with any 5d fields that carry the corresponding
conserved charge.  All this is to say that there is no conformal gauge
theory without three-point functions and operator product expansion
(OPE) coefficients --- except at $N\to\infty$ --- and that $T_{\mn}$
and conserved currents $J_\mu$ always have a nontrivial OPE.

Once these interactions are introduced, we no longer expect a tower of
extremely narrow states decaying to standard model particles.  Any
state with high mass will decay via these interactions.  The states
may still be relatively narrow, but nowhere near as narrow as
predicted in \cite{unSteph}.

\subsubsection{Another necessary condition}

Let us now estimate the widths of the excited states.  (We assume
$d_\OO$ is not very close to $d_{min}$; otherwise special treatment is
required.)  The width of the $n^{th}$ state ($n>1$) to other hidden
sector states, may be very roughly estimated as
\be
\sum_{n',n''}\Gamma_{\ket{n}\to \ket{n',n''}} \sim   
{g(n) m_n\over 8\pi N^2}
\ee
where $g(n)$ characterizes the growth in the number of decay channels
and monotonically grows with $n$.  The states become narrow rather
quickly with $N$, for fixed $n$, but conversely their widths grow with
$n$.  Suppose as $n$ becomes large that $m_n\sim n^{\sigma}\Lambda$
and $g(n)\sim C n^\beta$.  Typically $\sigma\leq 1$ (it is $\frac12$
in QCD and in string theory and is 1 in many gauge-gravity duality
examples.) Meanwhile one might naively expect $\beta\sim 2$,
accounting for the scaling of available channels with $n$, but in most
computable theories all but $n$ couplings are small, as in
\cite{haduniv}, so to be conservative let us only assume $\beta\geq 1$.
Finally $C>1$ accounts for the presence of multiple towers of states
in the theory, which provide multiple classes of decay channels.  If
$N_f$ is large, then $C\sim N_f$; the situation for small $N_f$ is
less clear, but $C$ is certainly larger than 1.  The states bleed
together when
\be 
m_n-m_{n-1}\sim n^{\sigma-1} \Lambda 
\ee
is of order
\be
 \Gamma_n\sim {C
n^{\sigma+\beta}\over 8\pi N^2} \Lambda
\ee
and thus occurs at 
\be
n\sim\left[ {8 \pi\over C} N^2\right]^{1/(\beta+1)} \ .
\ee
Since $\beta+1\geq 2$, we expect at most the first $N$ states to be
narrow relative to their separation.  Actually this is often a large
overestimate, due to the fact that we have ignored stringy effects;
we will return to this in the
Sec.~\ref{subsec:finitelambda}.

The inverse of these widths puts an upper limit on the lifetimes of
the excited states.   Unless
\be
{g(n)\over 8\pi N^2}\lsim 10^{-9}\  \ ({\rm decay\ to\ electrons})
\ee\be
{g(n)\over 8\pi N^2}\lsim 10^{-11}\  \ ({\rm decay\ to\ muons})
\ee
the lifetimes will be too short for displaced vertices.
Thus, except for the lowest one or two states, for which $g(n)=0$,
this condition requires $N\sim 10^4$ for electrons and $10^5$ for
muons.  And this is generous, because we assumed the lowest possible
mass for the decaying state.  Also, we would hope to see at least four
or five states, in order to measure a power law, and $g(n)\sim
Cn^\beta$ is often large compared to one and grows with $n$.

{\it In short, we do not expect a tower of states with displaced vertices unless
both $N\gg 10^4$ and $\alpha_{{\rm eff}}^{(n)}$ is very small.}
What, then, is the phenomenology more likely to be?  

There are two possibilities, depending on whether the decays of the
hidden sector states are to other hidden sector states or to standard
model particles.  Decays to standard model states will dominate only
if
\be
\label{whowins}
\alpha_{{\rm eff}}^{(n)} > {g(n)\over 8\pi N^2} 
\ee
But $\alpha_{{\rm eff}}^{(n)}$ cannot be large, or effects from the
hidden sector would already have been seen.  (In fact, if
$\alpha_{{\rm eff}}^{(n)}$ is of the same order as $g(n)\over 8\pi
N^2$, then constraints on $\alpha_{{\rm eff}}^{(n)}$ are even stronger
than is often realized, because of effects that we will discuss in
Sec.~\ref{subsec:unproduceN}.)  Again, we are forced to take large $N$
--- not as large as required for a tower of displaced vertices, but
still very large.  Being very generous, we would require $N\sim 300$
in almost any conceivable situation; much larger $N$ is required, for
instance, in the case of a vector unparticle mixing with the $Z$
boson.

\subsubsection{If decays to standard model particles dominate}

If $N$ is of order 100 or more and the unparticle coupling is as
large as is allowed by experiment, then there is a narrow window in which
\begin{itemize}
\item the widths of the states are determined by the unparticle coupling 
and
\item the widths of the states are large enough to measure.
\end{itemize}
Assuming resolutions in the few MeV range, and masses in the few GeV
range, one might imagine that if $\alpha_{{\rm eff}}\gsim 10^{-3}$ then one
could measure the scale-invariant prediction of \cite{unSteph} through
a tower of states with growing widths.

However, if $\alpha_{{\rm eff}}$ is too small, the states simply won't have
a measurable width, making the prediction untestable.  In this case
one can only measure the masses $m_n$.  But recall that these
are not determined by scale invariance, and cannot be used to measure $d_\OO$.

One might hope that since of order $N$ states may have narrow widths,
higher states in the tower might always have $\alpha_{{\rm
eff}}^{(n)}$ so large that the widths are large enough to measure.
But a little thought shows there is no guarantee of such a regime.
The higher states move closer together as $n$ increases, if
$\sigma<1$, so instead of the widths growing to measurable size, the
distance between adjacent states may shrink to unmeasurable size.
Also, as $n$ increases so does $g(n)$, so it is possible that the
higher states do not decay preferentially to the standard model.  And
worse, the logic used in the estimates in this entire section breaks
down when $n$ is large enough that stringy effects must be accounted
for; see Sec.~\ref{subsec:finitelambda}.

\subsubsection{If decays to standard model particle do not dominate}

If $N$ is less than 100 or so, or if $\alpha_{{\rm eff}}^{(n)}$ is small,
the decays within the hidden sector dominate.  {\it In this case the
lifetimes are not determined by scale invariance; they are determined
by $g(n)$, which depends on the details of the hidden sector.}
Moreover, the partial widths to lepton pairs are typically very small,
unless $N$ is very large, making the line-shape of the resonances
difficult to observe.  This is by far the most likely scenario for a
hidden sector!

{\it Thus we are not very likely to observe dilepton pairs from a
tower of states}.  For reasonable values of $N$, and for $\Lambda >
2m_e$, { we will find at most $N$ rather narrow states, decaying too
rapidly to other hidden sector states for a displaced vertex, and with
tiny branching fractions to dileptons.}  The lifetimes of the states
will not be set by the dimension of the unparticle.  {\it The lightest
state or states in some towers, which are the only states that cannot
decay within the hidden sector, will have much longer lifetimes, and
may decay to standard model particles with displaced vertices} (though
possibly outside our detectors.)  Only in a narrowly tuned case ---
small $N_f$, very large $N$, and a coupling large enough that the
widths of the first few states are rather large and are determined by
the unparticle coupling, might the lifetimes be both measurable and
detemined by scale invariance.

\subsubsection{Summary}

Let us summarize the implications of this section.  Reference \cite{unSteph}
includes a correct computation of the {\it partial width} for each
state in the tower to decay to standard model particles and become
observable.  But one cannot then assume that this partial width is the
total width, and invert it to infer long lifetimes for all the states
in the tower.  Instead, at reasonable $N$, almost every state in the
tower, except the lowest one or two, will decay predominantly and
rather rapidly to other states in the tower, or to states in other
towers associated to other operators.  The lifetimes will be much
shorter than estimated in \cite{unSteph}, so there will be no
displaced vertices. Moreover, the branching fraction to standard model
states will be tiny, and it will be very difficult to measure the
partial widths.  This is simply the statement, familiar from QCD
itself, that most hadrons in a gauge theory decay rapidly to other
hadrons, and have large widths to do so; their branching fractions to,
say, $e^+e^-$ are very small.  Thus, unfortunately, the predictions of
\cite{unSteph}, while possible at extraordinarily large $N$, are
unlikely to be seen in nature.

\makefig{0.4}{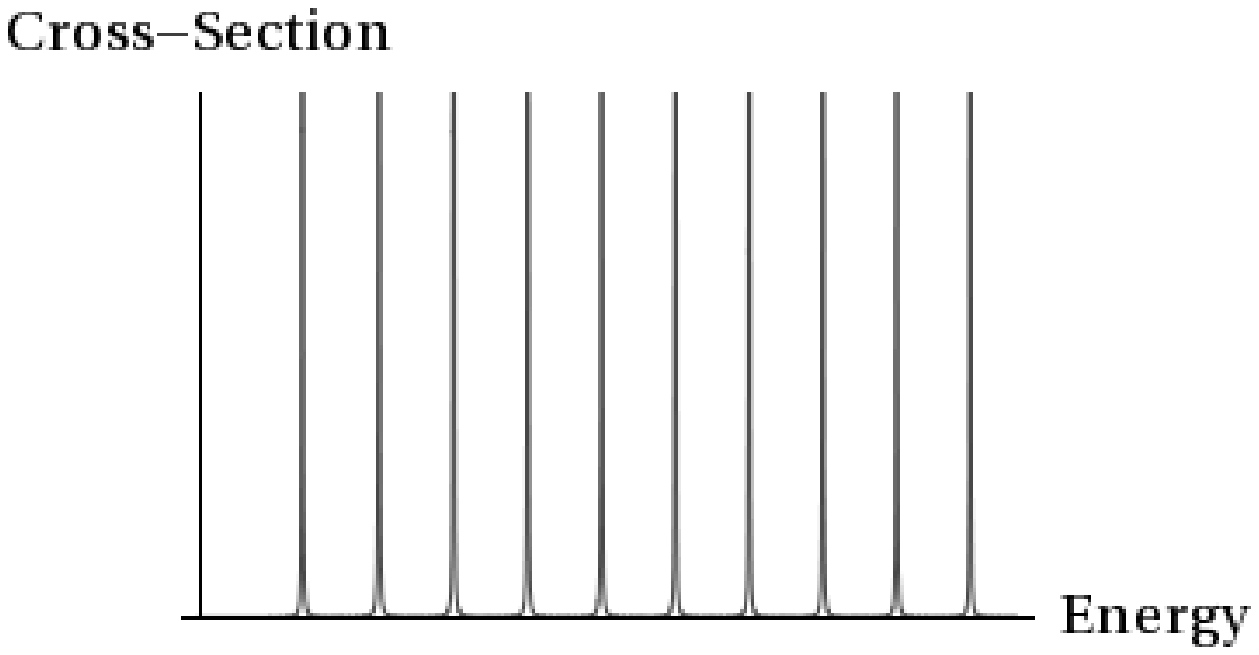}{The total cross section $\sigma(s)$ for
hidden-sector production in an $e^+e^-$ collider, in the narrow
tower model of an unparticle.}

\subsection{Unparticle Production at Finite $N$}
\label{subsec:unproduceN}


\makefig{0.4}{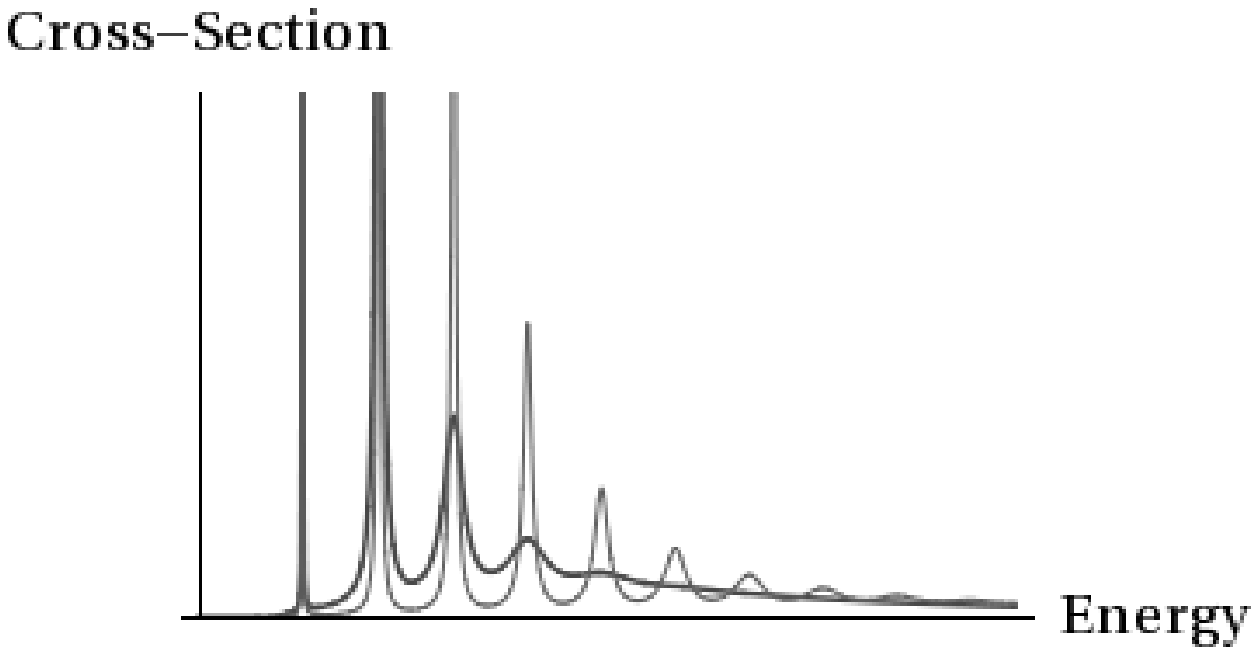}{The total cross section $\sigma(s)$ for
hidden-sector production in an $e^+e^-$ collider.  The thick curve is
for small $N$ and resembles QCD; the thin curve is for larger $N$ and
more resonances are visible.  The first resonance is much narrower
than the others, as it can decay only by emission of standard model particles; the others decay within the hidden sector.  See \cite{HVWis} for a study of a hidden valley with a light dilepton resonance.}

The cross-section in the narrow tower model takes the form of
\reffig{narrowtower}.  In \reffig{comparetowers} are shown possible
cross-sections for more realistic towers, for small $N$ and large $N$;
the small $N$ case resembles low-energy QCD in the $\rho$ channel, and
the large $N$ case resembles charmonium production without open charm.
But this is the total cross-section for hidden sector production,
whereas the cross-section for $e^+e^-\to \mu^+\mu^-$, both for small
and larger $N$, suffers from the low dilepton branching fraction for hidden-sector states; it is shown in \reffig{comparetowers2}. Only the first
resonance is potentially observable, and since it is small and very
narrow, it may easily have been missed up to now.

\makefig{0.4}{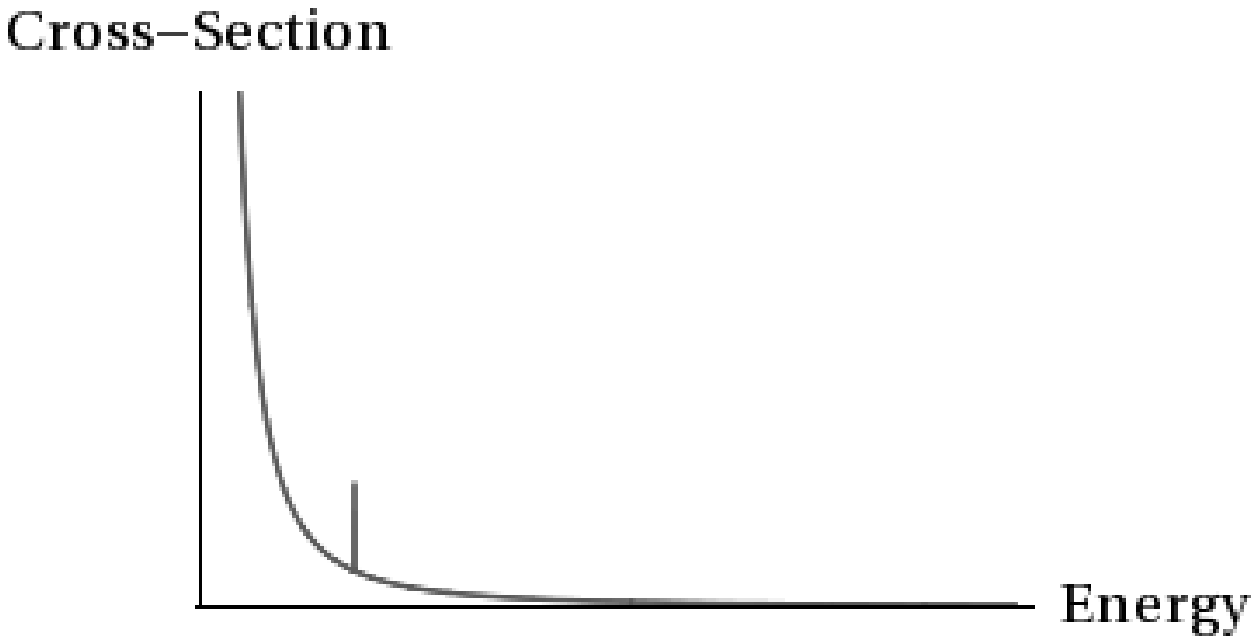}{A cartoon of the cross-section for
$e^+e^-\to \mu^+\mu^-$ in a model at small to moderate $N$.  The
falling standard model production rate is supplemented by a single
extremely narrow resonance, the lowest resonance in
\reffig{comparetowers}.  Its height has been exaggerated greatly for
clarity.  None of the other resonances have measurable branching
fractions to $\mu^+\mu^-$.  Only a handful of visible resonances are
expected in a generic hidden valley, at most one or two for each tower.}

In fact, QCD is an excellent model for the hidden sector.  From the
point of view of the leptons of the standard model, with
electromagnetism turned off, it {\it is} a hidden sector, coupled to
leptons only by the Fermi interaction.  It is no accident that AdS/QCD
methods, placed into the hidden sector and treated with care,
reproduce QCD-like physics in their effect on $e^+e^-\to \mu^+\mu^-$.

Now suppose that we do choose a theory with $N\sim 20$ and choose to
run an $e^+e^-$ collider on one of the excited hidden-hadron
resonances (say, $n=10$).  What will we see?  Certainly we will not
observe the process shown in \reffig{unrare}; the branching fraction is too
small.  Most of the time the resonance will undergo a
cascade decay to several light hidden hadrons, each stable against
decay to others, which {\it in turn} will decay with long lifetimes to
standard model particles.  This classic hidden valley signature is
shown in \reffig{uncommon}.

\makefig{0.4}{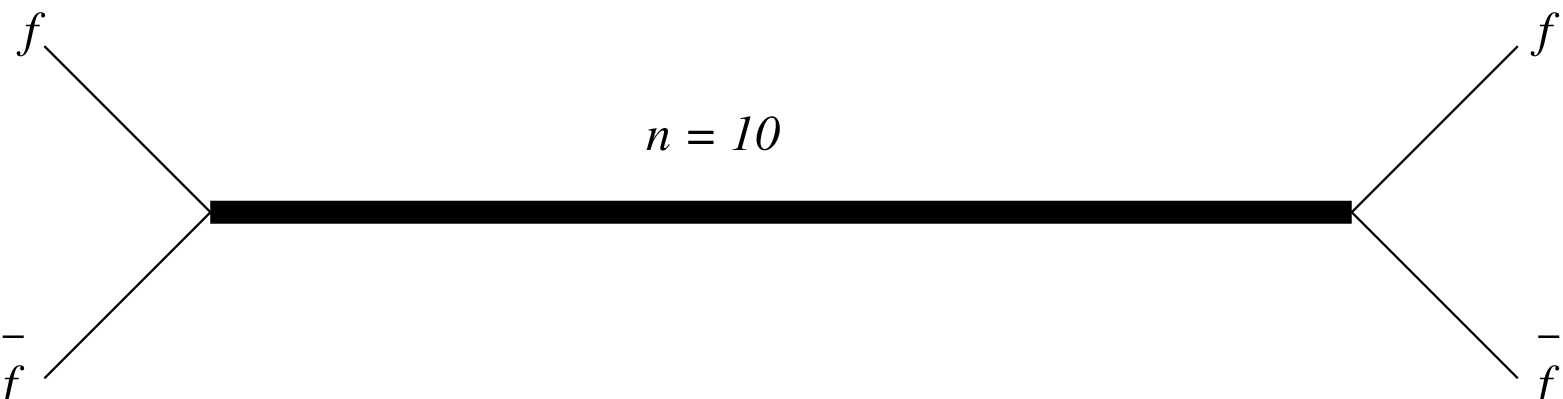}{An allowed but very rare process, in which the
tenth resonance is produced by and decays back to standard model fermion
pairs.}

\makefig{0.4}{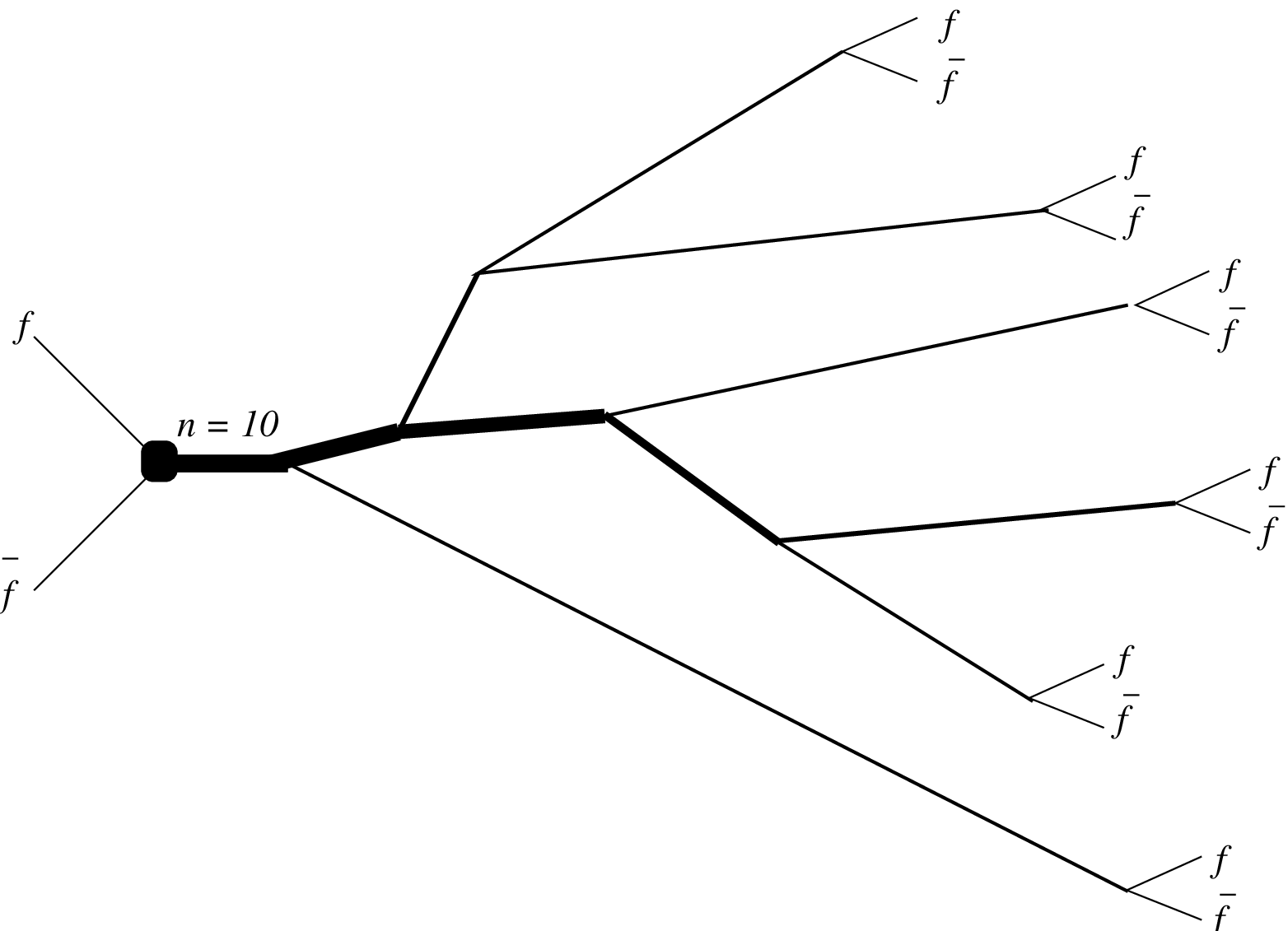}{A much more common process; the 10th
resonance, produced by standard model fermion pairs, decays through a
cascade.  In the final state appear several particles, each the lightest
$(n=1)$ resonance of a tower.  Each of these
then decays (possibly late) to standard model particles, here assumed
to be fermion pairs.}

For large $n/N$, the very language that we used in the previous
paragraph breaks down.  The sum in \Eref{twopointAdS} is modified in two
ways at finite $N$.  First, the poles at $m_n^2$ move off the real
axis to become resonances.  This already gives the two-point function
the form shown in \reffig{comparetowers}.  But we also must supplement
the formula \eref{twopointAdS} with cuts from multi-particle
production, which is suppressed by factors of $N$ but becomes
increasingly important at large $q^2$.  We need not produce hadrons
only by producing them resonantly, as in $e^+e^-\to \rho \to
\pi^+\pi^-$; we may simply produce them directly, as in $e^+e^-\to
\rho^+ \pi^-\pi^0$.  As $q^2$ increases, the cuts accumulate,
eventually stealing support from the resonances and building up the
continuum contribution to the two-point function.  The dominant
production at large $q^2$ is thus not even of single unstable heavy
hadrons, but rather of multiple light or moderately light hadrons,
whose phase space grows very rapidly as $q^2$ increases.  Any excited
hadrons among those produced will decay to the lightest ones, and a
large number of light hadrons may then decay with long lifetimes to
standard model particles.  The production process, absent in the
narrow tower model, is illustrated in \reffig{likely}; of course it
also gives a high-multiplicity final state.

{\it Importantly, even if the dilepton branching fractions of the
excited states are large, rare high-multiplicity events must not be
ignored, as they provide strong constraints on new hidden sectors.}
Even a small number of events with four or more leptons or photons
would have been easily observed, since standard model backgrounds fall
so rapidly with multiplicity.  In turn, this implies there are much
stronger constraints on $\alpha_{{\rm eff}}^{(n)}$, and thus on $c$,
the unparticle coupling, than have so-far appeared in the literature.
And in turn, because of \Eref{whowins}, this makes the likelihood of
observing excited states with measurably large dilepton branching
fractions even smaller, and the likelihood of high-multiplicity states
even larger.

\makefig{0.4}{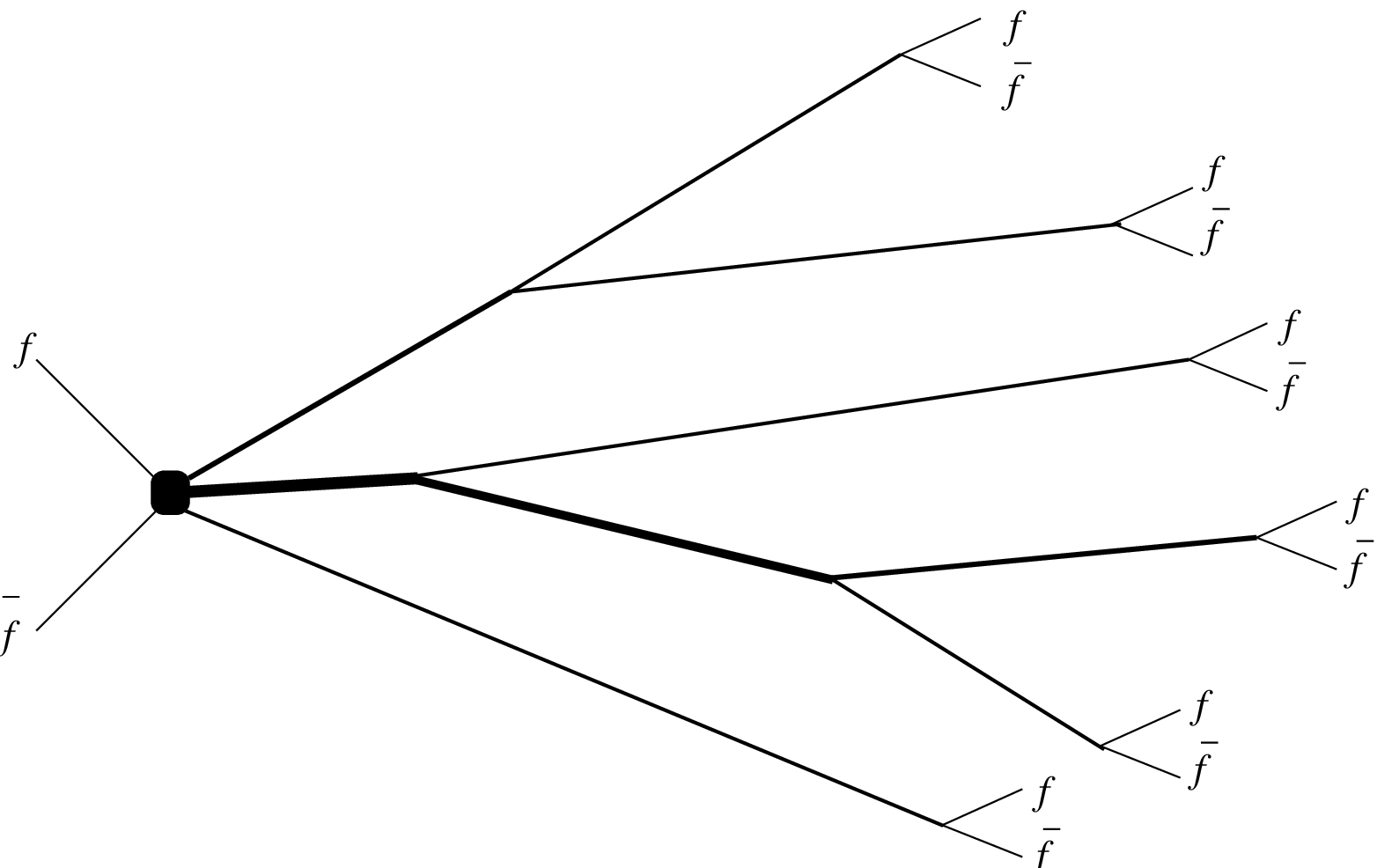}{At higher energy, multiple resonances are produced together; each undergoes a cascade decay as in \reffig{uncommon}.}

Thus, instead of a tower of neutral resonances with long lifetimes,
the prediction of the narrow-tower model at a reasonable $N$ is indeed
that of a typical hidden valley.  As in \cite{HV1}, events that access the
hidden sector will result in {\it light neutral resonances with long
lifetimes, produced in abundance, with large event-to-event
fluctuations and possibly large missing energy.}  Here the abundance
arises through the cascade decay of a heavy resonance, and/or through
nonresonant production of multiple particles.

\subsection{Effect of Finite $\lambda$}
\label{subsec:finitelambda}

Still we have not captured all the physics.  If $\lambda$ is infinite,
there are no states in the spectrum with spin higher than two.  But
any confining gauge theory has hadrons with arbitrary spin.  In
gauge/string duality it has been found \cite{GKP} that the ratio of
the masses of hadrons with spin $>2$ and higher to the lightest
hadrons of spin 0 through 2 is $(\lambda)^{1/4}$.  Thus there are
towers of states of spin 5/2 and above whose lowest states lie near
$(\lambda)^{1/4}\Lambda$.  Note this need not be that large; if
$N=100$ one cannot take $\lambda>100$ (see \cite{myunSteph} for more
details) and the fourth root of 100 is only about 3.  Thus it is only
legitimate to fully neglect these states, in this case, for processes
which have center-of-mass energy below, say, $5\Lambda$, or perhaps
$10\Lambda$.  At higher energy, the higher-spin states are present,
increasing the phase space for heavy resonance decays and the rate of
multi-hadron production at large $q^2$.  This drives us still further
from the model of a narrow tower.

In addition, even a low-spin tower itself is altered by mixing of the
original states in the tower with low-spin strings when the excitation
level $n\sim \lambda^{1/4}$.  This is a key breakdown of the gravity
approximation to the string theory.  Although the mixing is small, the
number of strings at any given excitation level grows exponentially,
so at large $n$ these states cannot be neglected.  Unparticle
production is then of massive strings in five dimensions, not of
low-mass scalars in five dimensions.  These massive strings then decay
to a number of light resonances that lie at the bottom of the various towers.

This is almost enough to recover our one missing piece: the parton
shower, and ensuing hadronization.  This is claimed in \cite{HV1} as
the dominant high-energy process in a hidden valley, and is certainly
the dominant process in QCD.  Where is it in this five-dimensional
langauge?  More details will be given in \cite{myunSteph}, and the
argument uses some subtle features of string theory, but I will simply
claim here that the parton shower at large $\lambda$ is dual to a
string falling in five dimensions toward the hard wall.  

What is happening from the five dimensional point of view is easy to
understand by looking carefully at what happens in the gauge theory.
Let us use the language of QCD, speaking of color, quarks, antiquarks
and gluons; we will then carry over the insight into the hidden
sector.  When a quark and antiquark are produced, they emit gluons,
whose color lines are correlated with the quarks and with each other,
as captured in 't Hooft's double-line notation.  As is well known, at
any moment in time we can draw a line from the quark to the gluon
whose anticolor is correlated with the quark's color, and from there
to the next gluon whose anticolor is correlated with the first gluon's
color, and so forth --- forming a string.  This is the same string
which is used at the moment of fragmentation in the Lund string model,
but before fragmentation occurs, during the parton shower, this is a
string falling in five dimensions.  (Note this is no classical string;
it is a quantum, fluctuating, string.)  It falls from $r\sim\sqrt{\hat
s}/R^2$ down to $r = \Lambda/R^2=r_{{\rm IR}}$, where it stops falling and
hadronization occurs.  As the string falls, an observer at a fixed
$r$, corresponding to a probe of the string at some momentum scale
$\propto r$, can resolve smaller and smaller structures within the
string \cite{Susskind}.  Thus the string contains more and more
resolveable partons as the scale decreases.

Is this claim, that the full string theory is needed to see the parton
shower, plausible?  As another line of argument, which is really the
same argument in a crossed channel, consider the following.  The same
operators which control parton splitting in DGLAP evolution control
the leading-order behavior of parton splitting in the parton shower.
We must keep the classically-twist-two 
operators of spin 4 and higher at weak coupling, if we
are to see the parton shower.  
If we then dial up the
coupling, and {\it drop} all operators of dimension $\sim
\lambda^{1/4}$, which include all operators of spin $>2$, we will be
throwing away the parton shower itself.  A more precise version of
this argument will be presented elsewhere.

Of course, when the string reaches the bottom at $r_{{\rm IR}}=\Lambda R^2$,
it may for an instant be viewed as a highly excited, highly tangled
hadron.  It cannot be viewed as one of the hadrons in the tower of a
simple five-dimensional scalar; such hadrons are five-dimensional
points and have no internal structure.  Here we have an extended
object, with many quantum numbers, fundamentally a string.
Interactions then cause the string to fragment, slowly if $N$ is
large, quickly if $N$ is smaller or if the number of light flavors
$N_f$ is nonzero.  The hadronization process is, as in the Lund string
model, the breaking apart of a string which is largely {\it
four-dimensional}, being localized in the fifth dimension near the
minimum value of $r_{{\rm IR}}$.  

In short, we must remember that $\lambda$ is finite, and include the
strings, if we are to see the physics of parton showers and the
process of hadronization.  This physics can play an important role in the
phenomenology, as outlined in \cite{HV1}.

\subsection{Conclusions}

In this section a more physical version of the (formerly)-narrow-tower
model has been reconstructed.  When modes of
the hidden sector are produced, we expect the following signatures:
\begin{itemize}
\item At low energy $\sim \Lambda$, one may find a few light stable hidden
hadrons of low spin, which decay back to standard model
particles with long lifetimes.  
\item
At higher energies one finds excited resonances, which decay rapidly
to two or three light stable hidden hadrons, which in turn decay
to standard
model particles in the final state.
\item Next the resonances become a continuum; production of several 
hidden hadrons becomes likely, leading to an increasing multiplicity
of light hidden hadrons and consequently of standard model particles.
\item 
At still larger energy, the parton shower begins to play a role in the
evolution of the final state, making a purely hadronic description of
the process inconvenient and indeed misleading, but without changing
the basic signature of a high-multiplicity event.
\item In all of these processes, the final
state consists of a number of light neutral long-lived hidden-sector
hadrons, with a multiplicity that grows with energy.
\item The lightest hadrons have lifetimes orders of magnitude longer
than most other hadrons, and may produce observable displaced
vertices or give missing-energy signals.
\end{itemize}
The specific signals observed will of course depend on the details of
its gauge and matter content of the hidden sector.  But having included
five-dimensional interactions and five-dimensional strings, we now see
that this model has all the features and signatures of a confining
hidden valley.  As we did not rely upon strict conformal invariance,
the result is largely independent of whether there are true
unparticles at energies above the confinement scale.

However, despite the universality of the result, there are some
important features of the hidden-valley signal which can be affected
by strong dynamics.  Let us now turn our attention to these.

\section{Other Effects of Strong Dynamics on Hidden Valley Phenomenology}
\label{sec:othereffects}

From the way I have presented things, the reader might be left with
the impression that the ultraviolet strong dynamics that is present in
conformal field theories with large anomalous dimensions has no impact
whatsoever on the infrared physics.  But this is not the case.  Even
if the low-lying states in the hidden sector are visible, giving
hidden valley signals and making unparticle methods less central to
the phenomenology, the strong dynamics can have a crucial impact on what we
will see.  Yet this comes not from the produced ``unparticle'', but from other
effects of strong coupling on the phenomenology: on resonances, 
on flavor symmetries within the hidden sector, on supersymmetry
breaking within the hidden sector, and on the hidden parton shower.

\subsection{Effect of Strong Dynamics on Resonances}
\label{subsec:resonances}

As we noted in Sec.~\ref{subsec:unbroken1}, one may well expect narrow
resonances just below the point where continuum production of the
hidden sector begins.  (There may also be resonances, narrow or wide,
within the continuum.) The spacing between the resonances, and the
number of resonances that precede the continuum, are an important
opportunity for learning about the hidden sector, just as charmonium
was an important probe of QCD.  Had QCD been more strongly coupled,
the spacing between the charmonium states would have been wider, and
the number of states below the open-charm continuum might have been
very large.  (Indeed in the large $\lambda$ limit, using
\cite{KarchKatz}, it was found \cite{wintersD7} that a number of
extremely deeply bound states may in some cases appear well below the
onset of a continuum.  The known examples require supersymmetric
cancellations, but perhaps there are other mechanisms to obtain
similar effects.)

In fact, if the resonances are from a massive field $\phi$ whose mass
does {\it not} strongly destabilize the conformal dynamics, then the
$\phi$-onium states will still be bound by exchange of
effectively-massless gauge bosons.  This can happen in the toy models
B and C of Sec.~\ref{sec:breakCFT} if only one of the $N_f\gg 1$
scalars becomes massive at $M$, the others remaining light. The
$\phi$-onium states will form an almost perfectly positronium-like
system (though possibly at much stronger coupling) and the strong
coupling can be measured from the positions of the resonances.  Note
this neither measures nor depends upon the dimension of any
unparticle; conformal exchange is always Coulomb exchange, and only its
coefficient depends on the coupling.

However, the central phenomenological question is whether and how the
resonances can be observed.  The branching fraction directly to two or
three standard model particles will be low in this case, because
annihilation to the effectively-massless hidden sector gauge bosons
will dominate; see \cite{HVWis} for a brief discussion of hidden
quarkonium.  If the low-energy theory has some of the signatures of
Sec.~\ref{subsec:broken} or \ref{subsec:unbroken2}, then the decays
will be observable.  Still, detailed kinematic reconstruction will have
poor resolution at a hadron collider, so it is unclear how much
information can be extracted without an ILC.

\subsection{Effect of Strong Dynamics on Global Symmetries}
\label{subsec:flavor}

In models with conformal dynamics and approximate global symmetries,
renormalization effects can often either enhance or destroy those
symmetries.  This is determined by whether global-symmetry-breaking
operators have larger or smaller dimensions than
global-symmetry-preserving operators; if they are larger, then the
global symmetry is accidentally restored at low energy, while if 
smaller then flavor is badly broken.  For instance,
in models A and B, a critical question is whether the adjoint or
singlet scalar-bilinear has a larger anomalous dimension.  

As an illustration, let us imagine how different the standard model might
have been had it been conformal at high energy.
Imagine for example that the standard model itself became strongly
interacting above the electroweak scale.  This is not a completely
consistent example, but still instructive.  A very important effect is
that the Yukawa operators coupling the Higgs boson to standard model
fermions would have nontrivial anomalous dimensions.  If the
dimensions of the Yukawa operators were all less than 4, than all
Yukawa couplings would be driven large, all masses would be driven
toward 100 GeV.  But if all the operators were irrelevant (as in the
simplest versions of technicolor) then all masses would end up small
(which is in part why simple technicolor has trouble with the top
quark Yukawa coupling.)  If all masses are driven very small then
there can be greatly enhanced symmetries: for example, light $c$, $b$
and $t$ quarks would give larger chiral symmetries below the QCD
scale.  This would then introduce stronger GIM-like suppressions into
flavor-changing processes, and organize the hadrons into larger multiplets.

Another more subtle effect is that mixing angles between quarks could
also be reduced by strong dynamics. If some Yukawa operators have very
different anomalous dimensions from others, then the matrix of Yukawa
couplings may tend to become highly structured and the CKM matrix is
driven diagonal.  Indeed, an analogous phenomenon was put to 
use in a realistic model of flavor in \cite{NSflavor}.

In a similar way, strong dynamics in the hidden sector can easily
drive physics into regimes where hidden flavor symmetries are
naturally approximate, rather than being either violated at order one or
exact.  Such a structure will in turn can have observable
effects, such as
\begin{itemize}
\item
A large multiplet of nearly degenerate hidden states which cannot decay to
one another, and thus must decay via emission of standard model
particles;
\item 
Increased lifetimes for decays between hidden-sector ``generations,''
due to reduced intergenerational mixing angles.
\end{itemize}
Both of these can have a major impact on the phenomena observed.  


Unfortunately, these signatures are not unique to, nor are they
required by, strong dynamics, so they are not smoking guns for large
anomalous dimensions.  Also, I know of no new urgent experimental
issues raised by this possibility that are not already under
discussion.  In short, though interesting, this consequence of strong
coupling is something that theorists might consider further, but
appears not to be urgent for experimentalists preparing for the LHC.

\subsection{Effect of Strong Dynamics on Supersymmetry Breaking}

It is well known that strong dynamics in a supersymmetric theory can
suppress part or all of supersymmetry breaking in that sector; for an
application to phenomenology, see \cite{NSSUSY} (which
contains a review of the basic physics in an appendix.)  Thus although
supersymmetry beraking in our sector may have a characteristic scale
of 100s of GeV, the hidden sector may be much more supersymmetric than
our own.  The spectrum of hidden sector particles is not easy to
determine, but might show approximate degeneracies amongst bosons and
fermions.  Indeed it is even possible that the discovery of
supersymmetric particles, or convincing verification of supersymmetry
in nature, will occur through the hidden valley phenomenology.

For example, if in addition to hidden-hadron decays to the standard model
there are supersymmetric hidden-hadrons (analogous to R-hadrons within QCD
\cite{farrar}), then the decay patterns of the various states may
reveal degeneracies, and perhaps relations among decay rates or
branching fractions, that would be characteristic of a supersymmetric
theory.  One amusing example is that of $\phi$-onium resonances within
a still-conformal sector, a case mentioned in
Sec.~\ref{subsec:resonances}.  There one could imagine predicting,
and observing, effects from the supersymmetric generalization of positronium.

Let us make two immediate cautionary remarks.  This scenario is not
unique to strong dynamics.  Supersymmetry breaking could be suppressed
in the hidden sector by other, purely perturbative means.  For
example, in gauge mediation the messenger sector coupling to the
hidden sector may be absent, or the relevant messengers very heavy, such that
the standard model sector receives a louder message and gets a larger
array of soft masses than does the hidden sector.  Also, very weak
gauge couplings in the hidden sector would result in rather weak
supersymmetry-breaking effects.  Furthermore, even if
supersymmetry-breaking effects are large for most particles,
approximate symmetries or dimensional counting arguments can easily
make gaugino masses very small, leaving a hidden sector whose low
energy physics might accidentally be an almost exactly supersymmetric
Yang-Mills theory, with massless hidden gluons and with hidden gluinos
light compared to the hidden confinement scale.

The other problem is that degeneracies in the hidden sector spectrum
can arise from ordinary bosonic global symmetries, or simply from
kinematics.  For example, in a hidden pure Yang-Mills theory, the
masses of metastable hidden glueballs which can only decay to the
standard model sector lie within a narrow band, just by kinematics.
In a model with light pions, the observable pions may have very
similar masses because they transform as a simple multiplet under an
accidental global symmetry --- possibly enhanced through strong
dynamics, as discussed above in Sec.~\ref{subsec:flavor}.

Thus, it is certainly possible that a hidden sector will reveal the
first superpartners and other features of supersymmetry at the LHC,
but it is likely to be far from obvious.  Fortunately, it appears that
there are no special experimental challenges in the supersymmetric
limit, so that detection and analysis techniques do not need to be
specially tuned.  It appears to be enough to make careful but standard
measurements of multiple processes.

\subsection{Effect of Strong Dynamics on the Parton Shower}

By contrast, the last effect I want to consider is of both theoretical
and experimental importance.  In hidden valley models with strong
dynamics far above the mass gap, the parton shower can be very much
more powerful than in QCD.  This potentially can turn hard jets into
soft spray, and make events more spherical, with much higher
multiplicities, than in hidden valleys with weaker couplings above the
mass gap.  This possibility poses significant experimental challenges,
whose details cannot be known without further theoretical development.

Let us first see why the parton shower is the most sensitive
ingredient in a hidden valley to strong conformal or near-conformal
dynamics in the 1--1000 GeV range.  The first element in a hidden
valley model is the coupling between the two sectors via some
communicator, can be impacted by the strong dynamics, but indirectly,
in a way that might not readily be observed.  (There could be a large
impact on the line shape of an easily observed particle, however.)
The third element, a mass gap,
explicitly involves the breaking of conformal invariance, and we have
seen how the phenomenology can emerge in many different ways that are
not especially sensitive to the dimensions of operators in the
conformal regime.  What of the second ingredient, the multi-particle
production mechanism?  Cascade decays explicitly involve breaking of
conformal symmetry, and
hadronization is a violent violation of conformal symmetry.  But the
parton shower {\it is} conformal dynamics in action.

The very form of the parton shower is determined by anomalous
dimensions of operators, as noted above.  These are not the low-spin
low-dimension operators which we might couple to the standard model in
the Lagrangian, but Wilson lines, or in the crossed channel, the
high-spin high-dimension classical-twist-two operators which are
always present in a gauge theory.  

If these operators have small anomalous dimensions, the parton shower
is inefficient.  In QCD the quark in a $q\bar q$ production process
loses only a moderate fraction of its energy to gluon emission, much
of which is collinear with the quark, maintaining a coherent jet.  For
this reason, $q\bar q$ production gives predominantly two-jet events,
sometimes three.  In a weakly-coupled hidden valley, this jetty
structure in the hidden sector is typically retained, though blurred,
as the hidden hadrons decay into standard model particles.  An example
\footnote{This event was generated using the HVMC Monte Carlo, version
  0.5 \cite{HVMC}. This Monte Carlo is based on Pythia 6.4, combining
  its routines to simulate $Z'$ decay to the hidden sector, showering
  and hadronization within the hidden sector, and decay of
  hidden-sector hadrons back to standard model particles, which is
  followed by standard model showering and hadronization.} is shown in
\reffig{weakvjets}.

However, if the DGLAP operators have large anomalous dimensions, as
occurs in gauge theories at large $\lambda$ with string theory dual
descriptions, then the parton shower is very efficient.  (See
\cite{DIS} for a related effect in deep inelastic scattering.)  The
quarks and gluons shower so quickly, through both collinear and soft
emission, that they all become soft.  Soft emission is not collinear
with the initial quark, and so the original direction of the quark's
motion is largely forgotten.  Moreover, hard emission is also not
supressed as it is in weak coupling, so the production process itself
is likely to have several gluons along with the quark and
antiquark. Altogether, the events at strong coupling are likely to be
more spherical than jetty, though I do not know how to calculate the
fluctuations away from perfect sphericity. \footnote{The statements
made here are clearest at large $N$; there will be $1/N^2$ corrections
to these statements that deserve more study, in which color singlet
combinations of partons are radiated off to begin their own, separate
parton shower.  Whether these could introduce strong fluctuations in
the appearance of the events is not yet clear.}  Also, the number of
hidden hadrons will be larger, and their $p_T$ distribution much
softer than at weak coupling.  A guess at the appearance of such an
event is shown in \reffig{strongvjets}; again I emphasize this is a
guess. \footnote{A $Z'$ decay to a hidden quark and antiquark was
dressed with a Pythia parton shower of hidden gluons, in which the
showering rate was enhanced by fixing the gauge coupling at a large
enough value that collinear effects largely vanished by the scale of
hidden confinement.  The result may bear only a passing resemblance
to the actual physics of a strongly coupled theory, but this guess
appears physically reasonable, and should at least be
thought-provoking.}  Notice how the preferred axis and the strongly
variable calorimeter signal in \reffig{weakvjets} are absent in
\reffig{strongvjets}.

\makefig{0.4}{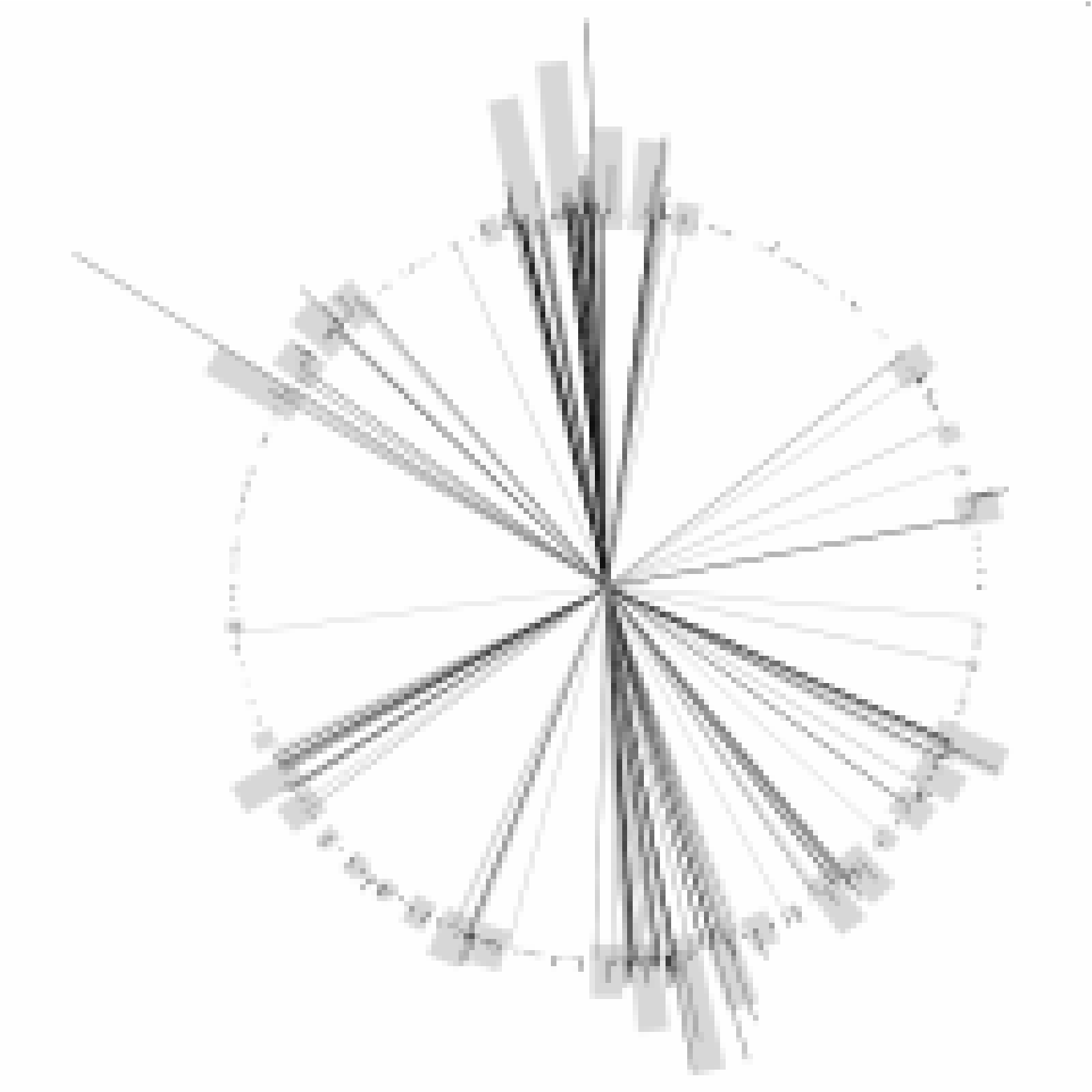}{An event (generated with HVMC 0.5) in which a
  3.2 TeV $Z'$ decays to 30 GeV v-pions (see \cite{HV1} for definitions)
  in a hidden sector which has a weak coupling above $\sim 100$ GeV.
  Notice the thrust axis is roughly vertical, though the events are by
  no means not pencil-like. The event shown contains roughly twenty
  bottom quarks and tau leptons.}

\makefig{0.4}{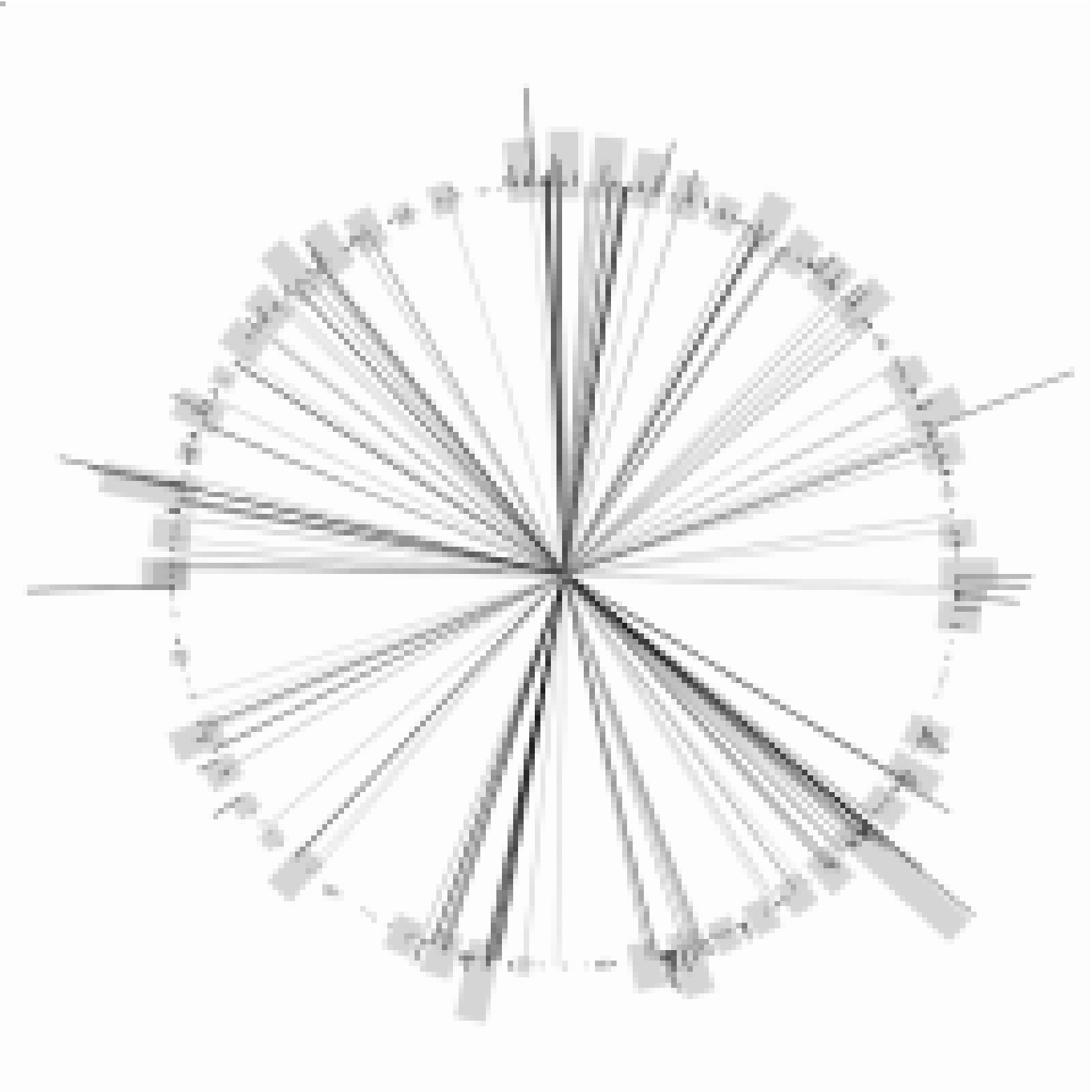}{An event (generated with HVMC 0.5) in which
a 3.2 TeV $Z'$ decays to 30 GeV v-pions (see \cite{HV1} for
definitions) in a hidden sector which has a strong coupling at all
energies.  
Notice the event
is now spherical. The event shown contains roughly fifty bottom quarks
and tau leptons.}

There are many challenges here which need to be addressed.  For
example, what will result from the application of jet algorithms to
these events?  How much energy must such events have such that they
are easy to see?  Are there any serious detector backgrounds which
could mimic such an effect?  What if most of particles produced are
long-lived; could the pattern recognition software become too confused
to operate properly?  And if such events are identified, what
questions can we ask of them to identify their source?  What
observables will most usefully allow analysis of such events?  And
finally, what is needed in formal theoretical development so that the
guesswork that goes into \reffig{strongvjets} can be replaced by
reliable prediction?

\section{Conclusion}

We have seen that unparticle models \cite{Un1, Un2,allunparticle} 
with mass gaps
\cite{unSteph, unFox, unQuiros} are typically examples of hidden valley
models \cite{HV1,HV2,HV3}.  Not all hidden valley models are
conformal; not all unparticle models have mass gaps which result in
particles with observable decays; but there are many ``hidden
valley/unparticle'' models with both features.  We have seen that in
these cases the dominant exclusive processes typically involve classic
hidden valley phenomenology: new neutral light particles with long
lifetimes, often produced with high multiplicity, along with
new Higgs decays (and also decays of LSPs and related particles,
though we did not study these here) to the hidden-valley particles.
Some of the hidden-valley particles themselves decay to standard model
particles, possibly with observably displaced vertices.  

To see this
required some tweaking of results in the literature, in particular,
clarifying what can and cannot be said about conformal symmetry
breaking using the language of unparticles \cite{unFox}, and adjusting the
narrow-tower model of unparticles \cite{unSteph}
to include various corrections
which though small make a huge change in the physical phenomena.  We
also saw that the assumption that the hidden sector physics is
invisible, as for example in \cite{unQuiros}, is often too pessimisstic;
the Higgs decays may not only be visible, they may be spectacular, as
in \cite{HV1,HV2}.

Still, in addition to this, inclusive studies using the
above-mentioned events, and certain rare processes, may be able to
reveal the special conformal kinematics associated to unparticles.  I
have also suggested that a dominant effect of strong coupling, and in
particular large anomalous dimensions for twist-two operators, is the
strong enhancement of the parton shower, and an increased sphericity,
higher multiplicity, and lower $p_T$ spectrum in high-energy hidden
valley events.  More work is needed to confirm this suggestion, and to
understand how strong dynamics, with or without conformal
invariance, can affect the phenomenology of hidden valleys in other
observable ways.

As in QCD, where both exclusive and inclusive questions have their
merits, and where approximate scale invariance plays an important role
in many analyses of QCD data, the study of a hidden sector should
proceed by combining information from both exclusive final states
(which, if visible, involve hidden valley phenomenology) and inclusive
final states (which are determined by unparticle computations in the
scale-invariant region, but not elsewhere.)  It may happen that all
exclusive phenomena are invisible, and then one can only discuss the
physics from the inclusive point of view.  But if the exclusive events
can be observed, they are typically more abundant, and are often more
spectacular, more easily separated from background, and more
informative.  They are also often very unusual, and in some cases may
pose serious challenges for the Tevatron and LHC experiments. These
challenges should be addressed (and in some contexts are already being
addressed)
in the immediate future.

In conclusion, conformal invariance and inclusive signatures are
powerful tools, but they are not powerful enough to address the
physics of conformal-symmetry breaking, where the full diversity and
complexity of quantum field theory may be found.  Two-point functions
of local composite operators, and the constraints of conformal
invariance, simply cannot capture the phenomenological richness so
often found in a hidden valley.

\

\

I am pleased to thank A.~De Roeck, J.L.~Feng, H.~Lubatti, M.~Graesser,
B.~ Mele, S.~Thomas, D.~Ventura and K.~Zurek for useful discussions.


\begin{thebibliography}{99}


\bibitem{HV1}
  M.~J.~Strassler and K.~M.~Zurek,
  Phys.\ Lett.\  B {\bf 651}, 374 (2007)
  [arXiv:hep-ph/0604261].

\bibitem{HV2}
  M.~J.~Strassler and K.~M.~Zurek,
  arXiv:hep-ph/0605193.

\bibitem{HV3}
  M.~J.~Strassler,
  arXiv:hep-ph/0607160.

\bibitem{HVWis}
  T.~Han, Z.~Si, K.~M.~Zurek and M.~J.~Strassler,
  arXiv:0712.2041 [hep-ph].

\bibitem{Un1}
  H.~Georgi,
  Phys.\ Rev.\ Lett.\  {\bf 98}, 221601 (2007)
  [arXiv:hep-ph/0703260].

\bibitem{Un2}
  H.~Georgi,
  Phys.\ Lett.\  B {\bf 650}, 275 (2007)
  [arXiv:0704.2457 [hep-ph]].



\bibitem{SeibergNAD}
  N.~Seiberg,
  Nucl.\ Phys.\  B {\bf 435}, 129 (1995)
  [arXiv:hep-th/9411149].

\bibitem{RS2}
  L.~Randall and R.~Sundrum,
  Phys.\ Rev.\ Lett.\  {\bf 83}, 4690 (1999)
  [arXiv:hep-th/9906064].

\bibitem{RS1}
  L.~Randall and R.~Sundrum,
  Phys.\ Rev.\ Lett.\  {\bf 83}, 3370 (1999)
  [arXiv:hep-ph/9905221].


\bibitem{HEIDI}
  J.~J.~van der Bij and S.~Dilcher,
  Phys.\ Lett.\  B {\bf 638}, 234 (2006)
  [arXiv:hep-ph/0605008].


\bibitem{TwinHiggs}
  Z.~Chacko, H.~S.~Goh and R.~Harnik,
  Phys.\ Rev.\ Lett.\  {\bf 96}, 231802 (2006)
  [arXiv:hep-ph/0506256].

\bibitem{FoldedSUSY}
  G.~Burdman, Z.~Chacko, H.~S.~Goh and R.~Harnik,
  JHEP {\bf 0702}, 009 (2007)
  [arXiv:hep-ph/0609152].


\bibitem{unFox}
  P.~J.~Fox, A.~Rajaraman and Y.~Shirman,
  Phys.\ Rev.\  D {\bf 76}, 075004 (2007)
  [arXiv:0705.3092 [hep-ph]].

\bibitem{unQuiros}
  A.~Delgado, J.~R.~Espinosa and M.~Quiros,
  JHEP {\bf 0710}, 094 (2007)
  [arXiv:0707.4309 [hep-ph]].

\bibitem{unSteph}
  M.~A.~Stephanov,
  Phys.\ Rev.\  D {\bf 76}, 035008 (2007)
  [arXiv:0705.3049 [hep-ph]].

\bibitem{CFW}
  S.~Chang, P.~J.~Fox and N.~Weiner,
  JHEP {\bf 0608}, 068 (2006)
  [arXiv:hep-ph/0511250].

\bibitem{CFW4}
  S.~Chang, P.~J.~Fox and N.~Weiner,
  Phys.\ Rev.\ Lett.\  {\bf 98}, 111802 (2007)
  [arXiv:hep-ph/0608310].



\bibitem{manyhiggs}
  J.~R.~Espinosa and J.~F.~Gunion,
  Phys.\ Rev.\ Lett.\  {\bf 82}, 1084 (1999)
  [arXiv:hep-ph/9807275].

\bibitem{NMSSM}
  U.~Ellwanger, J.~F.~Gunion, C.~Hugonie and S.~Moretti,
  arXiv:hep-ph/0305109,
  arXiv:hep-ph/0401228;
R.~Dermisek and J.~F.~Gunion,
  Phys.\ Rev.\ Lett.\  {\bf 95}, 041801 (2005)
  [arXiv:hep-ph/0502105].
U.~Ellwanger, J.~F.~Gunion and C.~Hugonie,
  JHEP {\bf 0507}, 041 (2005)
  [arXiv:hep-ph/0503203];
 R.~Dermisek and J.~F.~Gunion,
  arXiv:hep-ph/0510322.


\bibitem{JHU}
  L.~M.~Carpenter, D.~E.~Kaplan and E.~J.~Rhee,
  arXiv:hep-ph/0607204.

\bibitem{Morningstar}
  C.~J.~Morningstar and M.~J.~Peardon,
  Phys.\ Rev.\  D {\bf 60}, 034509 (1999)
  [arXiv:hep-lat/9901004];
  C.~Morningstar and M.~J.~Peardon,
  Nucl.\ Phys.\ Proc.\ Suppl.\  {\bf 83}, 887 (2000)
  [arXiv:hep-lat/9911003].


\bibitem{malda}
  J.~M.~Maldacena,
  Adv.\ Theor.\ Math.\ Phys.\  {\bf 2}, 231 (1998)
  [Int.\ J.\ Theor.\ Phys.\  {\bf 38}, 1113 (1999)]
  [arXiv:hep-th/9711200].


\bibitem{GKP}
  S.~S.~Gubser, I.~R.~Klebanov and A.~M.~Polyakov,
  Phys.\ Lett.\  B {\bf 428}, 105 (1998)
  [arXiv:hep-th/9802109].

\bibitem{WittenAdS}
  E.~Witten,
  Adv.\ Theor.\ Math.\ Phys.\  {\bf 2}, 253 (1998)
  [arXiv:hep-th/9802150].


\bibitem{WittenADSQCD}
  E.~Witten,
  Adv.\ Theor.\ Math.\ Phys.\  {\bf 2}, 505 (1998)
  [arXiv:hep-th/9803131].

\bibitem{ReyTheisenYee}
  S.~J.~Rey, S.~Theisen and J.~T.~Yee,
  Nucl.\ Phys.\  B {\bf 527}, 171 (1998)
  [arXiv:hep-th/9803135].

\bibitem{Sonnen}
  A.~Brandhuber, N.~Itzhaki, J.~Sonnenschein and S.~Yankielowicz,
  JHEP {\bf 9806}, 001 (1998)
  [arXiv:hep-th/9803263].


\bibitem{Csaki2}
  C.~Csaki, J.~Russo, K.~Sfetsos and J.~Terning,
  Phys.\ Rev.\  D {\bf 60}, 044001 (1999)
  [arXiv:hep-th/9902067].

\bibitem{Nonestar}
  J.~Polchinski and M.~J.~Strassler,
  arXiv:hep-th/0003136.

\bibitem{cascade}
  I.~R.~Klebanov and M.~J.~Strassler,
  JHEP {\bf 0008}, 052 (2000)
  [arXiv:hep-th/0007191].

\bibitem{allunparticle}
  K.~Cheung, W.~Y.~Keung and T.~C.~Yuan,
  Phys.\ Rev.\ Lett.\  {\bf 99}, 051803 (2007)
  [arXiv:0704.2588 [hep-ph]].
  K.~Cheung, W.~Y.~Keung and T.~C.~Yuan,
  Phys.\ Rev.\  D {\bf 76}, 055003 (2007)
  [arXiv:0706.3155 [hep-ph]];



  K.~Cheung, W.~Y.~Keung and T.~C.~Yuan,
  arXiv:0710.2230 [hep-ph].

\bibitem{multiun}
J.~L.~Feng, A.~Rajaraman and H.~Tu, in preparation.


\bibitem{lightscalars}
  M.~J.~Strassler,
  arXiv:hep-th/0309122.


\bibitem{Rizzo}
  T.~G.~Rizzo,
  JHEP {\bf 0710}, 044 (2007)
  [arXiv:0706.3025 [hep-ph]].


\bibitem{SchabWells}
  R.~Schabinger and J.~D.~Wells,
  Phys.\ Rev.\  D {\bf 72}, 093007 (2005)
  [arXiv:hep-ph/0509209].

\bibitem{BowenWells}
  M.~Bowen, Y.~Cui and J.~D.~Wells,
  JHEP {\bf 0703}, 036 (2007)
  [arXiv:hep-ph/0701035].




\bibitem{springloaded}
  A.~E.~Nelson and M.~J.~Strassler,
  Phys.\ Rev.\  D {\bf 60}, 015004 (1999)
  [arXiv:hep-ph/9806346].



\bibitem{quarkonium5th}
  S.~Hong, S.~Yoon and M.~J.~Strassler,
  JHEP {\bf 0404}, 046 (2004)
  [arXiv:hep-th/0312071].



\bibitem{KLN}
  J.~Kang, M.~A.~Luty and S.~Nasri,
  arXiv:hep-ph/0611322; J.~Kang and M.~A.~Luty, unpublished.


\bibitem{HW}Z.~Chacko, R.A.~Harnik and T.~Wizansky, in preparation.




\bibitem{hardscat}
  J.~Polchinski and M.~J.~Strassler,
  Phys.\ Rev.\ Lett.\  {\bf 88}, 031601 (2002)
  [arXiv:hep-th/0109174].

\bibitem{DIS}
  J.~Polchinski and M.~J.~Strassler,
  JHEP {\bf 0305}, 012 (2003)
  [arXiv:hep-th/0209211].

\bibitem{softwall}
  A.~Karch, E.~Katz, D.~T.~Son and M.~A.~Stephanov,
  Phys.\ Rev.\  D {\bf 74}, 015005 (2006)
  [arXiv:hep-ph/0602229].
\bibitem{Pomeron}
  R.~C.~Brower, J.~Polchinski, M.~J.~Strassler and C.~I.~Tan,
  arXiv:hep-th/0603115.



\bibitem{KarchKatz}
  A.~Karch and E.~Katz,
  JHEP {\bf 0206}, 043 (2002)
  [arXiv:hep-th/0205236].

\bibitem{wintersD7}
  M.~Kruczenski, D.~Mateos, R.~C.~Myers and D.~J.~Winters,
  JHEP {\bf 0307}, 049 (2003)
  [arXiv:hep-th/0304032].


\bibitem{Hovdebo:2005hm}
  J.~L.~Hovdebo, M.~Kruczenski, D.~Mateos, R.~C.~Myers and D.~J.~Winters,
  Int.\ J.\ Mod.\ Phys.\  A {\bf 20}, 3428 (2005).


\bibitem{Bander:2007nd}
  M.~Bander, J.~L.~Feng, A.~Rajaraman and Y.~Shirman,
  Phys.\ Rev.\  D {\bf 76}, 115002 (2007)
  [arXiv:0706.2677 [hep-ph]].


\bibitem{haduniv}
  S.~Hong, S.~Yoon and M.~J.~Strassler,
  JHEP {\bf 0604}, 003 (2006)
  [arXiv:hep-th/0409118].

\bibitem{Brodsky}
  G.~F.~de Teramond and S.~J.~Brodsky,
  Phys.\ Rev.\ Lett.\  {\bf 94}, 201601 (2005)
  [arXiv:hep-th/0501022].

\bibitem{rhouniv}
  S.~Hong, S.~Yoon and M.~J.~Strassler,
  arXiv:hep-ph/0501197.

\bibitem{BPST}
  R.~C.~Brower, J.~Polchinski, M.~J.~Strassler and C.~I.~Tan,
  arXiv:hep-th/0603115.



\bibitem{Erlich:2005qh}
  J.~Erlich, E.~Katz, D.~T.~Son and M.~A.~Stephanov,
  Phys.\ Rev.\ Lett.\  {\bf 95}, 261602 (2005)
  [arXiv:hep-ph/0501128].


\bibitem{NSSUSY}
  A.~E.~Nelson and M.~J.~Strassler,
  JHEP {\bf 0207}, 021 (2002)
  [arXiv:hep-ph/0104051].

\bibitem{NSflavor}
  A.~E.~Nelson and M.~J.~Strassler,
  JHEP {\bf 0009}, 030 (2000)
  [arXiv:hep-ph/0006251].


\bibitem{farrar}
  G.~R.~Farrar,
  arXiv:hep-ph/9408379.

\bibitem{GW}
  S.~Gopalakrishna, S. Jung and J.~D.~Wells,
in preparation.

\bibitem{Susskind}
  L.~Susskind,
  Phys.\ Rev.\  D {\bf 49}, 6606 (1994)
  [arXiv:hep-th/9308139].


\bibitem{SonSteph}
  D.~T.~Son and M.~A.~Stephanov,
  Phys.\ Rev.\  D {\bf 69}, 065020 (2004)
  [arXiv:hep-ph/0304182].

\bibitem{ACG}
  N.~Arkani-Hamed, A.~G.~Cohen and H.~Georgi,
  Phys.\ Rev.\ Lett.\  {\bf 86}, 4757 (2001)
  [arXiv:hep-th/0104005].



\bibitem{myunSteph}
M.J.~Strassler, unpublished.

\bibitem{HVMC}
M.J.~Strassler, unpublished.

\end{thebibliography}
\end{document}